\documentclass{aa}

\usepackage{graphicx}
\usepackage{txfonts}
\usepackage{color}

\newcommand{\ind}[1]{_{\mathrm{#1}}}

\newcommand{\Msun}{M_{\odot}}
\newcommand{\Rsun}{R_{\odot}}
\newcommand{\Teff}{\textsl{T}_{\mathrm{eff}}}
\newcommand{\TeffComp}{\textsl{T}_{\mathrm{eff},2}}
\newcommand{\Dnu}{\Delta \nu}
\newcommand{\DnuObs}{{\Delta \nu}_{\mathrm{obs}}}

\newcommand{\numax}{\nu_{\mathrm{max}}}
\newcommand{\numaxRef}{\nu_{\mathrm{max,ref}}}
\newcommand{\DnuRef}{{\Delta \nu}_{\mathrm{ref}}}
\newcommand{\TeffRef}{\textsl{T}_{\mathrm{eff,ref}}}
\newcommand{\Mdyn}{M_{\mathrm{dyn}}}
\newcommand{\Rdyn}{R_{\mathrm{dyn}}}

\newcommand{\Mpdyn}{M_{1,\mathrm{dyn}}}
\newcommand{\Rpdyn}{R_{1,\mathrm{dyn}}}
\newcommand{\Mpsis}{M_{1,\mathrm{seis}}}
\newcommand{\Rpsis}{R_{1,\mathrm{seis}}}
\newcommand{\Porb}{P_{\mathrm{orb}}}

\newcommand{\epsR}{\varepsilon_r}
\newcommand{\epsCrit}{\varepsilon_{\mathrm{crit}}}
\newcommand{\vmicro}{v_{\mathrm{micro}}}
\newcommand{\vsini}{v \sin i}
\newcommand{\vbroad}{v_{\mathrm{broad}}}
\newcommand{\FeH}{[\mathrm{Fe/H}]}
\newcommand{\Od}{\mathrm{O}_{2}}
\newcommand{\HdO}{\mathrm{H}_2\mathrm{O}}
\newcommand{\DPi}{{\Delta \Pi}}
\newcommand{\DPiAs}{{\Delta \Pi}_{1,\mathrm{asym}}}
\newcommand{\epsG}{\epsilon_{\mathrm{g}}}
\newcommand{\rhoSis}{\bar{\rho}_{\mathrm{seis}}}
\newcommand{\rhoDyn}{\bar{\rho}_{\mathrm{dyn}}}
\newcommand{\gSis}{g_{\mathrm{seis}}}
\newcommand{\gDyn}{g_{\mathrm{dyn}}}
\newcommand{\tcirc}{\tau_{\mathrm{circ}}}
\newcommand{\Dlne}{\Delta \ln e}
\newcommand{\Sph}{S_{\mathrm{ph}}}

\begin{document}

\title{Spectroscopic and seismic analysis of red giants in eclipsing binaries discovered by \textit{Kepler}}

\author{M.~Benbakoura\inst{\ref{inst:aim},\ref{inst:aim2}} \and P.~Gaulme\inst{\ref{inst:mps},\ref{inst:nmsu}} \and J.~McKeever\inst{\ref{inst:yale},\ref{inst:nmsu}} \and S.~Sekaran\inst{\ref{inst:kuleuven}} \and P.~G.~Beck\inst{\ref{inst:graz},\ref{inst:iac},\ref{inst:laguna}} \and F.~Spada\inst{\ref{inst:mps}} \and J.~Jackiewicz\inst{\ref{inst:nmsu}} \and S.~Mathis\inst{\ref{inst:aim},\ref{inst:aim2}} \and S.~Mathur\inst{\ref{inst:iac},\ref{inst:laguna},\ref{inst:ssi}} \and A.~Tkachenko\inst{\ref{inst:kuleuven}} \and R.~A.~Garc\'ia\inst{\ref{inst:aim},\ref{inst:aim2}}}

\institute{IRFU, CEA, Universit\'e Paris-Saclay, F-91191 Gif-sur-Yvette Cedex, France\\
  \email{rafael.garcia@cea.fr}\label{inst:aim}
  \and
Universit\'e Paris Diderot, AIM, Sorbonne Paris Cit\'e, CEA, CNRS, F-91191 Gif-sur-Yvette Cedex, France\label{inst:aim2}
  \and
Max-Planck-Institut f\"ur Sonnensystemforschung, Justus-von-Liebig-Weg 3, 37077 G\"ottingen, Germany\\
  \email{gaulme@mps.mpg.de}\label{inst:mps}
  \and
Department of Astronomy, New Mexico State University, P.O. Box 30001, MSC 4500, Las Cruces, NM 88003-8001, USA\label{inst:nmsu}
  \and
Deparment of Astronomy, Yale University, 52 Hillhouse Avenue, New Haven, CT 06511, USA\label{inst:yale}
  \and
Institute of Astronomy, KU Leuven, Celestijnenlaan 200D, B-3001 Leuven, Belgium\label{inst:kuleuven}
\and
Institute of Physics, Karl-Franzens University of Graz, NAWI Graz, Universit\"atsplatz 5/II, 8010 Graz, Austria.\label{inst:graz}
  \and
Instituto de Astrof\'isica de Canarias, 38200 La Laguna, Tenerife, Spain\label{inst:iac}
  \and
Departamento de Astrof\'isica, Universidad de La Laguna, 38206 La Laguna, Tenerife, Spain\label{inst:laguna}
  \and
Space Science Institute, 4750 Walnut Street Suite 205, Boulder, CO 80301, USA\label{inst:ssi}
}

   \date{Received Feb 20, 2020; accepted Jan 13, 2021}

\abstract
{
Eclipsing binaries (EBs) are unique targets for measuring accurate stellar properties and constraining stellar evolution models. In particular, it is possible to measure masses and radii at the few percent level for both components of a double-lined spectroscopic EB (SB2-EB). On the one hand, detached EBs hosting at least one star with detectable solar-like oscillations constitute ideal test objects to verify the ability of ensemble asteroseismology to derive stellar properties. On the other hand, the oscillations and surface activity of stars that belong to EBs offer unique information about the evolution of binary systems. This paper builds upon previous works dedicated to red giant stars (RG) in EBs; so far 20 known systems have been discovered by the NASA \textit{Kepler} mission. We report the discovery of 16 RGs in EBs, which are also from the \textit{Kepler} data, leading to a total of 36 confirmed RG stars in EBs from the original \textit{Kepler} mission. This new sample includes three SB2-EBs with oscillations, resulting in a total of 14 known SB2-EBs with an oscillating RG component. This sample also includes six close systems in which the RG display a clear surface activity and complete oscillation suppression. Based on dedicated high-resolution spectroscopic observations (Apache Point Observatory, Observatoire de Haute Provence), we focus on three main aspects. Firstly, from the extended sample of 14 SB2-EBs, we confirm that the simple application of the asteroseismic scaling relations to RGs overestimates masses and radii of RGs by about 15\,\% and 5\,\%. This bias can be reduced by employing either new asteroseismic reference values for RGs or model-based corrections of the asteroseismic parameters. Secondly, we confirm that close binarity leads to a high level of photometric modulation (up to 10\,\%) and a suppression of solar-like oscillations. In particular, we show that it reduces the lifetime of radial modes by a factor of up to 10. Thirdly, we use our 16 new systems to complement previous observational studies that aimed to constrain tidal dissipation in interacting binaries. We confirm the important role of the equilibrium tide in binary evolution, but we also identify systems with circular orbits despite relatively young ages, which suggests the need to explore complementary tidal dissipation mechanisms in the future. Finally, as a by-product, we report the measurements of mass, radius, and age of three M-dwarf companion stars.
}

\keywords{asteroseismology -- binaries: eclipsing -- binaries: spectroscopic -- stars: fundamental parameters -- stars: oscillations -- stars: evolution}

\maketitle

\section{Introduction}

Stellar astrophysics relies on precise and accurate measurements of stellar parameters. Over the past decade, high-precision photometric space missions such as CoRoT\footnote{CoRoT: Convection, Rotation et Transits plan\'etaires}, \textit{Kepler}, and TESS\footnote{TESS: Transiting Exoplanet Survey Satellite} \citep{Baglinetal2006,Boruckietal2010,Rickeretal2014} have led to significant progress in this direction. As a result of the unprecedented performance of these missions, tens of thousands of stars have been characterized through asteroseismology. Among the different types of pulsating stars \citep[e.g.,][]{Aertsetal2010}, solar-like oscillators are the most observed. The vast majority of solar-like oscillators observed by \textit{Kepler} are red giants (RGs) because most targets were observed at a 29.4244 min cadence, which does not allow us to detect oscillations of main-sequence (MS) and subgiant stars. In total, solar-like pulsations were detected for more than 20\,000 RG and 500 MS stars \citep{ChaplinMiglio2013,Honetal2019,GarciaBallot2019}.

The oscillation properties of a solar-like oscillator are unambiguously connected to its mass and radius. The simplest and most popular application of asteroseismology consists of comparing the global oscillation properties\footnote{The oscillation frequency at their maximum amplitude $\numax$ and the mean frequency spacing $\Dnu$ between consecutive overtones.} of a given star to those of the Sun and retrieving its mass and radius from asteroseismic scaling relations \citep{Brownetal1991,KjeldsenBedding1995}. For RGs, the evolutionary state (hydrogen shell or helium core burning) and core rotation can also be determined from the analysis of the mixed modes, which result from the interaction of gravity ($g$) modes in the radiative core and acoustic pressure ($p$) modes in the convective envelope \citep{Becketal2011,Becketal2012,Beddingetal2011,Mosseretal2011AA532,Mosseretal2012AA540,Deheuvelsetal2012,Deheuvelsetal2014,Deheuvelsetal2016,Gehanetal2018}.

Given the central role of solar-like oscillators as tests for stellar models, it is of critical importance to identify a set of benchmark stars. Such references are stars whose main physical properties are known with high precision, especially mass, radius, metallicity, and temperature. Eclipsing binaries (EBs) hosting at least one star with detectable solar-like oscillations constitute ideal test benches as they can be accurately characterized from the combination of both spectroscopic and photometric observations. It is possible to determine the mass and radius of each component of a double-line spectroscopic binary (SB2) from measurements of eclipse photometry and radial velocities. Hitherto, all solar-like oscillators belonging to EBs are RGs detected by the \textit{Kepler} mission \citep[][]{Hekkeretal2010, Gaulmeetal2013, Gaulmeetal2014, Becketal2014,Becketal2015AA573, Kuszlewiczetal2019,GaulmeGuzik2019}. So far, 11 wide SB2 EBs, including an oscillating RG, have been fully characterized with the help of ground-based radial velocity support \citep{Frandsenetal2013,Rawlsetal2016, Gaulmeetal2016, Brogaardetal2018,Themessletal2018}. We note that an equivalent number of RGs displaying oscillations have been detected in highly eccentric non-eclipsing binaries \citep[heartbeat systems, see][]{Becketal2014, Becketal2015AA573, Gaulmeetal2013, Gaulmeetal2014, Kuszlewiczetal2019}, but most do not show eclipses and are single-line spectroscopic binaries (SB1s).

In that context, oscillating RGs in SB2 EBs are perfect targets to verify the accuracy of masses and radii obtained with asteroseismology.
\citet{Gaulmeetal2016} find that the strict application of the scaling laws as defined by \citet{KjeldsenBedding1995} leads to overestimating masses and radii by about 25\% and 10\%, respectively.
Corrections of the asteroseismic scaling relations that were proposed at the time \citep{Kallinger_2010, Chaplin_2011c, Mosseretal2013, Whiteetal2011ApJ743,Epsteinetal2014,Sharmaetal2016, Guggenberger_2016} reduced these discrepancies to 15\% and 5\%.
Since then, new corrections to the seismic scaling relations have been proposed \citep[e.g.,][]{Rodriguesetal2017,Kallingeretal2018,SahlholdtSilvaAguirre2018}.
\citet{Huberetal2017} find excellent agreement between radii obtained through the asteroseismic scaling relations and those computed from TGAS parallaxes \citep{GaiaCollaborationetal2016}.
\citet{Brogaardetal2018} obtain additional radial velocities for three sources that were previously observed by \citet{Gaulmeetal2016} and show that the correction published by \citet{Rodriguesetal2017} improved the accuracy of the seismic estimates of mass and radius.
\citet{Themessletal2018} report the detection of an 11th pulsating red giant in an eclipsing binary and propose new ``solar'' reference values by showing that it leads to a better agreement. Theoretical studies based on RG in EBs have already been undertaken to constrain stellar models. For example \citet{Tanda_Li_2018} obtain a better agreement between the dynamical and asteroseismic masses by increasing the convective overshoot by about 14\,\%. Besides, \citet{Ball_2018} suggest that differences between the composition profile of the stellar models and the actual profile could explain part of the current mismatch between the dynamical and seismic masses and radii. More such benchmark systems are needed to go further.

In parallel to this, significant advances were made possible by \textit{Kepler} concerning observational studies of the tidal evolution of binary systems. First, the long-term photometric stability of \textit{Kepler} made the measurement of stellar rotation rates up to periods as long as half a year possible \citep{Ceillieretal2017,Gaulmeetal2020}. This opened the door to observational studies of tidal synchronization \citep{Lurieetal2017}. In particular, \citet{Gaulmeetal2014} observe that RGs in the most close-in eclipsing binary systems of their sample did not oscillate as expected. Among the 19 \textit{Kepler} RGs in EBs they studied, six displayed partially suppressed oscillations and the giants in the four closest systems showed no detectable oscillations.

\citet{Gaulmeetal2014} suggest that the mode suppression in binary systems originates from tidal interaction.
Red giant stars that normally rotate in several tens of days, sometimes in a few hundreds, are spun up during synchronization, which leads them to develop a stronger dynamo mechanism inside the convective envelope \citep[see also][]{Carlberg_2011,Auriereetal2015}.
The magnetic field generated by this dynamo reduces the excitation of pressure waves by partially suppressing convective motions.
Additionally, spots absorb part of the energy of the pressure modes, altogether leading to oscillation suppression.

\citet{Gaulmeetal2016} show that the absence of mode detection was not an observational bias but a true mode damping.
The shorter the orbital period and the closer the stars are, the weaker the oscillation modes become until complete mode depletion is reached.
These authors observe the damping of modes in systems with an orbital period shorter than approximately 40 days, for which most orbits are circularized and rotation periods are synchronized.
This suppression of mode power confirms the inverse relation between mode amplitude and magnetic activity observed for the Sun and other solar-like stars \citep[for more details, see, e.g.,][]{Garciaetal2010,Mathur_2019}.
This is further confirmed by \citet{Becketal2018}, who show that systems with short circularization timescales have suppressed oscillations.

The \textit{Kepler} light curves also enabled the study of the tidal circularization of binary systems by, for example, \citet{VanEylenetal2016} and \citet{Becketal2018}. While the former studied mainly MS stars, the latter focused on RGs. \citeauthor{Becketal2018} test the theoretical predictions of \citet{Zahn1966a,Zahn1966b,Zahn1966c,Zahn1977,Zahn1989}, \citet{Mathis2015}, and \citet{Galletetal2017AA604} for tidal evolution timescales by comparing them to observational constraints obtained on well-characterized \textit{Kepler} RGs belonging to binary systems. Doing so, Beck et al. continued the study led by \citet{VerbuntPhinney1995} by extending the sample to a total of 50 stars and pushing further the calculations. Beck et al. find that the most up-to-date theory of tidal dissipation mechanisms was able to predict the orbital configuration of the systems they considered. Consequently, extending their sample to find challenging cases or confirm their conclusion is necessary to reinforce the connection between observational and theoretical studies of tidal interactions. 

In this paper, we report the identification of 16 new RGs in EB systems from the \textit{Kepler} data, for which we lead a detailed photometric analysis, including oscillations properties, mixed dipolar modes, and surface rotation and activity.
We then present the results of three years of high-resolution spectroscopic observations, primarily obtained at the Apache Point Observatory, with complementary support from the Haute-Provence Observatory.
The objectives of the present paper are: 1) to test and improve the calibration of the asteroseismic scaling relations of RGs; 2) to study the properties of oscillations and surface activity\footnote{
  In this paper, "surface activity" refers to any signal in the light curve of a star that is not linked to stellar oscillations or eclipses.
} of stars under the influence of a close stellar companion; and 3) to provide new benchmark stars for tidal evolution theories.
We first present our sample and describe the photometric and spectroscopic data sets we use (Sect.~\ref{sec:data}).
We then detail the methods employed to perform the seismic, dynamical, and orbital analyses of each binary we considered (Sect.~\ref{sec:methods}).
Results and discussions about each objective of the paper are covered in Sect. \ref{sec:results}.
Finally, our conclusions are presented in Sect.~\ref{sec:conclusion}.

\section{Observations and data reduction}
\label{sec:data}

In total, we observed and analyzed the data of 17 stars: 16 were new dectections and one, KIC~4663623, had already been published by \citet{Gaulmeetal2016}.
We found these systems by crossing catalogs of variable stars \citep[e.g., the General Catalog of Variable Stars;][]{Samusetal2017}, EBs \citep[e.g., the \textit{Kepler} Eclipsing Binary Catalog;][]{Prsaetal2011,AbdulMasihetal2016}, and RGs \citep[e.g., the APOKASC catalog;][]{Pinsonneaultetal2014}.We used two different types of data: photometric light curves obtained by the \textit{Kepler} satellite and stellar spectra obtained at ground-based telescopes.
We processed the \textit{Kepler} data in three different ways to optimize the light curves to study the eclipses, stellar oscillations, and surface dynamics.
From the spectroscopic observations, we derived the atmospheric parameters, chemical composition, and radial velocities of the stars in our sample.

\subsection{\textit{Kepler} photometry}
\label{sec:kepler_photo}

We worked with the \textit{Kepler} target pixel files (TPF) to create our own aperture photometry and to produce the \textit{Kepler} light curves, following the same approach as \citet{Bloemen2014}. For each quarter, we defined a mask, that is, a set of pixels around the star, which maximizes the stellar signal while minimizing the noise and other sources of pollution. The resultant light curves of each quarter were then assembled following three different methods to obtain the seismic, eclipse, and rotation light curves. Light curves are used for three purposes: eclipse modeling, asteroseismic modeling, and surface rotation analysis. This entails modeling eclipse shapes by removing surface activity and measuring solar-like oscillations by removing eclipse features \citep{Gaulmeetal2013,Gaulmeetal2014,Gaulmeetal2016}.

For the seismic light curve, we first removed the eclipses using the timestamps from the \textit{Kepler} Eclipsing Binary Catalog \citep[KEBC;][]{AbdulMasihetal2016}. The light curves of each quarter were then corrected and stitched together following \citet{Garciaetal2011}. The first step of their method consists of correcting the light curves of each quarter for outliers in the point-to-point deviation function, sudden jumps of the mean value, and drifts. Then, to connect two light curves together at a quarter edge without discontinuity, two methods are considered: first, computing the mean value of the flux, at the end of Q(n) and at the beginning of Q(n+1), and adding or subtracting the difference to the Q(n+1) light curve; and second, fitting the end of Q(n) and beginning of Q(n+1) with first-order polynomials and linking the fits together at the midpoint of the gap between them. Both options are applied and that with the lowest $\chi^2$ is chosen. The light curves we obtained through this process still had long-period trends and periodic gaps owing to instrumental reasons and the eclipse removal. To reduce their impact in the time series, we applied a high-pass filter with a cutoff period of 20 days and filled the gaps using a multiscale discrete cosine transform of an in-painting algorithm \citep{Garciaetal2014inpaint,Piresetal2015}.

The preparation of light curves optimized to analyze the surface rotation follows the same steps as for asteroseismic analysis. However, a cutoff period of 20 days would not be suited to study the rotation of RGs because it would filter out most of the signal \citep{Garciaetal2014rotation}. Since the \textit{Kepler} quarters are approximately 90 days long, signals with a period longer than this duration are not reliable, thus we set the cutoff period to 80 days for the surface-rotation light curves.

For the study of the eclipses, we assembled the light curves in a different way. We computed the average of the flux at the end of Q(n) and at the beginning of Q(n+1) and we equalized them to ensure the best continuity in the time series. We then followed \citet[][see their Sect. 3.2]{Gaulmeetal2013}, who defined five categories of EBs, depending on their orbital period and surface activity, and described the methods they used to smooth them.

\begin{figure}
\includegraphics[width=\hsize]{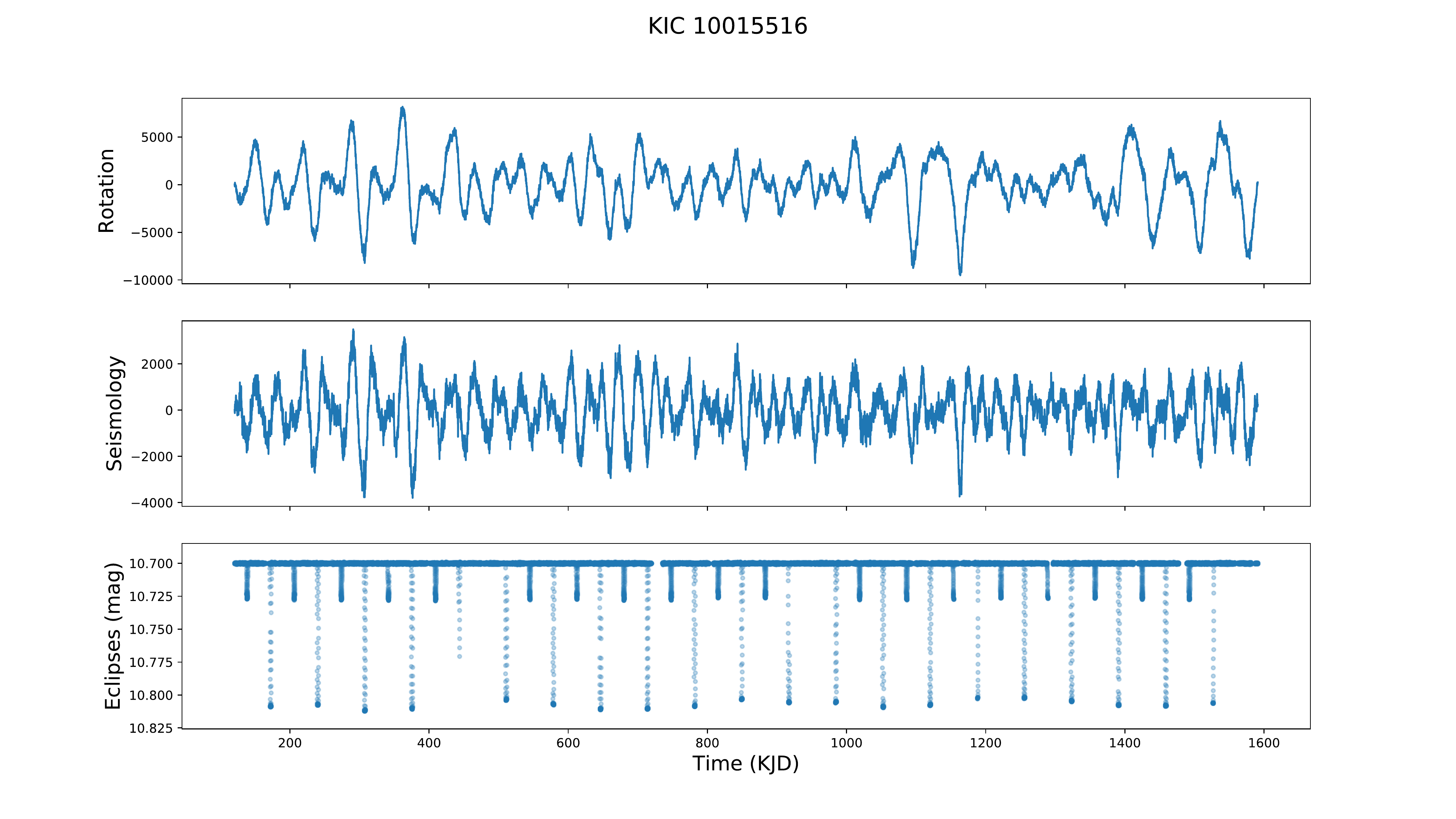}
\caption{Three light curves of KIC~10015516 computed from the \textit{Kepler} observations. Top panel: Light curve optimized to analyze the surface rotation. Middle panel: The same data optimized for the asteroseismic analysis. Bottom panel: The same for eclipse analysis.}
\label{fig:10015516_3_LCs}
\end{figure}

Figure~\ref{fig:10015516_3_LCs} shows the three light curves we produced for KIC~10015516. It should be noted that, in the middle panel, the modulations caused by the surface rotation are not totally erased in the seismic light curve. Suppressing these would imply to choose a shorter cutoff period for the high-pass filter. However, as explained in Sect.~\ref{subsec:seismo}, it is important to keep the signature of the granulation in the power spectrum to perform the seismic analysis, which is why we kept the value of 20 days for the cutoff period.

\subsection{Spectroscopic observations}

The majority of the spectra we used were obtained during a four-year campaign with the ARCES spectrograph \citep[][$R \sim 31\,000$]{Wangetal2003} mounted on the 3.5 m telescope at Apache Point Observatory (APO). We also used one other instrument, SOPHIE \citep[][$R \sim 40\,000$ in High-Efficiency mode]{BouchySophieTeam2006} at the 1.93 m telescope at the Haute-Provence Observatory (OHP). We used spectra from both instruments to compute the radial velocities, but only kept those obtained with ARCES to derive the atmospheric parameters.

\begin{table*}
\caption{Atmospheric parameters of the giant stars in our sample.}
\label{tab:atmosphericParam}
\centering
\begin{tabular}{clccccc}
\hline\hline
Label & KIC & $\Teff$  & $\log g$  & [Fe/H] & $\vmicro$    & $\vbroad$ \\
      &     &  (K)     &  (dex)    &  (dex) &  (km~s$^{-1}$) & (km~s$^{-1}$) \\
\hline
A & 4054905\tablefootmark{a} &  4790 $\pm$ 190  & ...               &  -0.72 $\pm$ 0.31  &  0.74 $\pm$ 0.75  &  57.6 $\pm$ 5.5  \\
B & 4360072                  &  5020 $\pm$ 210  &  2.79 $\pm$ 0.45  &  -0.14 $\pm$ 0.23  &  0.57 $\pm$ 0.58  &  38.0 $\pm$ 3.3  \\
C & 4473933                  &  4530 $\pm$ 220  &  2.89 $\pm$ 0.59  &  -0.41 $\pm$ 0.28  &  1.25 $\pm$ 0.80  &  49.8 $\pm$ 4.5  \\
D & 4663623\tablefootmark{b} &  4812 $\pm$ 92   &  2.7  $\pm$ 0.2   &  -0.13 $\pm$ 0.06  &  ...              &  ...             \\
E & 5193386\tablefootmark{c} &  4780 $\pm$ 100  &  3.284$\pm$ 0.014 &  -0.36 $\pm$ 0.06  &  ...              &  ...             \\
F & 5866138                  &  4960 $\pm$ 120  &  3.04 $\pm$ 0.24  &   0.05 $\pm$ 0.13  &  1.40 $\pm$ 0.33  &  5.4  $\pm$ 1.6  \\
G & 6307537\tablefootmark{a} &  4960 $\pm$ 240  & ...               &  -0.03 $\pm$ 0.43  &  2.0  $\pm$ 1.4   &  95.4 $\pm$ 8.1  \\
H & 6757558                  &  4590 $\pm$ 110  &  2.30 $\pm$ 0.31  &  -0.04 $\pm$ 0.13  &  1.24 $\pm$ 0.40  &  4.9  $\pm$ 1.9  \\
I & 7133286\tablefootmark{a} &  4500 $\pm$ 110  & ...               &  -0.60 $\pm$ 0.18  &  1.40 $\pm$ 0.58  &  31.1 $\pm$ 2.4  \\
J & 7293054\tablefootmark{a} &  4790 $\pm$ 160  & ...               &   0.11 $\pm$ 0.26  &  1.99 $\pm$ 0.86  &  26.1 $\pm$ 2.2  \\
K & 7768447                  &  4760 $\pm$ 160  &  2.96 $\pm$ 0.24  &   0.17 $\pm$ 0.25  &  3.0  $\pm$ 1.4   &  16.2 $\pm$ 2.8  \\
L & 8435232\tablefootmark{a} &  4460 $\pm$ 130  & ...               &  -0.15 $\pm$ 0.24  &  2.00 $\pm$ 0.85  &  30.9 $\pm$ 3.0  \\
M & 9153621                  &  4760 $\pm$ 190  &  2.34 $\pm$ 0.39  &  -0.35 $\pm$ 0.21  &  1.38 $\pm$ 0.55  &  16.6 $\pm$ 2.3  \\
N & 9904059                  &  4830 $\pm$ 160  &  2.83 $\pm$ 0.24  &   0.01 $\pm$ 0.26  &  1.05 $\pm$ 0.69  &  24.5 $\pm$ 2.3  \\
O & 10015516                 &  4830 $\pm$ 130  &  1.83 $\pm$ 0.22  &  -0.43 $\pm$ 0.15  &  1.27 $\pm$ 0.35  &  23.6 $\pm$ 1.4  \\
P & 10074700\tablefootmark{d}&  5070 $\pm$ 100  &  4.53 $\pm$ 0.20  &  -0.40 $\pm$ 0.10  &  ...              &  ...             \\
Q & 11235323\tablefootmark{a}&  4840 $\pm$ 200  & ...               &  -0.36 $\pm$ 0.25  &  2.6  $\pm$ 1.4   &  26.3 $\pm$ 3.0  \\
\hline
\end{tabular}
\tablefoot{For some systems, we added an extra step in the GSSP analysis or used different values.\\
\tablefoottext{a}{For these SB2 systems, we fixed the value of $\log g$ during the spectroscopic analysis to that provided by the dynamical analysis (see Table~\ref{tab:dyn_results_JKTEBOP}).}\\
\tablefoottext{b}{KIC~4663623: atmospheric parameters from \citet{Gaulmeetal2016}.}\\
\tablefoottext{c}{KIC~5193386: atmospheric parameters from APOGEE DR14 \citep{Abolfathietal2018}.}\\
\tablefoottext{d}{KIC~10074700: The spectroscopic analysis with GSSP did not converge; we used the values from the \textit{Kepler} Input Catalog, where no uncertainty is given. We thus adopt the typical error bar found for this sample.
}
}
\end{table*}

\subsubsection{Radial velocities}

From the reduced 1D spectra obtained at APO, we computed the radial velocities with the broadening-function technique \citep[BF;][]{Rucinski1999,Rucinski2002}. The fundamental hypothesis of this method is that the program spectra can be reconstructed as the convolution of a template spectrum and a broadening function. The latter can then be fitted by a Gaussian function whose central value is equal to the radial velocity of the star. For SB2 systems, the BF shows two peaks that correspond to the radial velocities of both components of the binary star.

For the BF template spectra, we used stellar atmosphere models generated by the PHOENIX BT-Settl code \citep{Allardetal2003}. These spectra were computed using solar abundances published by \citet{Asplundetal2009}. We used templates of MS stars with $\Teff$ = 5500~K or 5800~K to maximize the probability of detecting the radial velocity of the  companion. Moreover, the RG atmosphere models contain numerous lines that generate noise in the BF. As a consequence, even the radial velocity of the giant component is easier to determine using an atmosphere model of a MS star.

For each star in our sample, we defined a wavelength range on which we computed the BF. The portion of spectra we used are typically between 4\,500~Å and 5\,800~Å. The lower limit of the wavelength range cannot be less than 4\,300~Å because the APO spectra are too noisy below this value. We did not consider wavelengths above 6\,300~Å for two main reasons. First, we wanted to maximize the contribution of the companion in the spectrum used for the BF. Since our systems are composed of a RG and a hotter star, this implied we would only keep the bluest wavelengths possible. Second, telluric lines arising from the absorption of light by $\Od$ and $\HdO$ molecules in the terrestrial atmosphere pollute the stellar spectra. If we neglect the impact of the $\Od$ $\delta$ band, the first absorption lines of this molecule are located around 6\,288~Å, which corresponds to the $\Od$ $\gamma$ band \citep{NewnhamBallard1998}.

Prior to the BF computation, the spectra had to be evenly spaced in velocity. For each star in the sample, we used a specific step size in velocity, typically around 1~km~s$^{-1}$. The wavelength range and the velocity step size could vary from one star to another. For example, when the companion was not luminous enough, we considered bluer wavelength ranges to maximize our chances to detect its radial velocity.

For the SOPHIE data, an existing pipeline allows us to determine the radial velocities from the spectra. More details on the data reduction and radial velocity measurement from the SOPHIE spectra are given in, for example, \citet{Santerneetal2011AA528,Santerneetal2011AA536}, and references therein. The radial velocities we used in this work are compiled in Table~\ref{tab:radial_velocities}.

\subsubsection{Atmospheric parameters}

From our high-resolution spectra, we determined the stellar atmospheric parameters: the effective temperature $\Teff$, logarithmic surface gravity $\log g$, metallicity $\FeH$, microturbulent velocity $\vmicro$, and the total non-microturbulent velocity broadening\footnote{We quote the value for the total non-microturbulent velocity broadening $\vbroad$ instead of the more conventional projected rotational velocity along the line-of-sight $\vsini$, since most of the velocity broadening of the spectral lines of RGs is a result of macroturbulence, which is degenerate with the rotational broadening.}, $\vbroad$. Given that the targets are all EBs, we cannot directly derive the parameters from the optical spectra. We wanted to make use of all the spectra we had (about 15 spectra per star). However, we first needed to deal with the fact that each observed spectrum is a combination of the spectra of two stars that are in motion.

For our SB2 systems, we disentangled our observed spectra with the FDBinary software described by \citet{Ilijicetal2004}. Then, we analyzed the resulting spectra of the giant component with the Grid Search in Stellar Parameters (\textsc{GSSP}) code \citep{Tkachenko2015}.\ The \textsc{GSSP} code works by fitting synthetic spectra, generated using the \textsc{SynthV} radiative transfer code \citep{Tsymbal1996} combined with a grid of atmospheric models from the \textsc{LLmodels} code \citep{Shulyak2004}, to our disentangled spectra. For our SB1 systems, the spectra of both components could not be disentangled owing to the small light contribution of the secondary component, so we analyzed the spectra assuming that the contribution of the companion was independent of the wavelength. To estimate this contribution, we used the light ratios determined by the orbital-parameter analysis (see section~\ref{subsec:dynamical}). This method differs from that of \citet{Gaulmeetal2016}, who had determined the atmospheric parameters of their giants from averaged spectra without taking into account the presence of the companion.

The atmospheric parameters derived from the spectra are given in Table~\ref{tab:atmosphericParam}.
The effective temperatures of the secondary components are given later in the paper (see Table~\ref{tab:massesAndRadii}). These temperatures were obtained through our JKTEBOP dynamical analysis (see Sect.~\ref{subsec:dynamical}).

To ensure the reliability of these values, we compared stars in our sample with the results of the APOGEE \citep{Gunnetal2006,Blantonetal2017,Majewskietal2017,Wilsonetal2019} DR14 \citep{Abolfathietal2018,Holtzmanetal2018} and DR16 \citep{Ahumadaetal2020} catalogs where data was available.
The APOGEE values were measured from infrared spectra through an automated analysis.
The discrepancies between ARCES and APOGEE temperatures were on the order of 50~K, within the uncertainties, except for KIC~7293054, 7768447, and KIC~10015516.
For KIC~7293054 and KIC~10015516, we obtained a temperature smaller than the APOGEE value by 200~K.
Since the luminosity ratios of these binaries are among the highest in our sample, the contribution
of the secondary component to the total light of the system has most probably
caused the automatic APOGEE pipeline to overestimate the effective temperature
of the giant ($L_2/L_1$ = 10.8\,\% for KIC~10015516; see Table~\ref{tab:dyn_results_JKTEBOP}). For KIC~7293054, the luminosity ratio is unknown, but the companion is one of the most visible in the optical spectra of the system. For KIC~7768447, we measured a temperature higher than the APOGEE value by 160~K, which is not incompatible with the measurement uncertainty we got from the GSSP analysis. For the metallicity, ARCES and APOGEE values agreed within the measurement uncertainties, with a discrepancy on the order of 0.1 to 0.2\,dex, except for KIC~6757558 and KIC~10015516. For the surface gravity, the discrepancies between the two sets of measurements were also below the uncertainties reported in Table~\ref{tab:atmosphericParam} (typically on the order of 0.1 to 0.2\,dex), except for KIC~7293054 and KIC~10015516. While the comments we made on the effective temperature apply for the latter two binaries, the situation is different for KIC~6757558 since the companion is a MS M dwarf (Table~\ref{tab:massesAndRadii}). We note that our asteroseismic results suggest that $\log g$ for this giant is 3.01, as can be seen in Table~\ref{tab:massesAndRadii}, which is closer to the APOGEE value of 2.98. We decided, however, to report our parameters instead of those of APOGEE in Table~\ref{tab:atmosphericParam}, primarily to follow the most systematic approach possible and because the parameters from APOGEE data releases were obtained using an automatic pipeline that does not take binarity into account.

Finally, if we did not perform any abundance analysis for individual elements, we looked for a few special features in the optical spectra. We first checked whether some targets show lithium absorption, given that Li enrichment in giants could be caused by tidal interactions in binary systems \citep{Casey_2019}. It appears that none of the RGs in our study shows particular Li absorption at 6707.7\AA. Secondly, we ran a preliminary analysis of the Ca H \& K emission at 3968.5 and 3933.7 \AA, respectively, which are tracers of magnetic activity. The ARCES spectrometer is not optimized for monitoring this spectral feature because it is at the very edge of its spectral domain. Nevertheless, Ca H \& K emission is obvious for a few systems: KICs 4473933, 5193386, 6307537, 7133286, 8435232, 10015516, and 11235323. The seven systems correspond to those with detectable rotational modulation in the \textit{Kepler} light curves (Sect. \ref{subsec:rotation}), which is also an indicator of strong magnetic activity. At last, the shortest-orbit system KIC 11235323 also displays variable H$_\alpha$ emission at 6562.8 \AA. We added comments in Table \ref{tab:massesAndRadii} to report these peculiarities.

\begin{table*}
\caption{Asteroseismic properties and surface magnetic activity. The oscillation frequency at maximum amplitude $\numax$ and observed large frequency spacing $\Dnu_{\mathrm{obs}}$ are expressed in $\mu$Hz. The dipole mode period spacing $\Delta\Pi_{1,\mathrm{asym}}$ is expressed in seconds and $q$ stands for the coupling factor of the mixed modes. The parameters $A\ind{max}$ and $\Gamma\ind{max}$ are the oscillation amplitude and width of the closest $l=0$ mode to $\numax$. Rotational periods $P_{\mathrm{rot}}$ are expressed in days and photometric activity index $S\ind{ph}$ in percent.}
\label{tab:seismicParam}
\centering
\begin{tabular}{lccccccccc}
\hline\hline
KIC      & $\numax$ & $\Dnu_{\mathrm{obs}}$             &$\Delta\Pi_{1,\mathrm{asym}}$&$q$& $A\ind{max}$ & $\Gamma\ind{max}$ & $P_{\mathrm{rot}}$ & $S\ind{ph}$ \\
         & ($\mu$Hz)              &  ($\mu$Hz)             & (s)      &      & (ppm) & ($\mu$Hz) &    (days)  & (\%)\\
\hline
4054905 & $  48.13 \pm   0.21 $ & $   5.42 \pm   0.01 $ & $159.5$ & $0.10$ & 78.1 & 0.09 & ... & 0.05 \\ 
4360072 & $  31.81 \pm   0.06 $ & $   3.90 \pm   0.01 $ & $392.2$ & $0.45$ & 101.0 & 0.34 & ... & 0.05 \\ 
4473933 & ... & ... & ... & ... & ... & ... & $ 68.4 \pm  6.1 $ & 2.05 \\ 
4663623 & $  54.09 \pm   0.24 $ & $   5.21 \pm   0.02 $ & $363.6$ & $0.10$ & 57.7 & 0.16 & ... & 0.03 \\ 
        &                       &                       & $76.7^\dagger$  &   &   &   &   & \\
5193386 & ... & ... & ... & ... & ... & ... & $ 25.6 \pm  2.0 $ & 1.72 \\ 
5866138 & $  83.71 \pm   0.46 $ & $   7.25 \pm   0.01 $ & $272.4$ & $0.25$ & 24.5 & 0.20 & ... & 0.03 \\ 
6307537 & ... & ... & ... & ... & ... & ... & $ 77.5 \pm  5.9 $ & 3.00 \\ 
6757558 & $ 129.38 \pm   0.28 $ & $  11.28 \pm   0.01 $ & $80.1$ & $0.15$ & 52.0 & 0.65 & ... & 0.03 \\ 
7133286 & ... & ... & ... & ... & ... & ... & $ 38.0 \pm  2.8 $ & 3.60 \\ 
7293054 & $  42.58 \pm   0.27 $ & $   4.32 \pm   0.01 $ & ... & ... & 72.7 & 0.10 & ... & 0.05 \\ 
7768447 & $  57.79 \pm   0.52 $ & $   5.89 \pm   0.02 $ & ... & ... & 92.2 & 0.34 & ... & 0.06 \\ 
8435232 & ... & ... & ... & ... & ... & ... & $ 48.1 \pm  3.4 $ & 4.63 \\ 
9153621 & $  38.22 \pm   0.30 $ & $   4.28 \pm   0.01 $ & ... & ... & 76.5 & 0.18 & ... & 0.08 \\ 
9904059 & $ 140.61 \pm   0.45 $ & $  11.91 \pm   0.01 $ & $79.8$ & $0.15$ & 33.9 & 0.44 & ... & 0.03 \\ 
10015516 & $  66.85 \pm   0.67 $ & $   5.90 \pm   0.01 $ & $294.5$ & $0.25$ & 20.0 & 0.30 & $ 65.7 \pm  6.2 $ & 0.24 \\ 
10074700 & $ 232.00 \pm   2.00 $ & $  18.37 \pm   0.02 $ & ... & ... & 26.4 & 0.41 & ... & 0.04 \\ 
11235323 & ... & ... & ... & ... & ... & ... & $ 23.9 \pm  1.9 $ & 1.72 \\ 
\hline
\end{tabular}
\tablefoot{For KIC~4663623, we used the values of $\Dnu$ and $\numax$ from \citet{Gaulmeetal2016}. $^\dagger$The period spacing $\Delta\Pi_1 = 76.7$ s was estimated for the \citet{Gaulmeetal2014} paper, but was omitted in the manuscript.}
\end{table*}

\section{Methods}
\label{sec:methods}
\subsection{Asteroseismic analysis}
\label{subsec:seismo}

Asteroseismology provides information on the properties of stellar interiors. In this work, we focus on three aspects: estimating stellar masses and radii from the asteroseismic scaling relations, measuring the amplitude and lifetime of radial modes, and determining the stellar evolutionary states from the frequency distribution of the mixed dipolar $p$ and $g$ modes.

\subsubsection{Global parameters}
\label{sect:globa_param}

We computed the global seismic parameters with the method employed by \citet{Gaulmeetal2016}. In this process, we started by fitting the power spectra with a sum of two super-Lorentzian functions for the convection \citep[e.g.,][]{2011ApJ...741..119M,Kallingeretal2014}, a Gaussian envelope for the modes, and a constant function for the white noise. The frequency of maximum power $\numax$ was determined as the central value of the Gaussian function. Then, we computed an initial $\Dnu$ from the autocorrelation function of the frequency range of the power spectrum containing oscillation modes as described by \citet{MosserAppourchaux2009}. The large frequency spacing obtained through this method may be biased because of stellar and stochastic realization noises, as was pointed out by \citet{Hekkeretal2009,Hekkeretal2011AA525}, \citet{Huberetal2009}, \citet{Mathuretal2010}, and \citet{MosserAppourchaux2009}. Thus, we used the universal pattern of RGs introduced by \citet{Mosseretal2011AA525} to correct it. The principle of this method is to compare the measured oscillation frequencies to a theoretical law, the so-called universal pattern, predicting the variations of these frequencies as a function of $\Dnu$ and the radial order.

To ensure the reliability of our measurements, we performed a second independent analysis using the \textit{A2Z} pipeline developed by \citet{Mathuretal2010}. This method is based on three packages run consecutively: p-mode range search, background fitting, and characterization of the p-mode envelope. First, the large frequency separation $\Dnu$ is determined through the analysis of the Fourier transform of small chunks of the power spectrum in a continuous blind way. The resultant value is used to estimate the frequency of maximum oscillation power $\numax$ and the p-mode range, i. e., the interval of frequency where the power excess due to stellar oscillations is visible. Then, the background of the power spectrum is fitted by a Gaussian function in the same manner as above. This last step provides a second measurement of $\numax$.
The results of the two methods agree within 2\% for all targets with only two exceptions: the values of $\Dnu$ of KIC~10015516 and $\numax$ of KIC~5866138 given by A2Z were 6\% and 8\% greater, respectively, than those given by our reference pipeline.
We decided to use the seismic parameters provided by the same method as \citet{Gaulmeetal2016} because the associated uncertainties are smaller.

The reference values are given in Table~\ref{tab:seismicParam}.
It should be noted that we report the observed value of $\Dnu$.
Since all the oscillating stars considered in this work are RGs, the asymptotic frequency spacing can be obtained by multiplying this value by 1.038, as shown by \citet{Mosseretal2013}.

\citet{Sekaranetal2019} simulate that if $\numax$, for a binary with one or two oscillating components differs from that of a single star. If this were the case, oscillating RGs in binaries would not be suited for calibrating the scaling relations. The authors show that the presence of a companion star does not generate systematic biases. However, stellar multiplicity as well as otherwise unrelated contaminants affect the signal-to-noise of the oscillations with respect to the background due to photometric dilution of the oscillation signal, as discussed by, for example, \citet{Becketal2018AA}.

\subsubsection{Individual oscillation frequencies}
\label{sect:peak_bagging}
To study the suppression of oscillations in the closest systems (Sect. \ref{subsub:nonOscRG}), we need to determine the properties of the radial ($l=0$) modes. We note that dipolar ($l=1$) modes are not suitable for this because they are very often split into multiple components by interference with inner $g$ modes, which makes the measurement of their amplitude more challenging. Moreover, it has been shown that $l=1$ modes are often depleted in RG stars \citep[e.g.,][]{2014A&A...563A..84G,Fuller_2015,Stello_2016, Mosser_2017, Gaulmeetal2020}, which would add another dimension to the problem.

We fit the oscillation spectra with a maximum a posteriori method, which was originally described by \citet{Gaulme_2009} and regularly updated since then. The updates mainly consist of an improved approach for mode identification. The routine starts by fitting the background noise of the power spectral density (PSD) as described in Sect. \ref{sect:globa_param}. Then a null hypothesis statistical significance test is performed around $\nu\ind{max}$ to detect significant peaks with respect to the local background model. Next, the expected $l=0$ frequency pattern is computed according to the solar-like oscillation universal pattern ``UP'', which was originally developed to analyze the oscillations of RGs by \citet{Mosseretal2011AA525} and extended to MS stars by \citet{Mosseretal2013}, as follows:
\begin{eqnarray}
\nu\ind{up,l=0} &=& \left[n + \epsilon + \frac{\alpha}{2} \left(n - \frac{\nu\ind{max}}{\Delta\nu}\right)^2\right] \Delta\nu \label{eq:echelle_mosser_l0} ,\\
\nu\ind{up,l=1} &=& \left[n + \frac{1}{2} + \epsilon - d_{01} + \frac{\alpha}{2}\left(n - \frac{\nu\ind{max}}{\Delta\nu}\right)^2\right] \Delta\nu \label{eq:echelle_mosser_l1} ,\\
\nu\ind{up,l=2} &=& \left[n + 1 + \epsilon - d_{02} + \frac{\alpha}{2}\left(n - \frac{\nu\ind{max}}{\Delta\nu}\right)^2\right] \Delta\nu \label{eq:echelle_mosser_l2}
,\end{eqnarray}
where  $\alpha = 2\times0.038/n\ind{max}$ for RGs and $\alpha = 2\times0.57/n\ind{max}^2$ for MS stars \citep{Mosseretal2013}. The separations relative to $\Delta\nu$ are measured to be about $d_{01} = (-0.056 -0.002\log \Delta\nu)$ and $d_{02} = (0.131 -0.033 \log \Delta\nu)$ \citep{Mosseretal2011AA525}.

The code then cuts the spectrum into $\pm0.6 \Delta\nu$ wide ranges around the expected $l=0$ position and it picks the highest peak, assuming that it corresponds with the actual position of $l=0$. It eventually searches for the mean position of the $l=1$ ridge with the assumption of no mixed modes, which is expected to be separated by $d_{01}$. The routine then looks for the actual closest significant peaks to the expected ridge, which in other words allows for only one $l=1$ mixed mode per overtone. As a final step it looks for $l=2$ modes by searching for significant peaks near the $l=0$ ridge in a separate H0 testing.

Once the modes are identified in terms of degree and order, each individual frequency is fitted with a Lorentzian function. Since our goal is to measure the properties of radial modes, we do not include rotational splitting or mixed modes: one Lorentzian is used to model each $(n,l)$ mode. Optimization is performed as described by \citet{Gaulme_2009}, with loose priors on the frequency positions based on the mode identification process above. The amplitude and mode width of the $l=0$ mode closest to $\numax$ is reported for each oscillating RG in Table \ref{tab:seismicParam}.


\subsubsection{Mixed modes}
\label{subsub:mixed_modes}
In stars with an inner radiative zone and an outer convective envelope, pressure (p-) waves propagate in the whole interior and gravity (g-) waves propagate in the radiative zone. These waves carry information on the interior layers in which they travel. In RGs, the frequency range of the p modes partially coincides with that of the g modes so that the two types of modes interact. The modes resulting from this interaction are called \textit{mixed} modes: they are tracers of the physical conditions in the stellar core \citep[e.g.,][]{Becketal2011,Beddingetal2011}.

In particular, mixed modes are used to determine the evolutionary status of a RG through the measurement of the period-spacing of gravity modes. As shown by \citet{Tassoul1980}, gravity modes of equal degree can be considered evenly spaced in period in the asymptotic regime. We denote $\DPi_{l}$ the period-spacing of gravity modes of degree $l$. Since modes of degree $l = 1$ travel deeper into the stellar interior than those of higher degree, studies of mixed modes have focused on determining the value of $\DPi_1$ \citep[e.g.,][and references therein]{Buysschaertetal2016}. \citet{Beddingetal2011} showed that this period spacing allowed one to distinguish between hydrogen-shell-burning stars, located on the red giant branch (RGB) of the Hertzsprung-Russell diagram and helium-core-burning stars of the horizontal branch, also called red clump (RC).

To determine $\DPi_1$, we implemented a method similar to that described by \citet{Buysschaertetal2016}. They compared two approaches: one based on a Lorentzian fitting of the observed period spacing $\Delta P$ and the other based on an asymptotic relation derived by \citet{Mosseretal2012AA540}. In the latter approach, the authors fitted three parameters, the asymptotic period spacing $\DPiAs$, the coupling factor $q$, which quantifies if the mixed modes are dominated by pressure or by gravity, and the phase offset $\epsG$, through a $\chi^2$ minimization. In this work, we used the asymptotic-relation technique with $\epsG$ fixed to 0 to determine the value of $\DPiAs$ and $q$.

As shown in the last two columns of Table~\ref{tab:seismicParam}, mixed modes were not detected for KIC~7293054, KIC~7768447, KIC~9153621, and KIC~10074700\footnote{
Given that we only used $\DPi_1$ to determine the evolutionary state of our stars, we did not need to compute measurement uncertainties for this quantity \citep[for more details, see, e.g.,][]{Beddingetal2011}.
}.
Two of the other stars, KIC~6757558 and KIC~9904059, are burning hydrogen in a shell.
The other five giants are burning helium in their core, as indicated by their period spacing larger than 100~s \citep{Beddingetal2011}.

\subsection{Surface activity and rotation}
\label{subsec:rotation}

We analyzed the surface rotation and magnetic activity following the methods described by \citet{Garciaetal2014rotation} and \citet{Ceillieretal2016}.
This method compares the periodicities obtained by three independent techniques applied to the light curve.
The first technique is the periodogram.
In active stars\footnote{
  In this case, "active" refers to stars with starspots.
  This phenomenon is described in greater detail in Sect.~\ref{subsub:nonOscRG}, where we study the link between tidal interactions, surface activity, and oscillation suppression.
}, the highest peak of the power spectrum of the time series is often the signature of the rotation due to the variability associated with starspots.
These spots induce important modulations in the light curve that often dominate the low-frequency part of the spectrum with periods of tens of days for RGs \citep{Ceillieretal2016}.
However, instrumental noise and drifts can also introduce peaks in the spectrum and a more detailed study of the nature of these low-frequency peaks is required.
This is why we also used a time-frequency analysis.
The principle of this method is to compute a two-dimensional wavelet power spectrum (WPS) containing the evolution with time of each periodicity \citep{TorrenceCompo1998}.
This method allows us to distinguish sustained periodic signals such as rotation from intermittent signals that only last for a few days, which could be induced by instrumental perturbations.
From the WPS, we computed the global WPS (GWPS), which corresponds to the WPS integrated along the temporal dimension.
Finally, we computed the autocorrelation of the light curve to have a third estimation of the rotation period.
We also computed a composite spectrum, which corresponds to the GWPS multiplied by the autocorrelation function \citep{Ceillieretal2016}.
Because instrumental noise affects each technique differently, combining these techniques increases the reliability of our detection \citep[for a benchmark comparison of various methods; see][]{Aigrainetal2015}.

\begin{figure}
\includegraphics[width=\hsize]{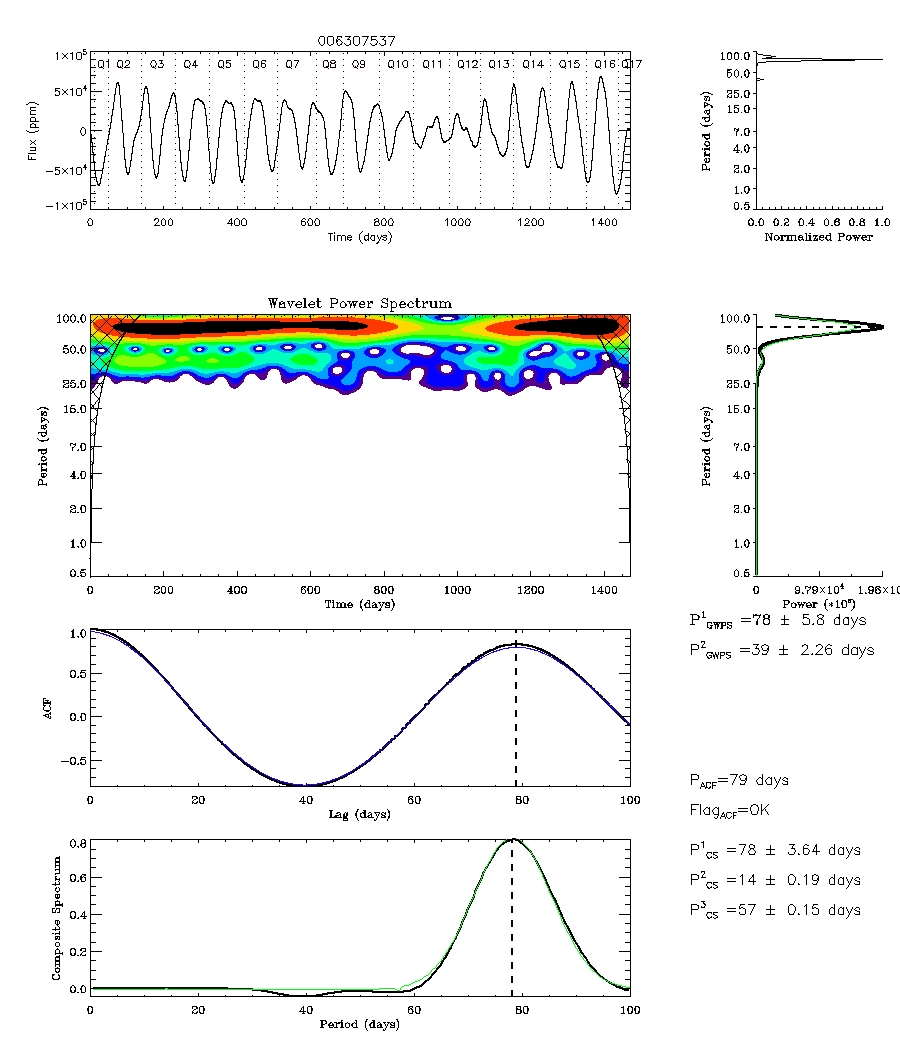}
\caption{Analysis of the rotation of KIC~6307037. \textit{Left:} From top to bottom, \textit{Kepler} light curve, WPS, autocorrelation function, and composite spectrum. \textit{Right:} From top to bottom, power spectrum, global WPS, and results of the analysis.}
\label{fig:006307537_rotation}
\end{figure}

Figure~\ref{fig:006307537_rotation} illustrates the different techniques applied to the light curves to one star in our sample, KIC 6307537. In this case, all techniques agree on the rotation period. Whereas the low-frequency periodogram peaks at a period around 80 days, the WPS shows that this periodicity is sustained over the whole length of the observation, which reinforces this detection. As can be seen in the WPS, periodicities in the stellar signal do not show up as sharp peaks at a well-defined value. Periodicities rather appear as scattered distributions of rotational velocities, which result in a Gaussian distribution once projected on the frequency axis. This dispersion is due to complex, poorly constrained effects such as the variable size of starspots and differential rotation associated with the presence of spots at several latitudes. Even though the uncertainty on the central periodicity of the Gaussian observed in the GWPS is very small, the width of this distribution has to be taken into account to enable fine analysis of the stellar rotation. This is why the provided uncertainty corresponds to the half width at half maximum (HWHM) of a Gaussian fit to the highest peak in the GWPS instead of the statistical error (Table~\ref{tab:seismicParam}).

\subsection{Dynamical analysis}
\label{subsec:dynamical}

\begin{figure*}
\centering
\begin{tabular}{ccc}
\includegraphics[width=.3\linewidth]{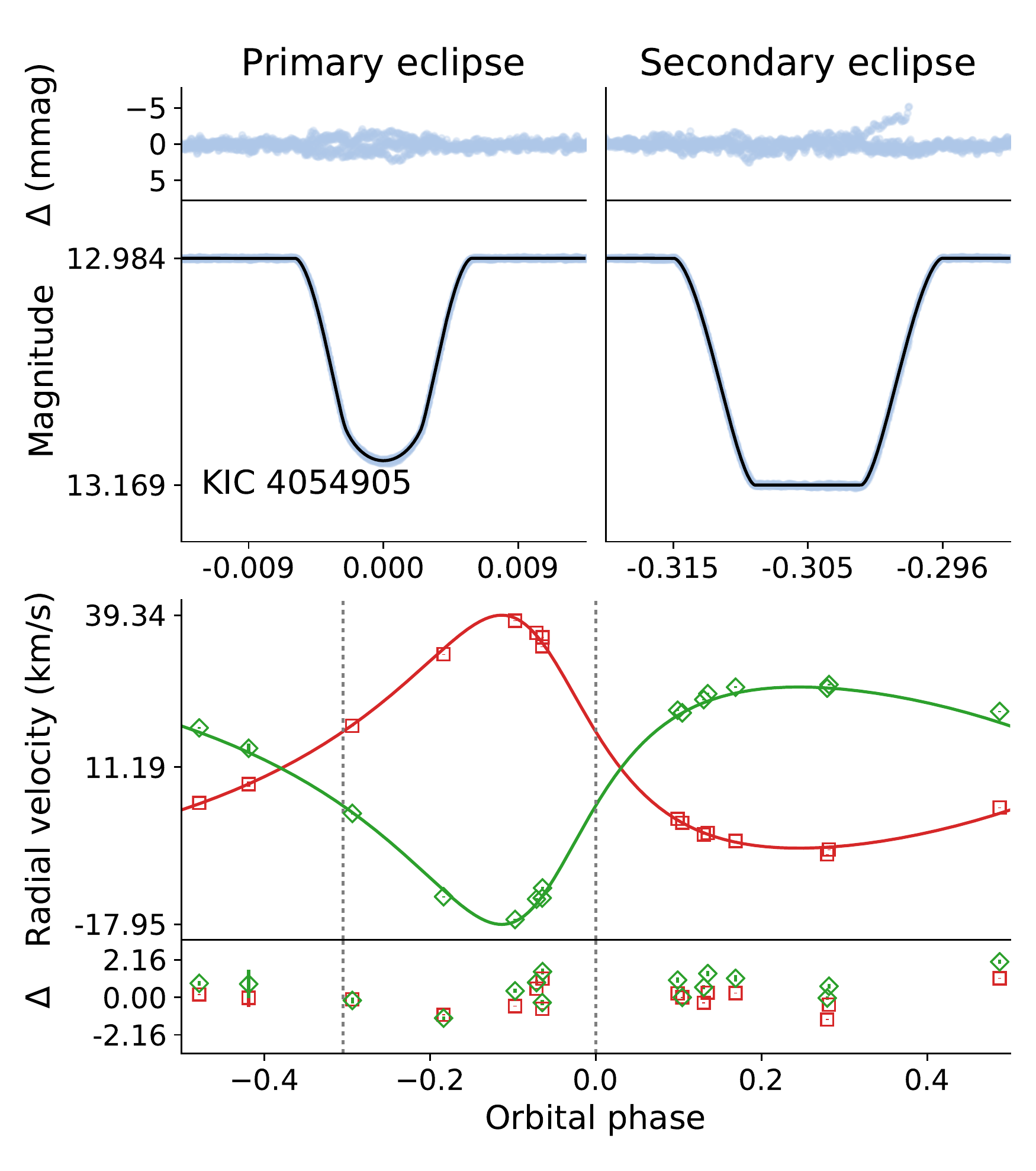} &
\includegraphics[width=.3\linewidth]{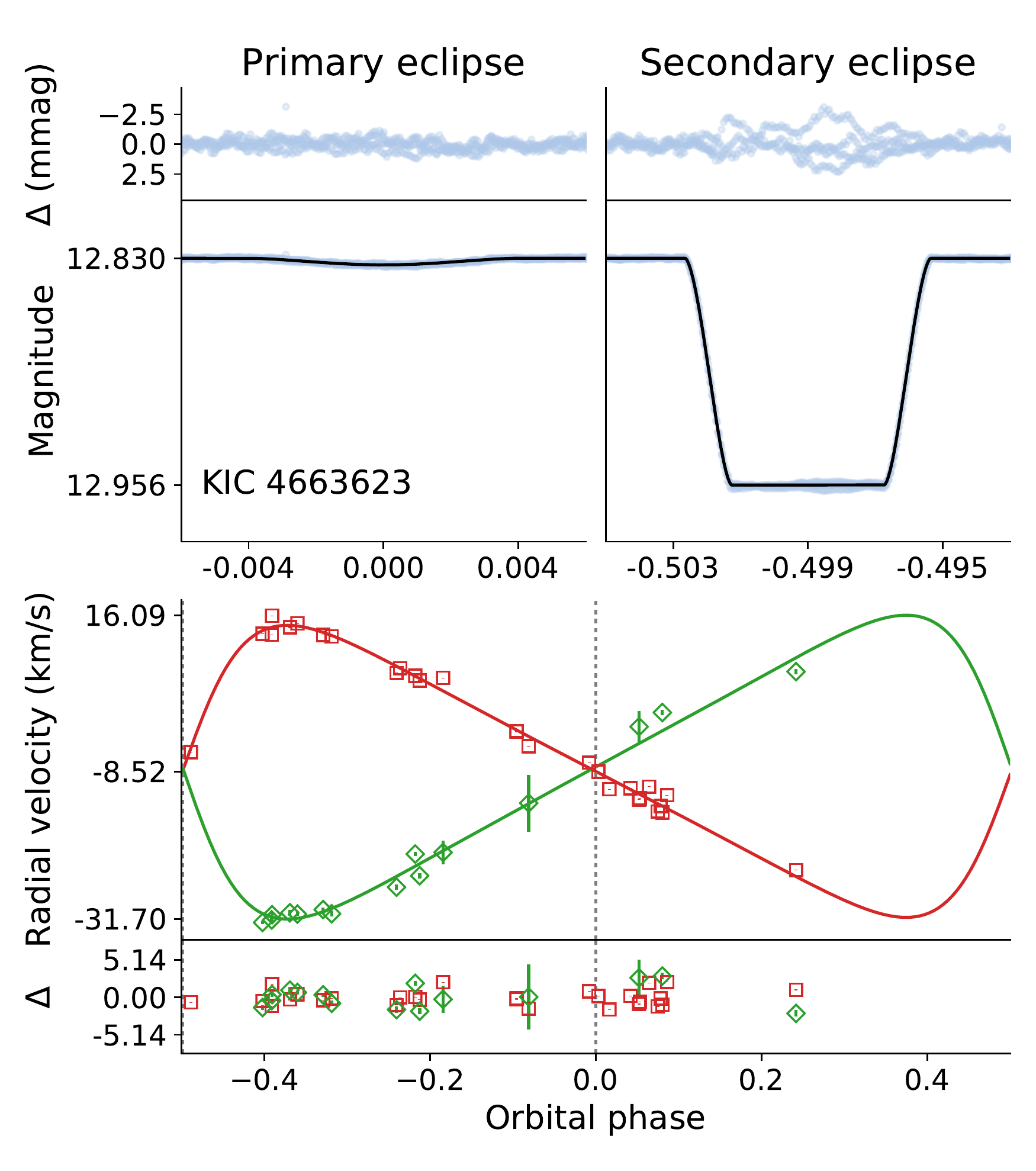} &
\includegraphics[width=.3\linewidth]{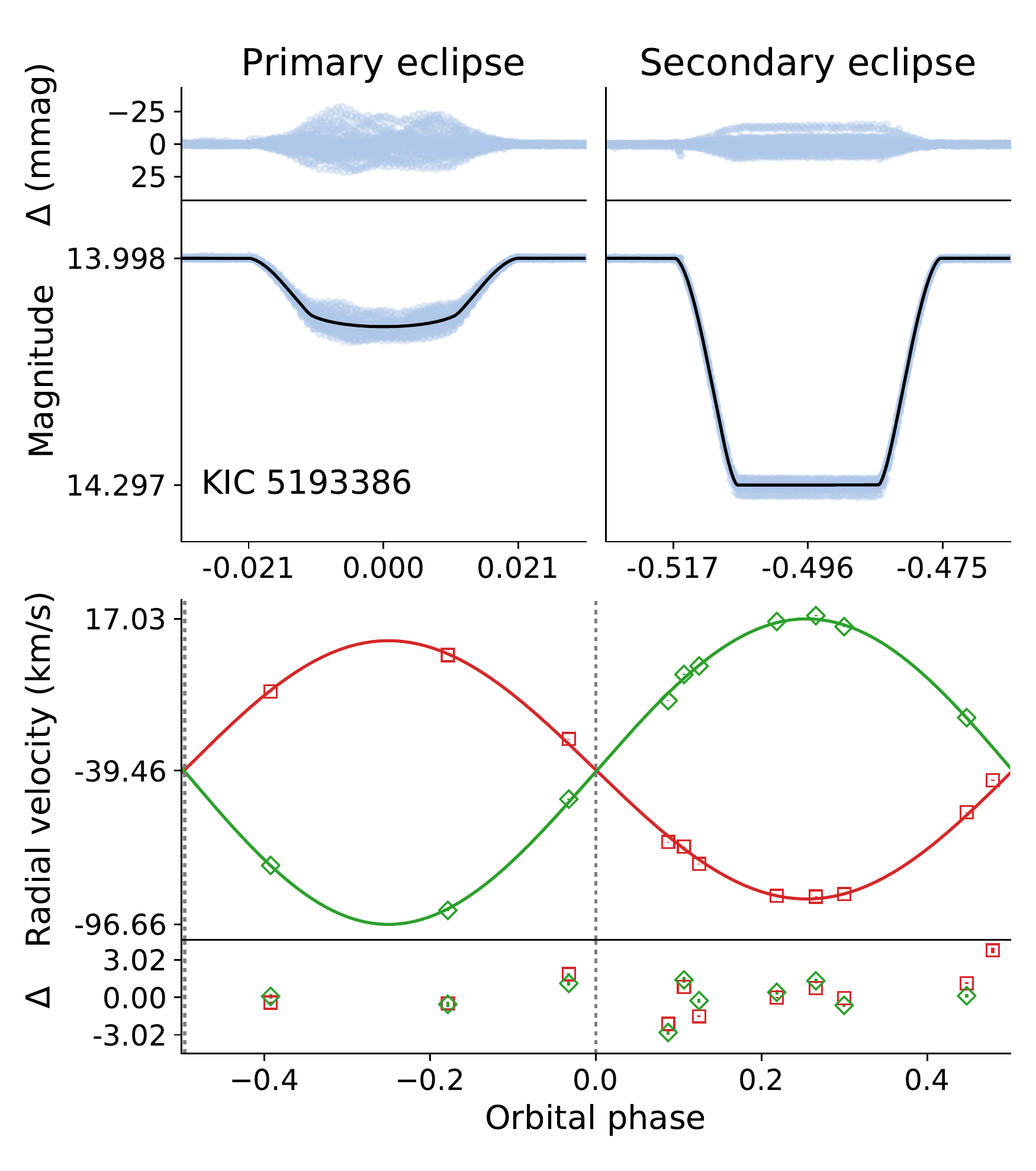} \\
\includegraphics[width=.3\linewidth]{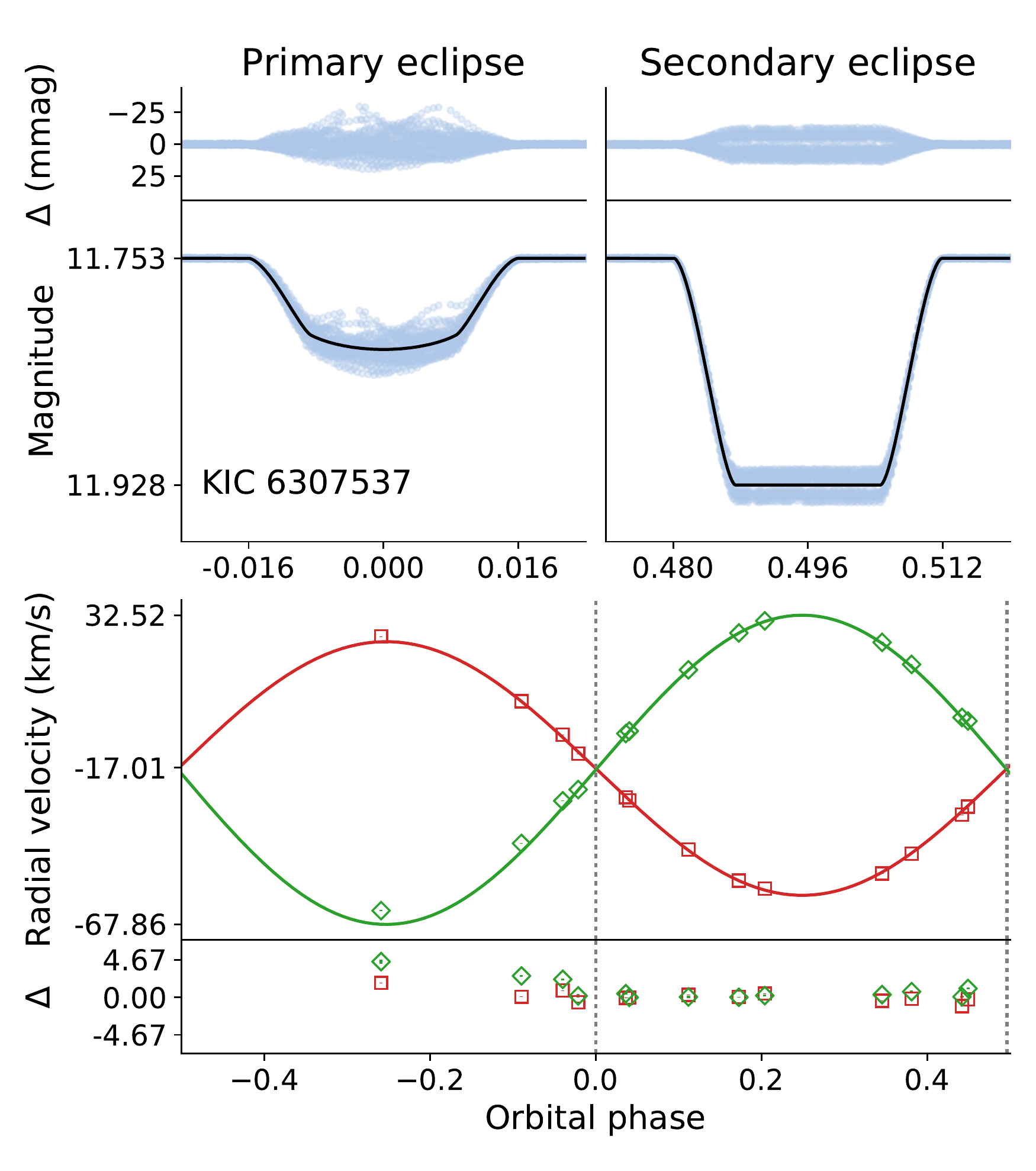} &
\includegraphics[width=.3\linewidth]{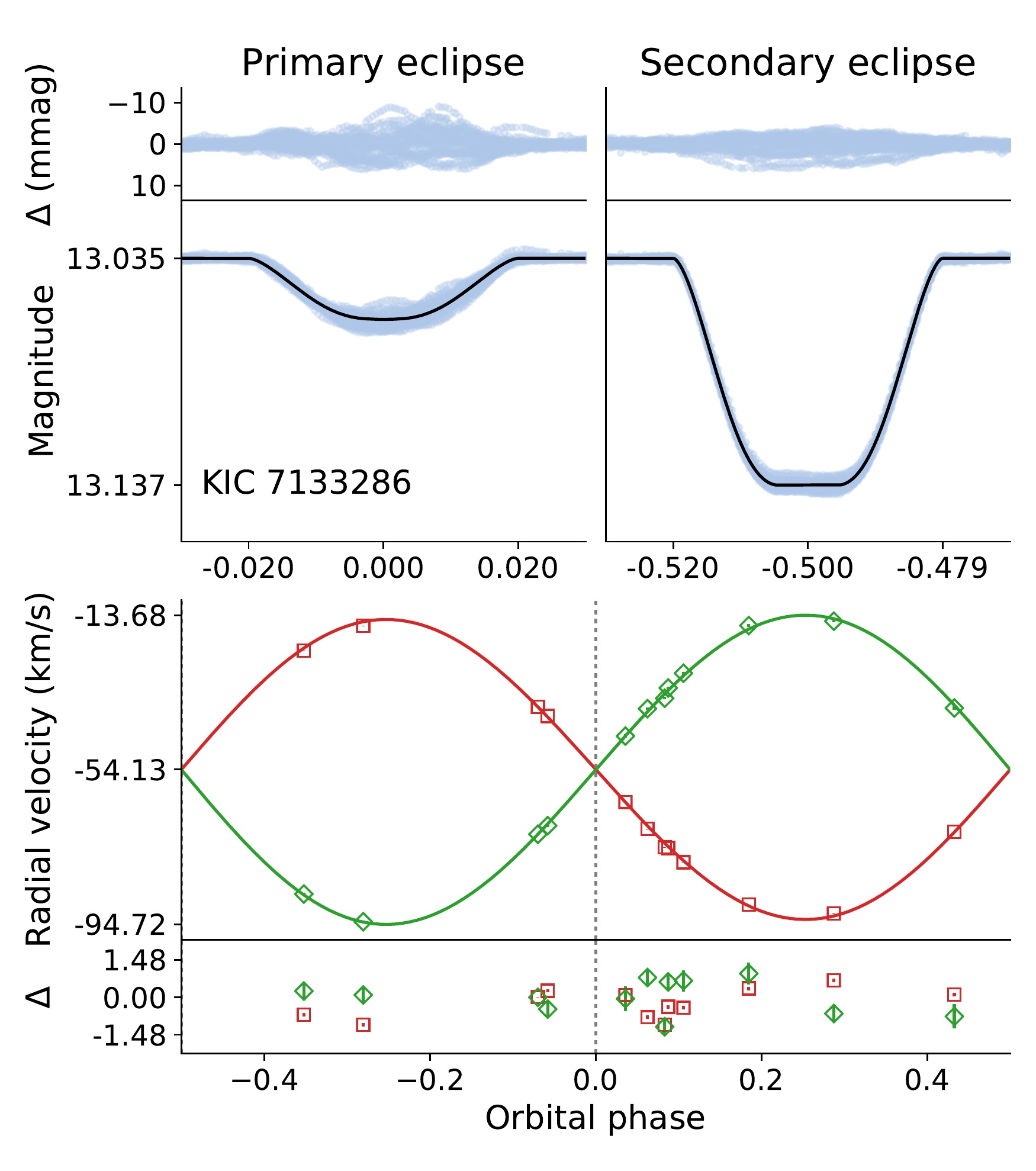} &
\includegraphics[width=.3\linewidth]{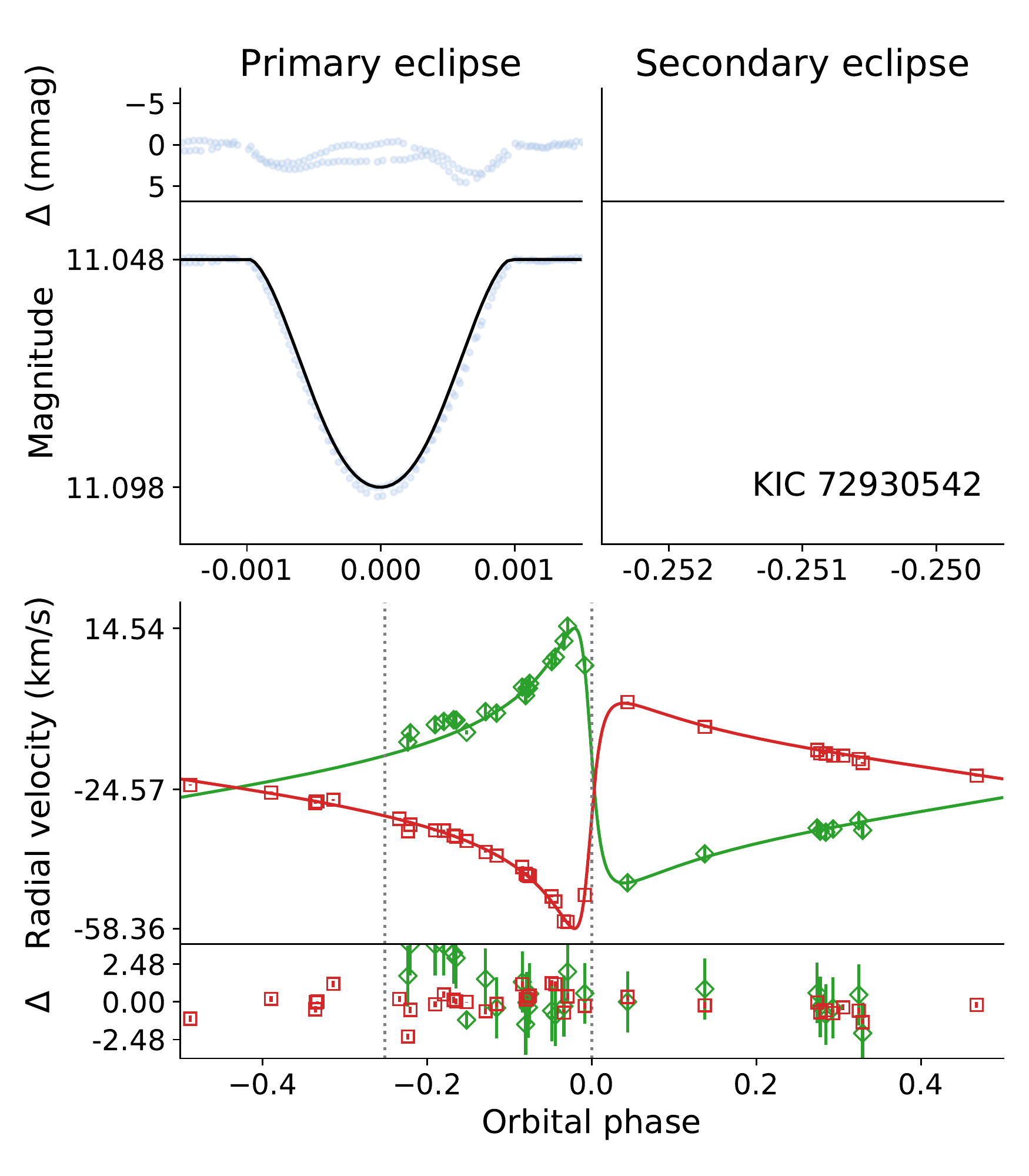} \\
\includegraphics[width=.3\linewidth]{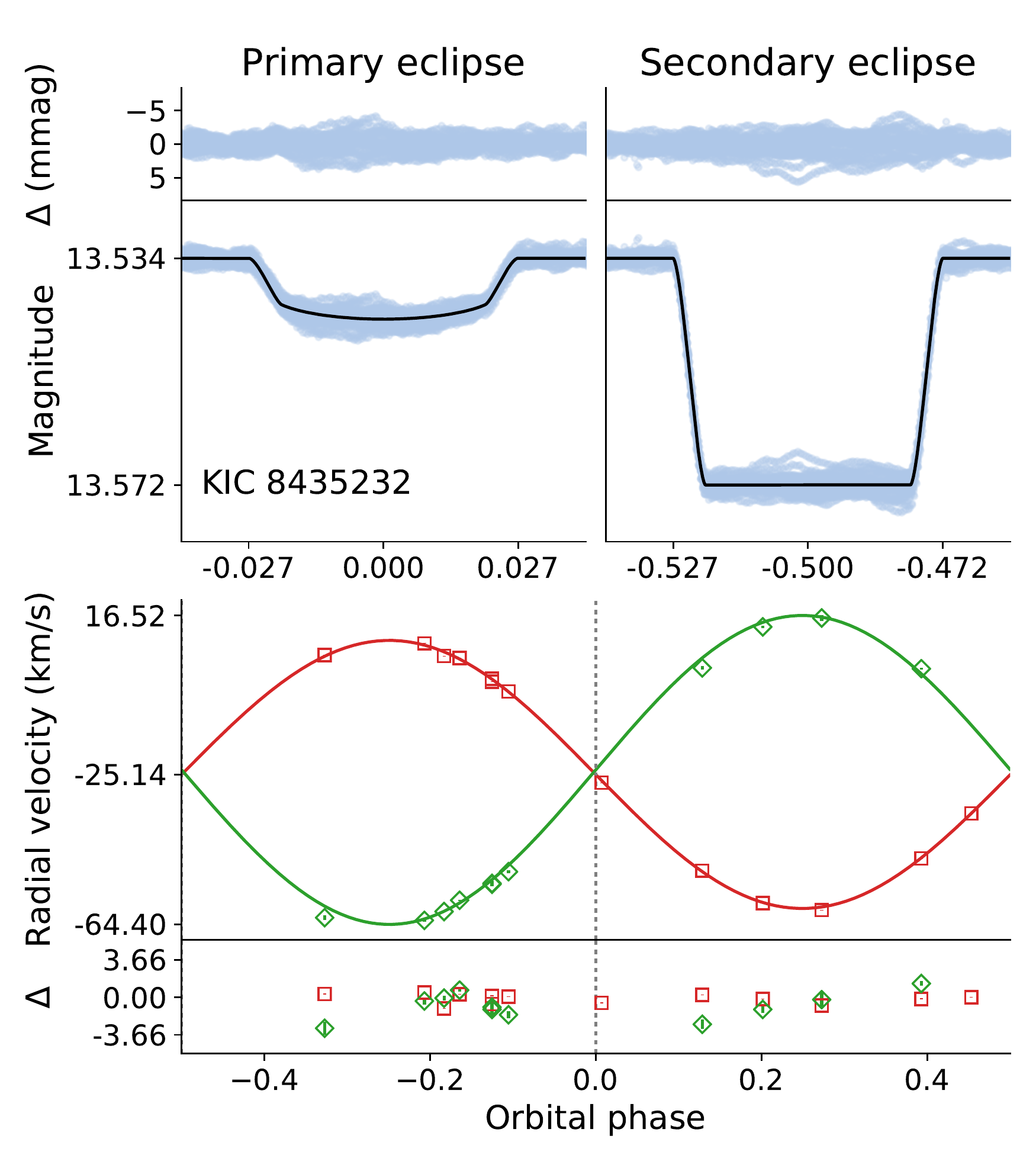} &
\includegraphics[width=.3\linewidth]{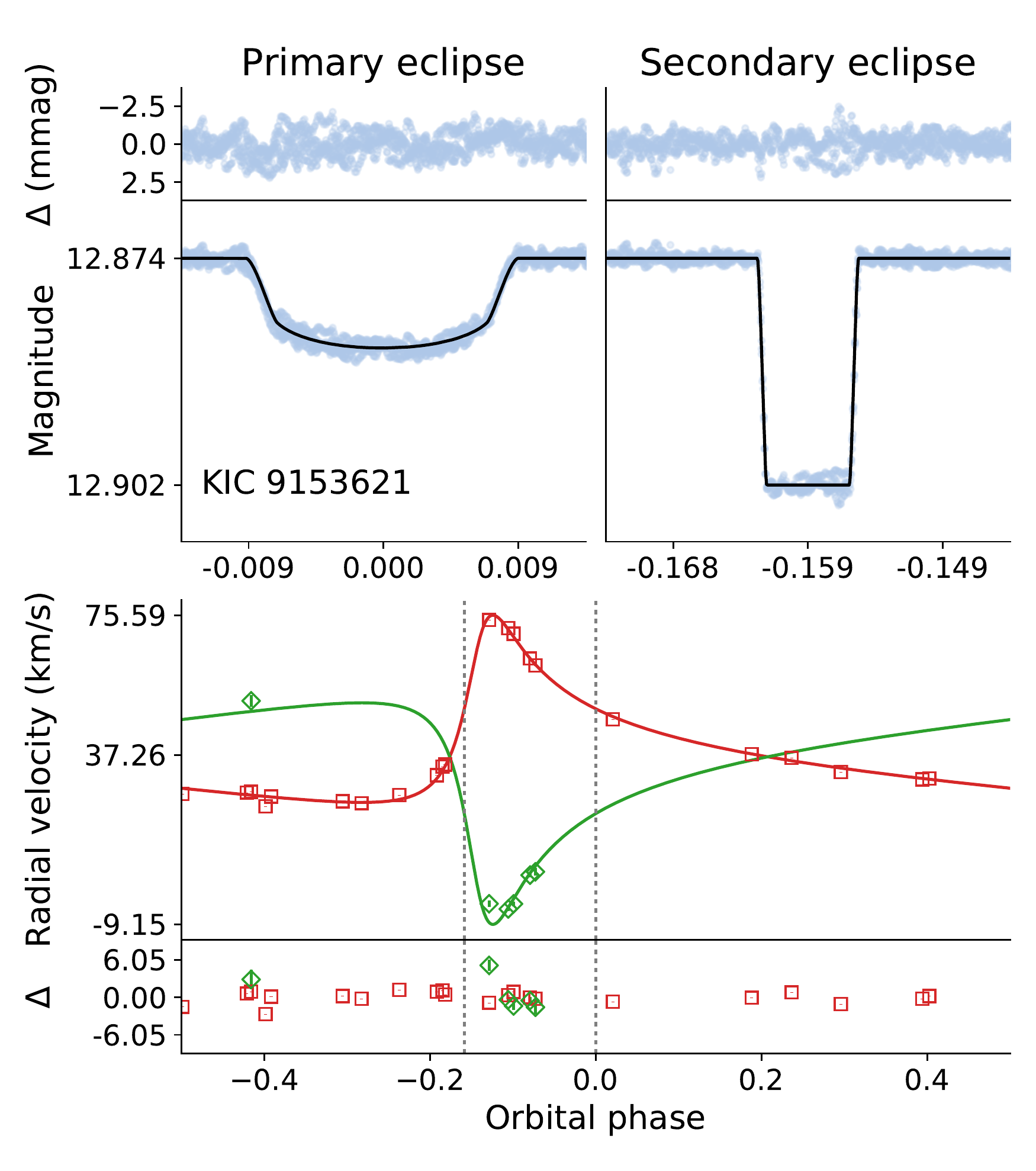} &
\includegraphics[width=.3\linewidth]{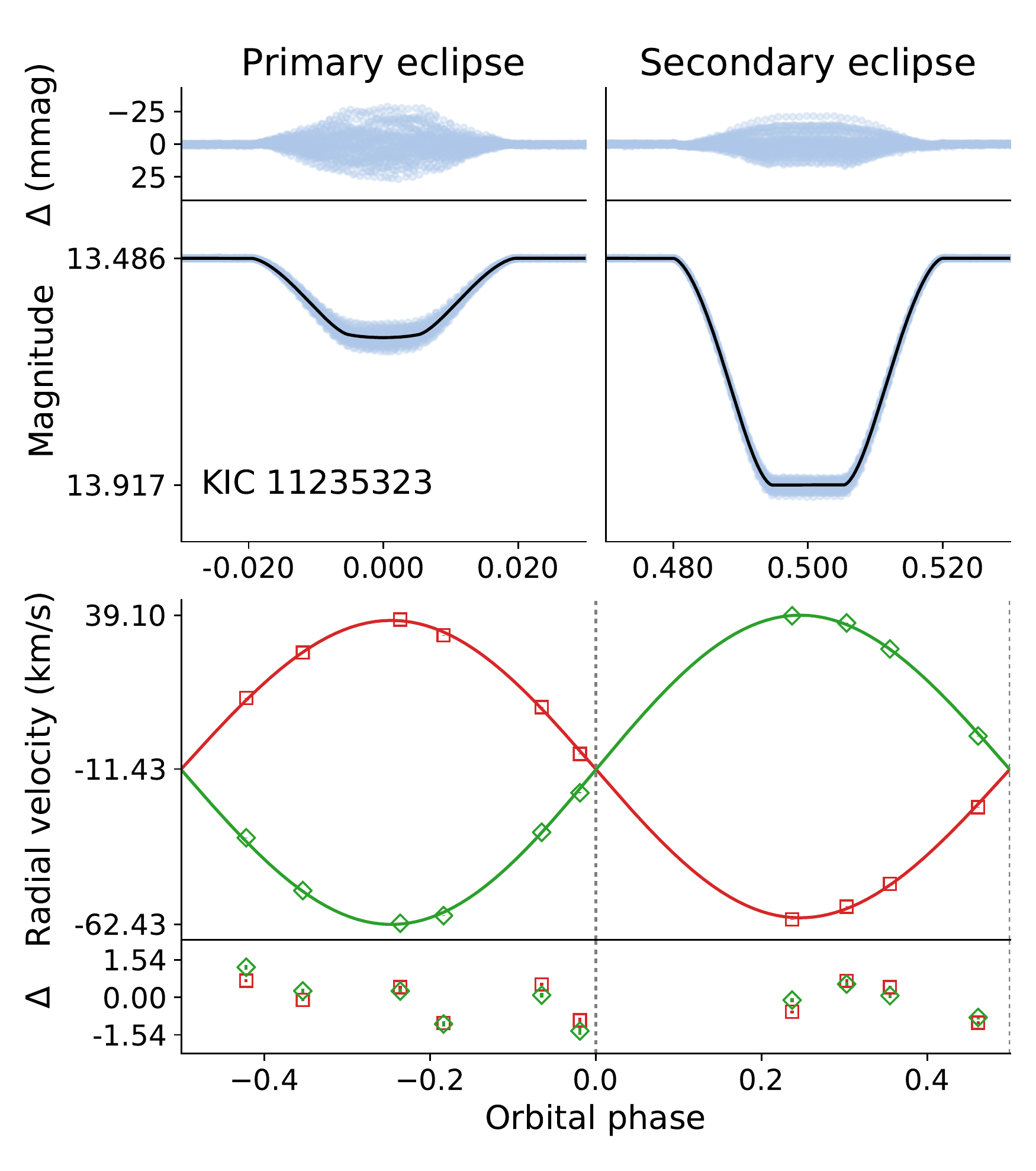}
\end{tabular}
\caption{Phase-folded eclipses and radial velocities for SB2 stars in our sample. For both eclipses, the processed \textit{Kepler} data is shown in light blue dots and the solid black line corresponds to the fit from JKTEBOP. For the radial velocities, red squares (resp. green diamonds) represent the observed radial velocity of the giant (resp. the companion) and the solid red line (resp. the green line) indicates the fit from JKTEBOP.}
\label{fig:RV_eclipses_SB2}
\end{figure*}

\begin{figure*}
\centering
\begin{tabular}{ccc}
\includegraphics[width=.3\linewidth]{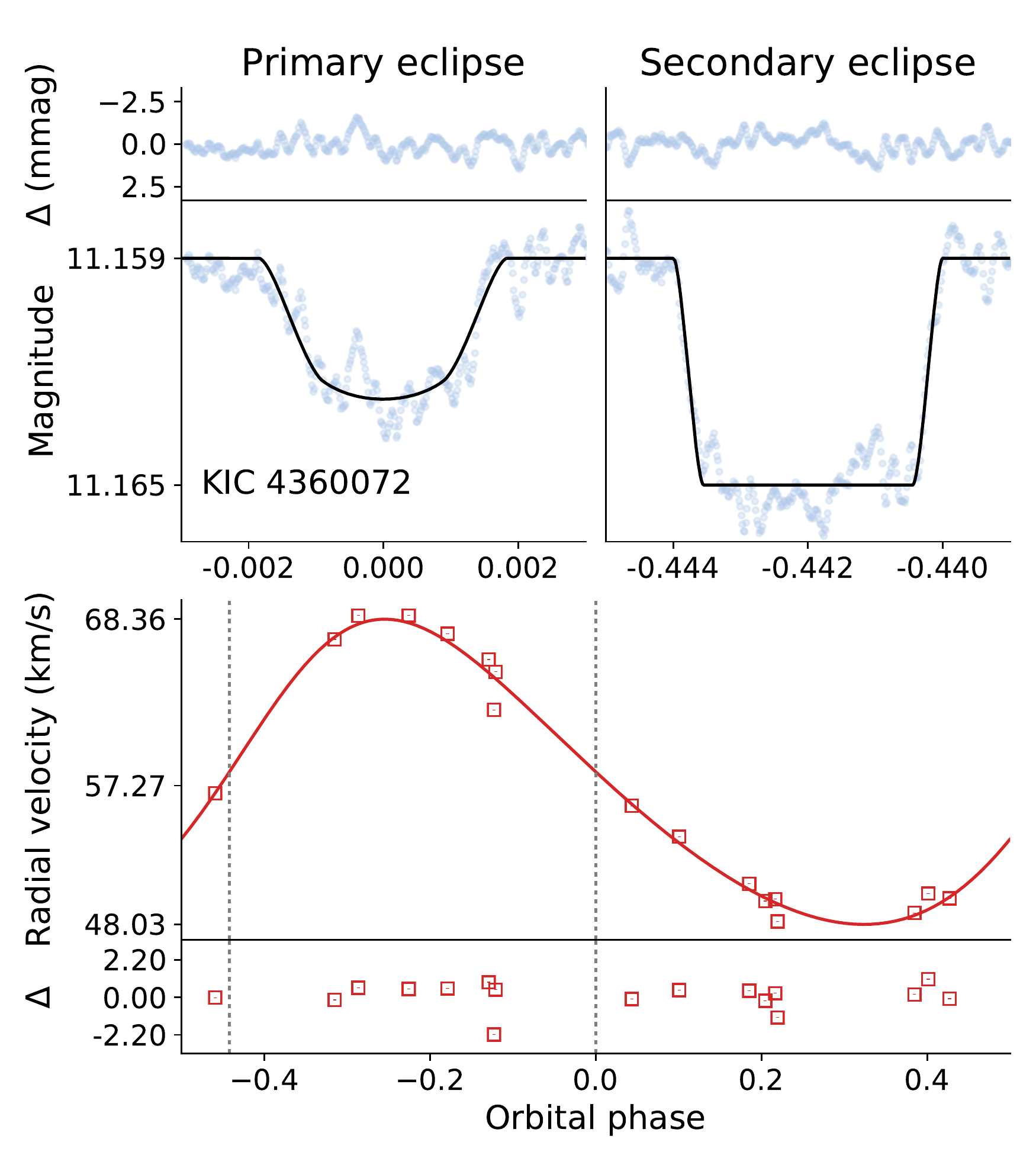} &
\includegraphics[width=.3\linewidth]{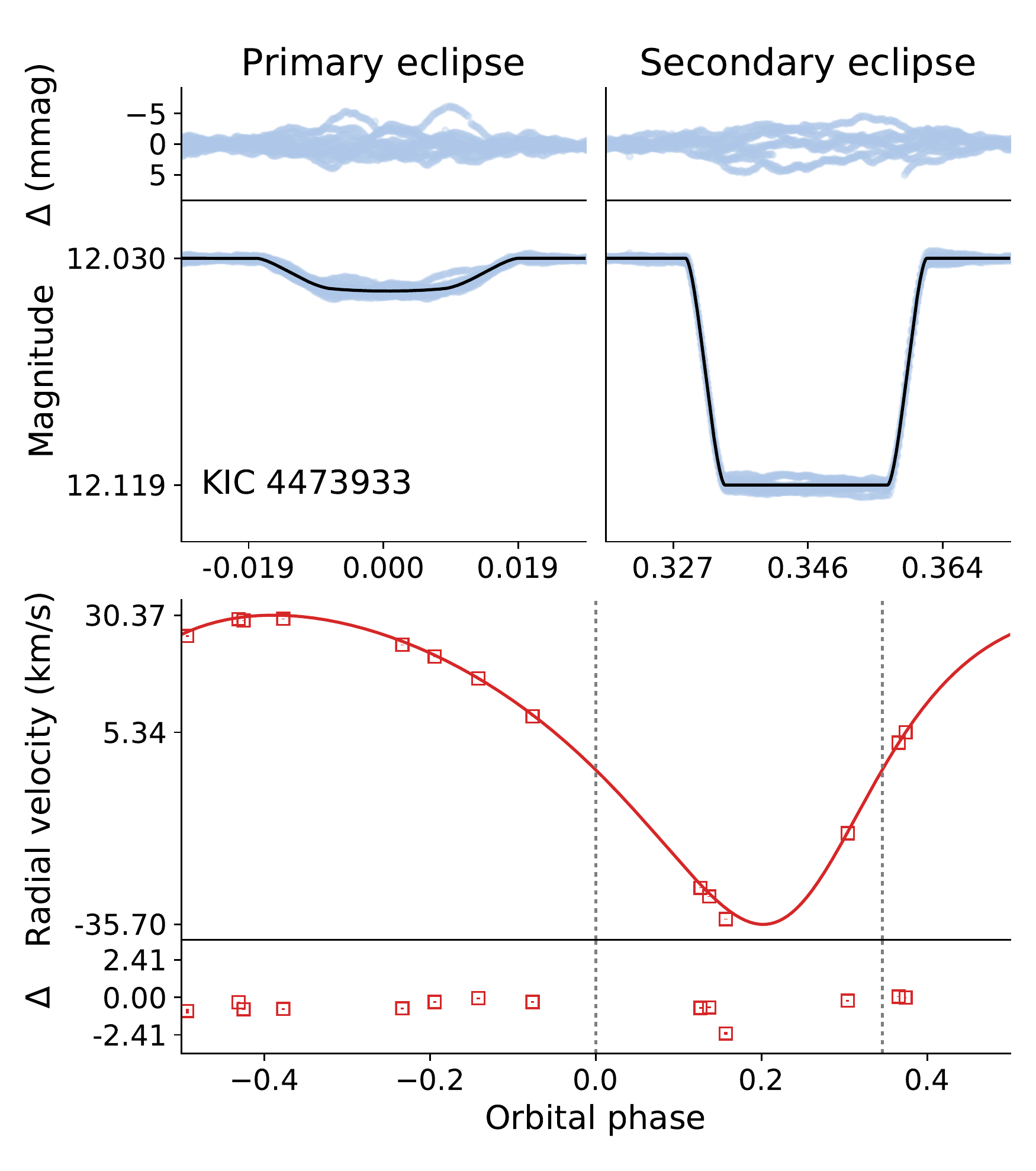} &
\includegraphics[width=.3\linewidth]{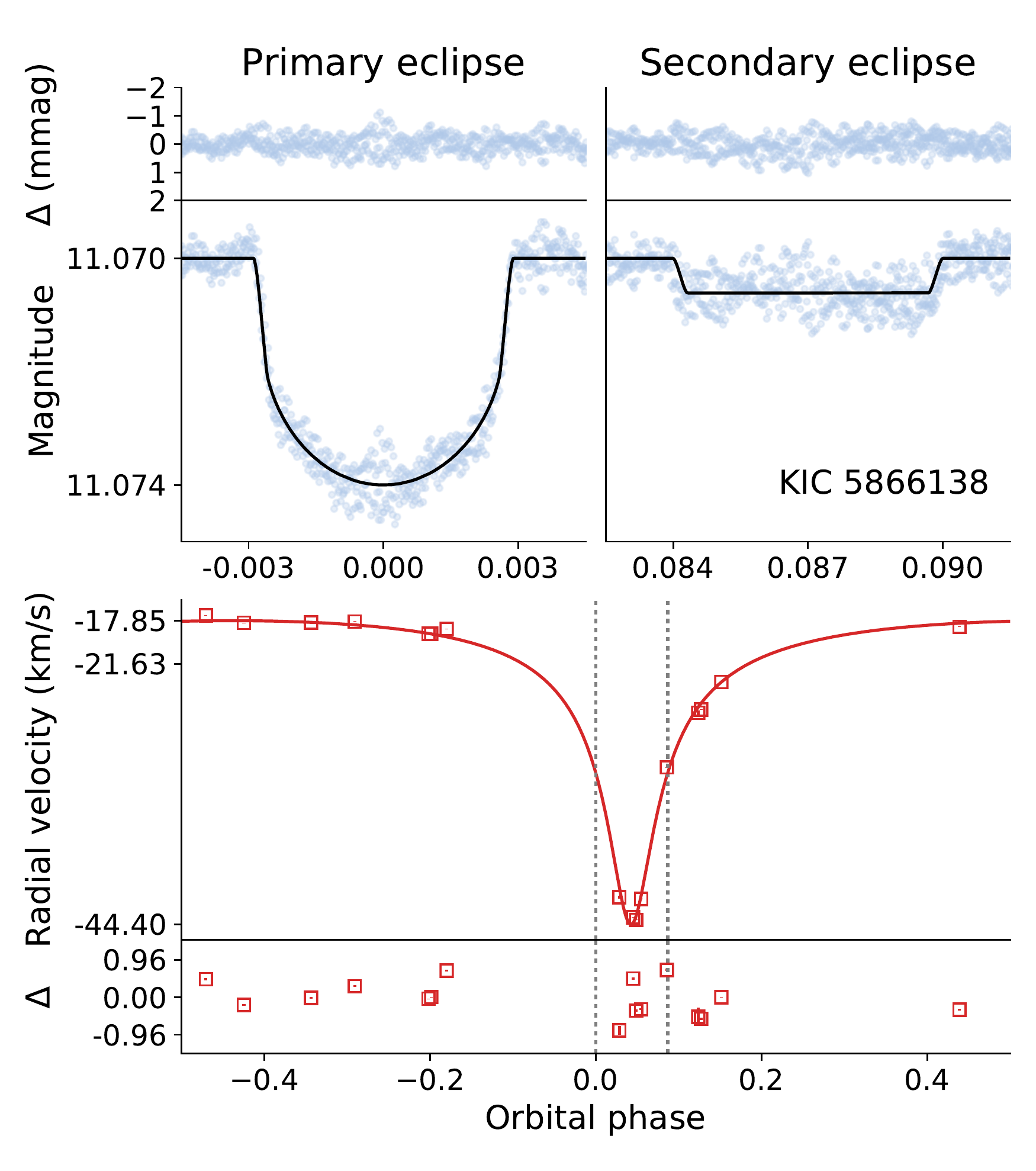} \\
\includegraphics[width=.3\linewidth]{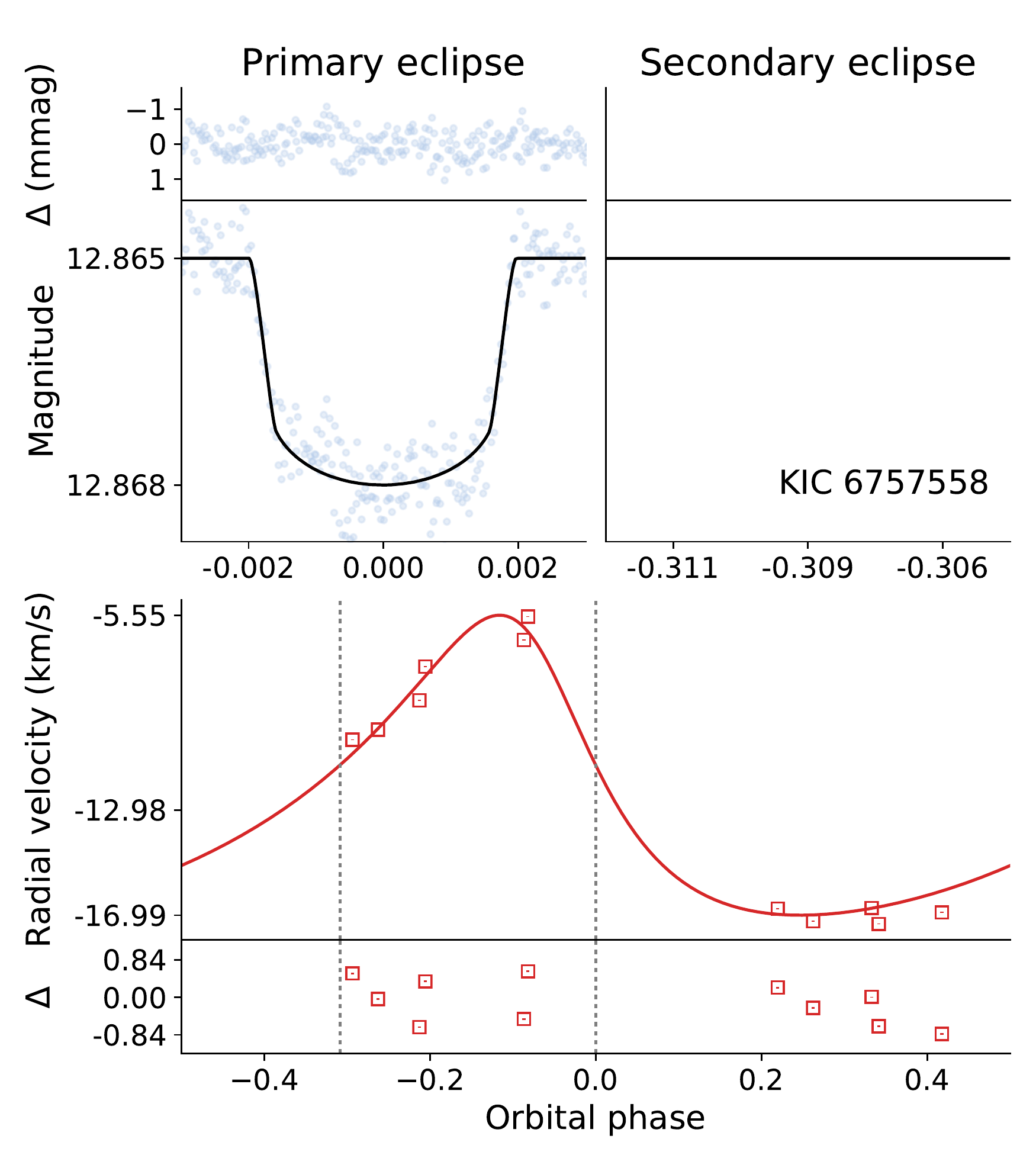} &
\includegraphics[width=.3\linewidth]{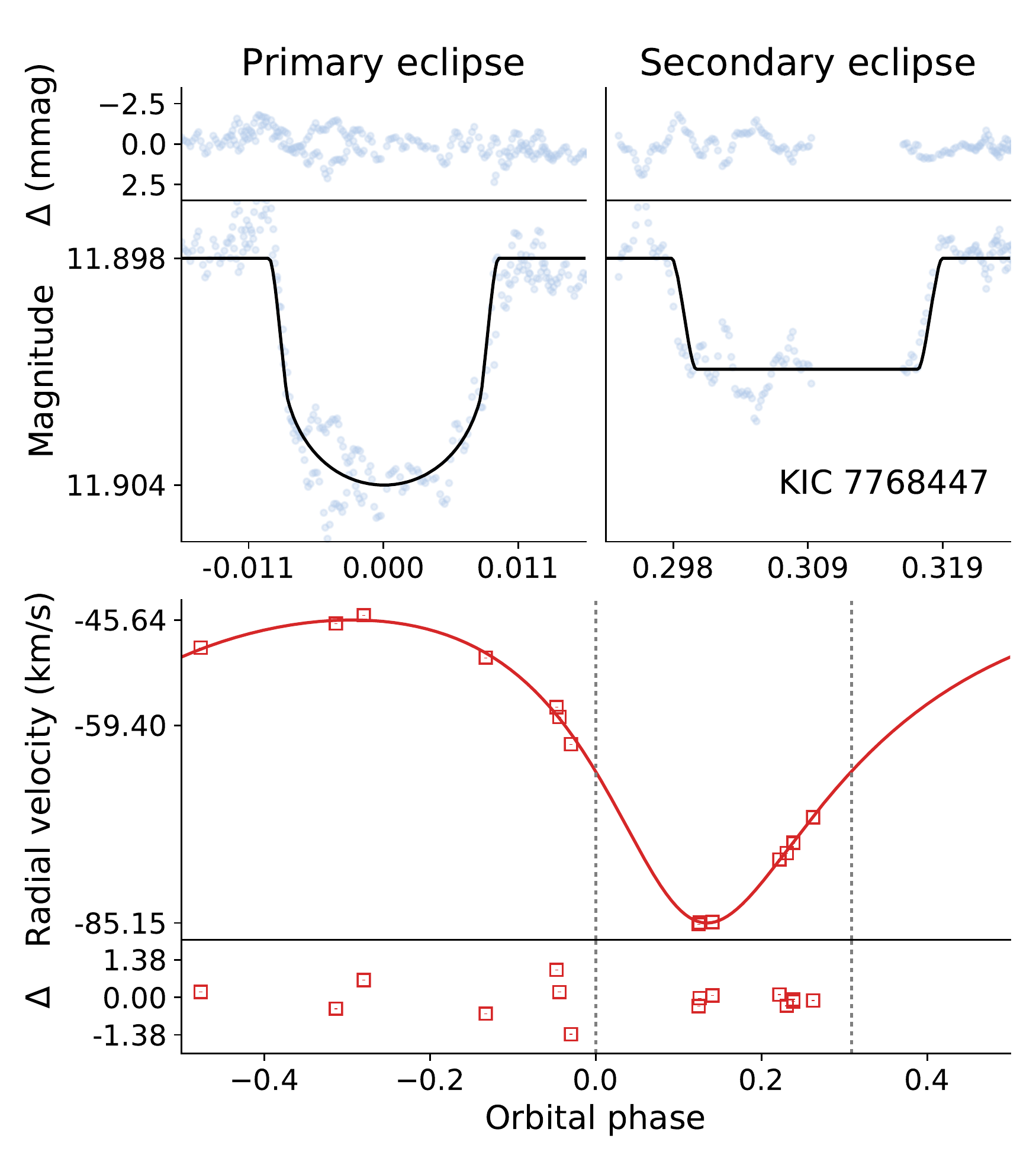} &
\includegraphics[width=.3\linewidth]{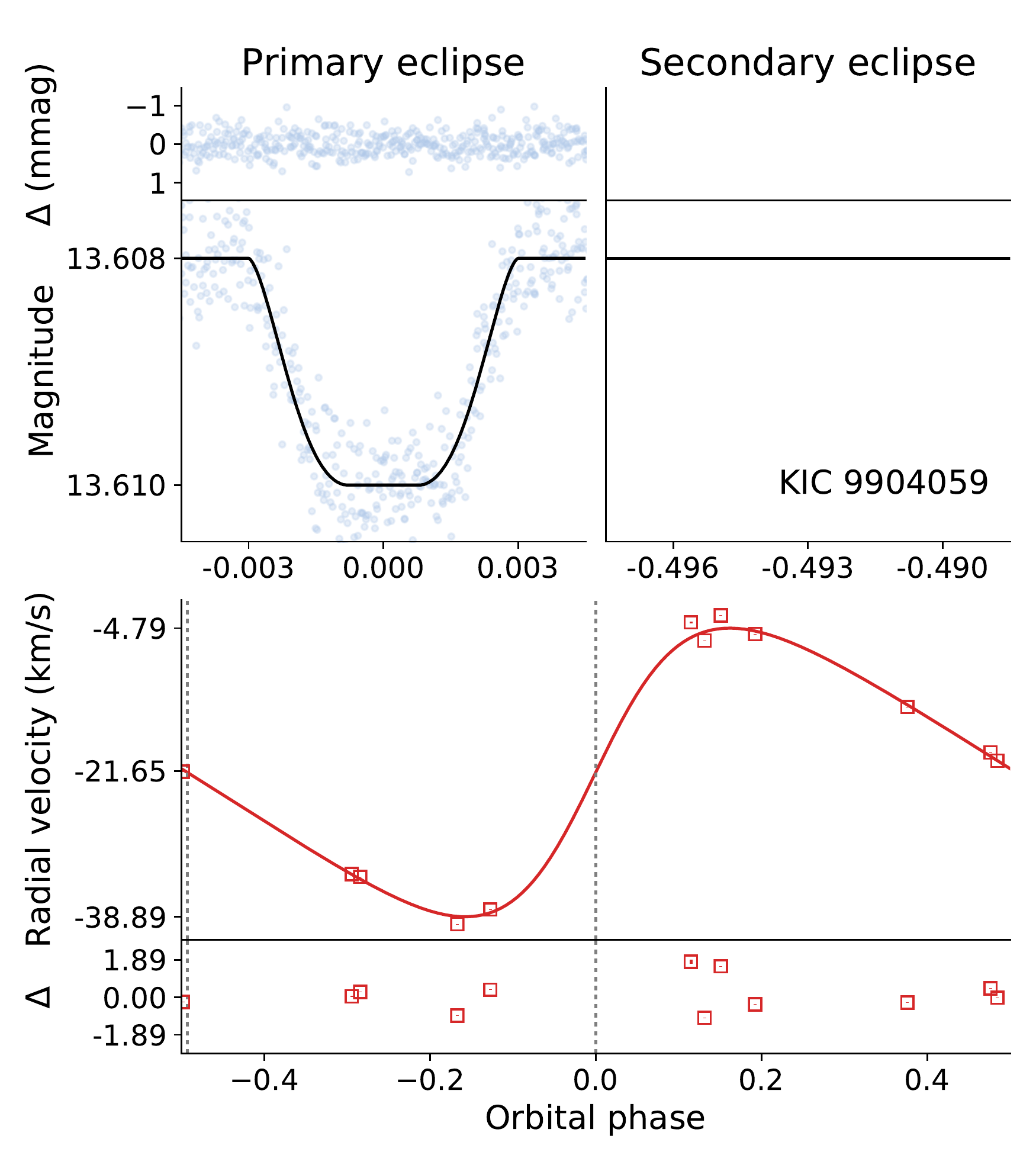} \\
\includegraphics[width=.3\linewidth]{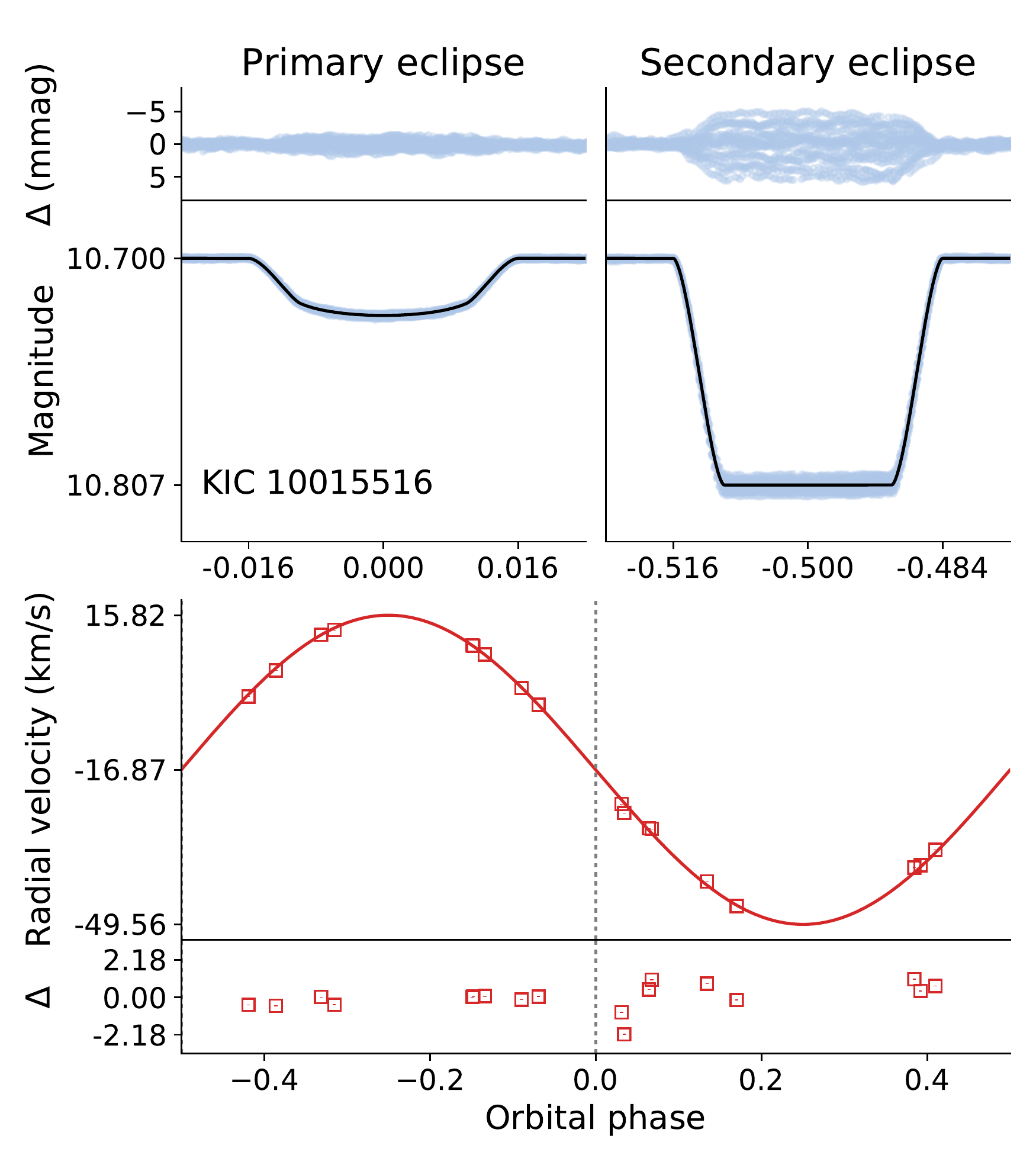} &
\includegraphics[width=.3\linewidth]{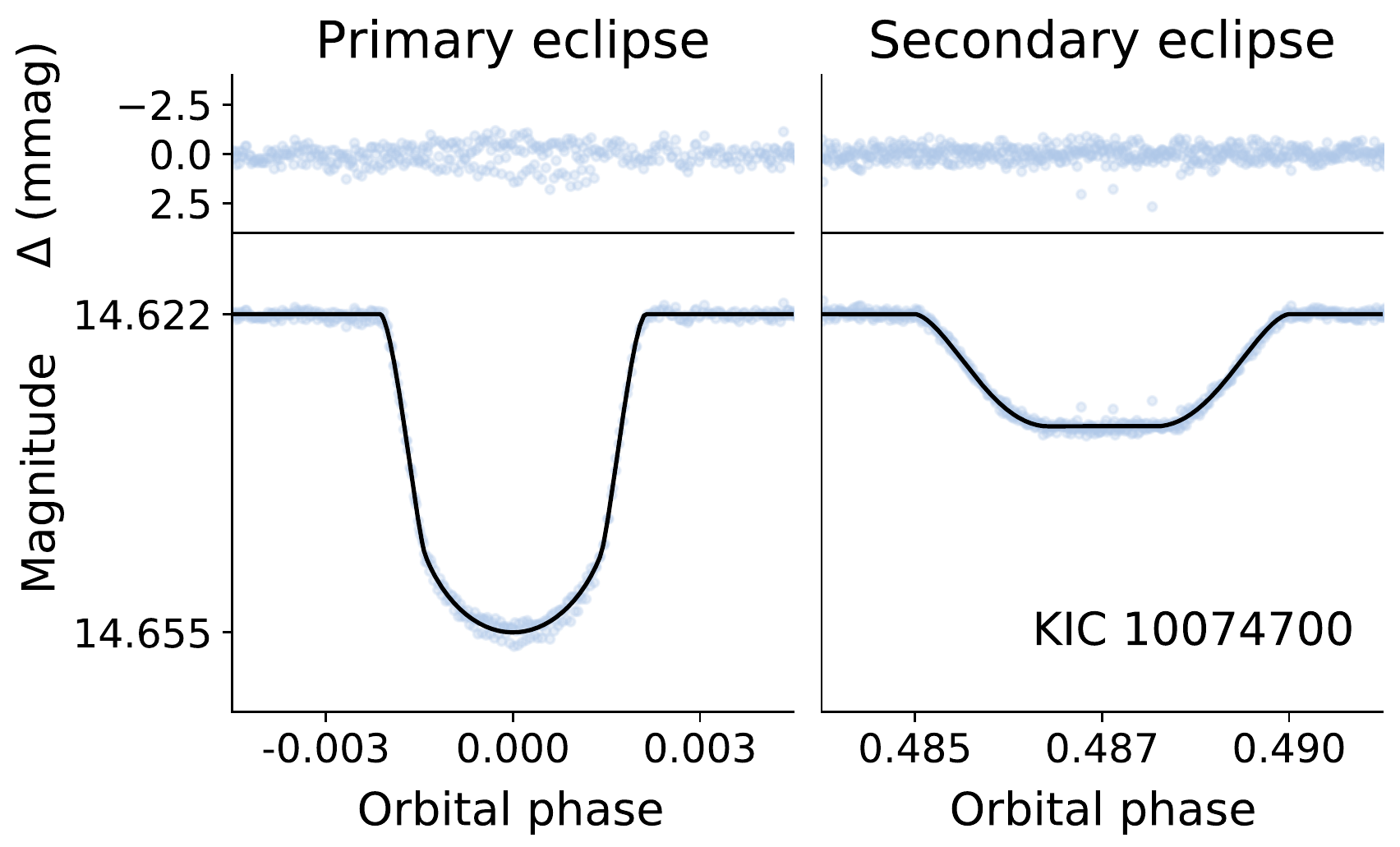} &
\end{tabular}
\caption{Phase-folded eclipses and radial velocities for SB1 stars in our sample. For both eclipses, the processed \textit{Kepler} data is shown in light blue dots and the solid black line corresponds to the fit from JKTEBOP. For the radial velocities, red squares represent the observed radial velocity of the giant and the solid red line indicates the fit from JKTEBOP.}
\label{fig:RV_eclipses_SB1}
\end{figure*}

\begin{sidewaystable*}
\centering
\caption{
Orbital parameters from dynamical modeling with JKTEBOP. The quantity 
$T_{\mathrm{prim,ecl}}$ stands for the reference time of the primary eclipses$^*$ in \textit{Kepler} Julian date$^\star$, $\omega$ the argument of periastron, $e$ the eccentricity, $i$ the orbital plane inclination, $(R_1, T_1, L_1),$ and $(R_2,T_2,L_2)$ the RG and companion's radii, effective temperatures, and luminosities. The quantities $K_1, K_2$ are the RV semi-amplitudes and $\gamma_1$, $\gamma_1$ are the radial velocity offsets. The last two columns contain the limb-darkening coefficients of the RG component used in the fit. The least significant digit in brackets after the value indicates the uncertainty. 
}
\label{tab:dyn_results_JKTEBOP}
\scalebox{0.87}{
\begin{tabular}{lllllllllllllllll}
\hline\hline
KIC & $\Porb$  & $T_{\mathrm{prim,ecl}}$  & $\omega$  & $e$ & $i$  & $\frac{R_2}{R_1}$ & $\frac{R_1 + R_2}{a}$ & $\left(\frac{T_2}{T_1}\right)^4$ & $\frac{L_2}{L_1}^\dagger$& $K_1$ & $K_2$  & $\gamma_1$  & $\gamma_2$ & $\alpha_{LD,\mathrm{lin}}$ & $\alpha_{LD,\mathrm{quad}}$ \\
    & (days) & (KJD) & ($^\circ$) &  &($^\circ$) &   &  & &  & (km s$^{-1}$) & (km s$^{-1}$) & (km s$^{-1}$) & (km s$^{-1}$) \\
\hline
 4054905 & 274.7306(4)  & 670.4562(9) & 35.1(5)  & 0.372(2)  & 89.387(7) & 0.380(2)  & 0.0515(1) & 1.287(5) & 0.1962(2)  & 21.6(3) & 22.0(5) & 11.2(3)   & 10.7(6)   & 0.56 & 0.15 \\
 4360072 & 1084.76(1)   & 812.17(8)   & 307(1)   & 0.152(1)  & 89.05(1)  &  0.052(1) & 0.0215(2) & 1.70(6)  & 0.0046(7)  & 10.2(5)  & ...    & 57.3(5)   & ...       & 0.55 & 0.18\\
 4473933 & 103.6014(4)  & 199.09(1)   & 210(4)   & 0.279(9)  & 84.27(9)  & 0.118(2)  & 0.147(2)  & 6.3(2)   & 0.0874(2)  & 33.0(7) & ...     & 5.3(4)    & ...       & 0.50 & 0.00\\
 4663623 & 358.0903(1)  & 308.18(8)   & 270(1)   & 0.399(1)  & 88.60(1)  & 0.186(3)  & 0.0389(5) & 4.1(2)   & 0.14400(8) & 22.9(5) & 23.9(5) & -8.5(5)   & -7.8(5)   & 0.51 & 0.18\\
 5193386 & 21.378310(9) & 136.4272(9) & 312(6)   & 0.010(1)  & 89.8(2)   & 0.3048(7) & 0.1323(3) & 3.59(2)  & 0.3334(5)  & 48(1)   & 56.8(4) & -39.5(3)  & -39.5(1)  & 0.40 & 0.21\\
 5866138 & 342.259(8)   & 436.13(1)   & 178(3)   & 0.7158(4) & 90.0(7)   & 0.056(1)  & 0.030(2)  & 0.18(1)  & 0.00064(5) & 13.3(2) & ...     & -21.6(2)  & ...       & 0.70 & 0.00\\
 6307537 & 29.74451(3)  & 142.643(1)  & 209.5(1) & 0.0067(8) & 88.47(2)  & 0.2630(7) & 0.1046(2) & 2.74(1)  & 0.18956(5) & 41(1)   & 50(1)   & -17.0(5)  & -18(1)    & 0.50 & 0.00\\
 6757558 & 421.190(9)   & 357.977(6)  & 47(7)    & 0.22(3)   & 88.6(1)   & ...       & ...       & ...      & ...        & 5.7(5)  & ...     &  -13.0(5) & ...       & 0.59 & 0.14\\
 7133286 & 38.51073(5)  & 136.332(2)  & 271.9(8) & 0.009(3)  & 82.99(2)  & 0.1717(9) & 0.1763(3) & 3.32(4)  & 0.0980(2)  & 39.3(6) & 40.5(6) & -54.1(4)  & -54.2(4)  & 0.50 & 0.00\\
 7293054 & 671.806(3)   & 311.384(2)   & 70.5(2.6) & 0.80(1) & 87.3(2)   & ...       & ...       & ...      &  ...       & 32(4)   & 38(4)   & -24.2(2)  & -23.3(5)  & 0.50 & 0.00\\
 7768447 & 122.32(4)    & 130.45(4)   & 160(8)   & 0.322(9)  & 88(2)     & 0.071(3)  & 0.07(2)   & 0.6(3)   & 0.003(2)   & 19.8(2) & ...     & -59.4(3)  & ...       & 0.55 & 0.13\\
 8435232 & 49.5708(2)   & 152.430(5)  & 77.5(9)  & 0.003(3)  & 84.94(8)  & 0.0993(5) & 0.191(1)  & 4.10(4)  & 0.0404(2)  & 35.1(3) & 40.5(7) & -25.1(3)  & -23.9(4)  & 0.50 & 0.00\\
 9153621 & 305.792(5)   & 218.24(3)   & 314.9(6) & 0.700(3)  & 88.82(9)  & 0.0978(6) & 0.0472(5) & 2.82(4)  & 0.027(2)   & 25.7(4) & 30(4)   & 37.26(8)  & 36(3)     & 0.45 & 0.16\\
 9904059 & 102.963(1)   & 158.078(4)  & 90(3)    & 0.32(1)   & 86.2(1)   & ...       & ...       & ...      & ...        & 17.0(5) & ...     & -21.7(5)  & ...       & 0.55 & 0.16\\
10015516 & 67.69217(9)  & 138.533(3)  & 271.2(9) & 0.00(1)   & 86.91(7)  & 0.1584(5) & 0.113(1)  & 4.18(9)  & 0.1080(2)  & 32.7(7) & ...     & -16.9(4)  & ...       & 0.47 & 0.21\\
10074700 & 365.6340(6)  & 238.984(1)  & 94(2)    & 0.29(6)   & 89.5(1)   & 0.1602(7) & 0.017(1)  & 0.4(2)   & 0.011(6)   & ...     & ...     & ...       & ...       & 0.60 & 0.00\\
11235323 & 19.668407(8) & 142.3589(7) & 90.4(3)  & 0.010(1)  & 87.71(1)  & 0.4315(6) & 0.1323(2) & 2.90(1)  & 0.5056(7)  & 48.8(2) & 50.8(4) & -11.43(8) & -11.67(8) & 0.23 & 0.20\\
\hline
\end{tabular}
}
\tablefoot{\\
$^*$  We note that the reference time is the epoch of the primary eclipse, whereas the reference time used in \citet{Gaulmeetal2016} is the epoch of the periastron.\\
$^\star$ \textit{Kepler} Julian dates KJD are related to barycentric Julian dates BJD by KJD = BJD $-$ 2,454,833 days.\\ 
$^\dagger$  All the values provided in this table were directly fitted on the light curves with JKTEBOP, except the luminosity ratio which was computed as $L_2 / L_1 = (R_2 / R_1)^2 (T_2 / T_1)^4$.\\
}
\end{sidewaystable*}

We used the JKTEBOP software \citep{Southworth2013} to compute the orbital parameters of the binary systems. This code uses the Levenberg-Marquardt algorithm to determine the best fit to the data. Error bars are then estimated with a Monte Carlo algorithm (``task 9'' of JKTEBOP). We simultaneously fitted both the eclipse-photometry light curves and the radial velocities obtained in section~\ref{sec:data}. We adopted the same convention as \citet{Gaulmeetal2016}, who defined the primary component as the giant and the secondary as its companion. This is why the radii ratios we obtained are smaller than 1. We had to make two exceptions to this rule for the eccentric systems 7293054 and 9904059, for which only the eclipse of the companion is visible.

Following \citet{Brogaardetal2018}, we assumed a quadratic limb-darkening law for the giant components of our binary systems.
We used the JKTLD code \citep{Southworth2015} to determine the quadratic coefficient
\footnote{
  We remind that a quadratic limb-darkening law has two coefficients: one quadratic and one linear. More information can be found in  \citet[][Eq. 2]{Sing2010}.
}.
This code inputs atmospheric parameters, which we got from the APO spectra, and outputs two coefficients for the spectral passband specified by the user.
In JKTEBOP, we fitted the linear coefficient and used the quadratic coefficient from JKTLD because of the poor constraints we have on this parameter.
We found that the limb darkening of the companion has negligible influence on the eclipse photometry.
This is most probably because the contribution of this component to the luminosity of the system is less than 10\% and its radius is three to ten times less than that of the giant.
For this reason, we assumed a linear limb-darkening law for all the companions and fixed its coefficient to 0.5, which is consistent with the values given by \citet{Sing2010}.
A more detailed analysis is required to better characterize the limb darkening of the companion, which is beyond the scope of the present paper.
The limb-darkening coefficients $\alpha_{LD,\mathrm{lin}}$ and $\alpha_{LD,\mathrm{quad}}$ of the giants are given in Table~\ref{tab:massesAndRadii}.

We used the amount of third light from the contamination values given on the online \textit{Kepler} catalog. Four values of contamination are provided, including one for each possible configuration of the satellite. We computed the average of these values to obtain the amount of third light and fixed it for the orbital-parameter determination. Since the stars in this study are not rotating fast, we did not take gravity darkening into account and fixed the associated coefficients to 1. We also fixed the reflection effect coefficients to 0.

For SB2 systems, the radial velocity of each component is the sum of two contributions: a constant offset $\gamma$ due to the systemic radial velocity of the binary star and a phase-dependent term due to the orbital motion of amplitude $K$. 
However, the gravitational redshift \citep{Einstein1952} and convective blueshift \citep{Gray2009} lead to different radial velocity offsets for each component. For instance, the gravitational redshift induces a wavelength shift,
\begin{equation}
\frac{\Delta \lambda}{\lambda} \simeq \frac{G M_\star}{R_\star c^2},
\end{equation}
where $\lambda$ is the optical wavelength, $G$ the gravitational constant, $M_\star$ the stellar mass, $R_\star$ the stellar radius, and $c$ the speed of light. In the visible domain, the gravitational redshift generates a radial velocity offset of $v_{\mathrm{offset},\odot} = 650$ m~s$^{-1}$ for the Sun and $v_{\mathrm{offset},\mathrm{RG}} = 65$ m~s$^{-1}$ for a 10-$\Rsun$ solar mass giant. Therefore, we fit one systemic radial velocity term per component, $\gamma_1$ and $\gamma_2$.

Besides, we do not take into account the ellipsoidal variability of the giants arising from the tidal interaction with their companion.
Modeling this phenomenon would first require us to disentangle the contributions from starspots, reflection or beaming effects, which is almost impossible. Then it would require us to develop and use more complex tools \citep[e.g.,][]{FaiglerMazeh2011,Prsa2019}, which could be the aim of future works. Taking this effect into account would allow us to derive more precise orbital parameters, especially for systems with $(R_1 + R_2) / a \ge 0.1$ \citep{Russell1939,Beech1985,Morris1985}. The systems concerned by such ellipsoidal effects are the closest systems, which are not used to test the asteroseismic scaling relations.

The dynamical models obtained with JKTEBOP are shown in Fig.~\ref{fig:RV_eclipses_SB2} for SB2 and Fig.~\ref{fig:RV_eclipses_SB1} for SB1 systems. The orbital parameters are reported in Table~\ref{tab:dyn_results_JKTEBOP}. The models show an overall good agreement with the data. However, we observe an enhanced dispersion of the photometric data during the eclipses of the systems with short periods and large magnetic activity (mainly KIC 5193386, 6307537, 7133286, or 11235323). This artifact originates from the preparation of the eclipse light curves and was also reported by \citet{Gaulmeetal2014}. 
We do not to correct these depth variations to avoid altering the eclipse shapes. This problem actually has very little impact on our study because the data points are symmetrically distributed around the photometric models, meaning that it causes no systematic biases on the ratio of radii. Nevertheless, this dispersion increases the relative error on the radii since errors are computed with a Monte Carlo algorithm that accounts for data dispersion. This is why the errors on radii for the shortest-period systems ($P\ind{orb} \leq 30$ d) are not smaller in average than for the systems with $P\ind{orb} \geq 200$ d, despite a much larger number of eclipses.
In any cases, we note that the systems that show enhanced eclipse-depth dispersion are not used for testing asteroseismology. They are all very close binaries in which the RG component shows strong magnetic activity and no oscillations. So, even by considering that our radii error bars could still be a little underestimated, it is irrelevant for testing asteroseismology.

\subsection{Ages from stellar evolution models}
\label{subsec:age_stellar_model}

We provide proxies of the system ages by fitting the measured parameters of each binary system individually against theoretical isochrones from the YaPSI database \citep[see][for details]{Spadaetal2017}.
The YaPSI database covers a wide range of metallicity and initial helium abundance; the isochrones were constructed from stellar evolution tracks for stars in the mass range 0.15 $\Msun$ to 5 $\Msun$, covering the evolutionary phases from pre-MS to the tip of the RGB.
It should be noted that the YaPSI tracks and isochrones do not cover the helium burning phases (horizontal branch and asymptotic giant branch).
Consequently, the uncertainty on the age of the stars whose evolutionary state is undetermined or classified as RC might be underestimated by our methodology.
Indeed, the observed mass of a RC star may be lower than its initial mass owing to mass loss near the tip of the RGB \citep[see for instance][]{Casagrandeetal2016}.
This introduces an uncertainty on the age of the concerned stars. Quantifying this effect in binary stars is out of the scope of the present paper.

The fit was performed in the mass-radius plane.
For the 12 systems for which a determination of the radius of both the primary and the secondary components is available, we determined the age range of isochrones that are compatible, within the error bars, with both stars for a given chemical composition.
This process was repeated for the central value of the metallicity [Fe/H] and for its minimum and maximum values as determined from the error on the [Fe/H] measurement.
For the remaining systems, our isochrones provide an estimate of the radius of the secondary component.
We note that, since in all these systems the secondary is a low-mass star on the MS, the radius estimate obtained in this way is very stringent.
The results of this fitting procedure are summarized in Figs.~\ref{fig:iso_fit1}, and ~\ref{fig:iso_fit2}.
The age ranges are given in Table~\ref{tab:massesAndRadii}.
We obtained the values of this table by computing the arithmetic mean of the smallest and largest age possible for each system (e.g., for KIC 4054905, 6.32 Gyr and 9.4 Gyr, see Fig.~\ref{fig:iso_fit1}).
The uncertainties reported correspond to the half difference between these two boundaries.

\section{Results and discussions}
\label{sec:results}

In this section, we first give an overview of the variety of systems that we identified. We then check the accuracy of the seismic scaling relations using our extended sample and we propose new reference values. Next, we present the oscillation properties of the giants we characterized and analyze how they are impacted by close binarity. Finally, we study the tidal circularization of the systems presented in this work and compare these to the other binaries hosting a RG from the literature.

\subsection{Nature of the systems}
\label{subsec:natureSystems}

As shown in Figs.~\ref{fig:RV_eclipses_SB2} and~\ref{fig:RV_eclipses_SB1}, among the 17~binaries studied in this work nine are SB2s and eight are SB1s.
To determine the mass and radius of the components of SB2 systems, we used the dynamical results from the analysis of Sect.~\ref{subsec:dynamical}.
In the SB2 case, determining the masses of the components and semimajor axis of the system is straightforward through the following relations:
\begin{align}
M_1^{\mathrm{(SB2)}} &= \frac{\Porb}{2 \pi\ G \sin^3 i} \left(1 - e^2\right)^{3/2} (K_1 + K_2)^2 K_2, \label{eq:determine_m1_dyn} \\
M_2^{\mathrm{(SB2)}} &= \frac{\Porb}{2 \pi\ G \sin^3 i} \left(1 - e^2\right)^{3/2} (K_2 + K_1)^2 K_1, \label{eq:determine_m2_dyn} \\
a^{\mathrm{(SB2)}} &= \frac{\sqrt{1 - e^2}}{2 \pi \sin i} (K_1 + K_2) \Porb, \label{eq:determine_a_dyn}
\end{align}
where $K_1$ and $K_2$ are the radial velocity semi-amplitudes of the giant and its companion, respectively.
The radii of the components can then be determined from the semimajor axis, the sum of the fractional radii, and the radius ratio of the system.
For the SB1 systems, we computed the mass, radius, and effective temperature of the companion by combining the RG mass and radius obtained through seismic analysis (see Sect.~\ref{subsec:seismo} and~\ref{subsec:testScalingLaws2}) with the orbital parameters given by JKTEBOP.
In this case, the value of $K_2$ is unknown so it is impossible to apply the relations~\eqref{eq:determine_m1_dyn}, \eqref{eq:determine_m2_dyn}, and \eqref{eq:determine_a_dyn}.
The mass function $f(m)$ of the system can however be determined.
This quantity is defined as
\begin{equation}
f(m) = \frac{\left(M_2 \sin i\right)^3}{\left(M_1 + M_2\right)^2},
\end{equation}
and is linked to the observable orbital parameters of the system through the following relation:
\begin{equation}
f(m) = \frac{\Porb K_1^3}{2 \pi\ G} \left(1 - e^2\right)^{3/2}.
\end{equation}
The mass ratio of the system, $q$, can then be determined by solving the following third-degree equation:
\begin{equation}
q^3 - \alpha q^2 - 2 \alpha q - a = 0,
\end{equation}
where
\begin{equation}
\alpha = \frac{f(m)}{M_1 \sin^3 i}.
\end{equation}
Finally, the radius of the secondary component can be computed from the seismic radius of the giant and the radius ratio of the binary star.
The equations given in this paragraph are obtained by studying the dynamics of a two-body system in gravitational interaction.
A clear presentation of this computation can be found in the lecture notes of \citet{Benacquista2013}.
The masses and radii of the stars in our sample are given in Table~\ref{tab:massesAndRadii}.
Overall, we identify four classes of systems: 1) four SB2s with a pulsating giant; 2) five SB2s with a non-oscillating giant; 3) two SB1 with a hot MS companion; and 4) six SB1s with a red-dwarf companion.
We now describe each group in more detail.

\begin{sidewaystable*}
\caption{Physical properties of the 17 systems studied in this paper. 
}
\label{tab:massesAndRadii}
\centering
\scalebox{0.81}{
\begin{tabular}{llllllllrllllllllllr}
\hline\hline
& \multicolumn{9}{c}{Red Giant} & &\multicolumn{3}{c}{Companion} &&\multicolumn{3}{c}{Spin/orbit} \\
\cline{2-10} \cline{12-14} \cline{16-18}
KIC & Evol & $\Mpdyn$ & $\Mpsis$ & $\Rpdyn$  & $\Rpsis$ & $\log g\ind{dyn}$  &  $\log g\ind{seis}$ &  $T\ind{eff,1}$ & [Fe/H] && $M_2$   & $R_2$ & $\TeffComp$ &&  $e$ & $P\ind{orb}$ & $P\ind{rot}$ & Age & Notes\\
    && ($\Msun$)  & ($\Msun$)  & ($\Rsun$)   & ($\Rsun$)  & (dex) & (dex) & (K) &  (dex) & &($\Msun$) & ($\Rsun$) & (K) && & (days) & (days) & (Gyr) & \\
\hline 
\multicolumn{20}{c}{Double-line spectroscopic binaries (SB2)}\\ 
\hline 
4054905 & RC & 0.95(4) & 0.91(6) & 8.19(8) & 8.2(2) & 2.589(9) & 2.571(9) & 4790(190) & -0.7(3) && 0.93(1) & 3.11(3) & 5100(197) & & 0.37 & 274.7 & ... & 7.9(1.6) &   \\ 
4663623 & ... & 1.41(8) & 1.52(6) & 9.8(2) & 10.0(1) & 2.60(1) & 2.623(5) & 4812(92) & -0.13(6) && 1.41(10) & 1.83(5) & 6827(140) & & 0.40 & 358.1 & ... & 2.0(0.2) & $l=1$ depleted \\ 
5193386 & RGB & 1.39(3) & ... & 4.49(5) & ... & 3.274(4) & ... & 4780(100) & -0.36(6) && 1.17(5) & 1.37(1) & 6622(127) & & 0.01 & 21.4 & 26(2) & 2.5(0.6) & Flares, Ca H\&K em \\ 
6307537 & RGB & 1.29(7) & ... & 4.45(6) & ... & 3.25(1) & ... & 4960(240) & -0.0(4) && 1.06(4) & 1.17(2) & 6387(306) & & 0.01 & 29.7 & 78(6) & 5.3(3.0) & Flares, Ca H\&K em \\ 
7133286 & ... & 1.05(3) & ... & 9.21(8) & ... & 2.532(7) & ... & 4500(110) & -0.6(2) && 1.02(3) & 1.58(1) & 6075(148) & & 0.01 & 38.5 & 38(3) & 5.4(0.9) & Flares, Ca H\&K em \\ 
7293054 & ... & 1.6(1) & 1.56(10) & ... & 11.4(2) & ... & 2.518(8) & 4790(160) & 0.1(3) && 1.4(1) & 4.0(3) & 5900(374) & & 0.80 & 671.8 & ... & 0.9(0.3) & Cmp eclipsed \\ 
8435232 & ... & 1.20(4) & ... & 12.9(2) & ... & 2.297(8) & ... & 4460(130) & -0.1(2) && 1.04(3) & 1.28(2) & 6347(174) & & 0.00 & 49.6 & 48(3) & 5.3(1.7) & Ca H\&K em \\ 
9153621 & ... & 1.1(2) & 1.16(8) & 10.4(6) & 10.4(2) & 2.45(4) & 2.470(9) & 4760(190) & -0.3(2) && 0.93(10) & 1.02(6) & 6170(235) & & 0.70 & 305.8 & ... & 7.8(5.2) &   \\ 
11235323 & RGB & 1.03(1) & ... & 3.578(9) & ... & 3.342(4) & ... & 4840(200) & -0.4(2) && 0.989(2) & 1.544(4) & 6319(254) & & 0.01 & 19.7 & 24(2) & 7.2(1.7) & Flares, H$_\alpha$ em.,\\
&&&&&&&&&&&&&&&&&&& Ca H\&K em \\ 
\hline 
\multicolumn{20}{c}{Single-line spectroscopic binaries (SB1)}\\ 
\hline 
4360072 & RC & ... & 1.05(8) & ... & 10.7(2) & ... & 2.402(9) & 5020(210) & -0.1(2) && 0.71(3) & 0.5(1) & 5923(530) & & 0.15 & 1084.8 & ... & 11.5(5.0) &   \\ 
4473933 & ... & ... & ... & ... & ... & ... & ... & 4530(220) & -0.4(3) && ... & ... & 7172(344) & & 0.28 & 103.6 & 68(6) & ... & Ca H\&K em \\ 
5866138 & RC & ... & 1.57(7) & ... & 8.1(1) & ... & 2.820(6) & 4960(120) & 0.1(1) && 0.49(2) & 0.456(10) & 3252(93) & & 0.72 & 342.3 & ... & 2.0(0.6) &   \\ 
6757558 & RGB & ... & 0.88(3) & ... & 4.96(6) & ... & 2.992(5) & 4590(110) & -0.0(1) && 0.217(9) & ... & ... & & 0.22 & 421.2 & ... & 12.1(3.5) & RG eclipsed \\ 
7768447 & RGB & ... & 1.12(7) & ... & 8.3(2) & ... & 2.650(8) & 4760(160) & 0.2(2) && 0.63(3) & 0.59(3) & 4115(555) & & 0.32 & 122.3 & ... & 8.8(3.2) &   \\ 
9904059 & RGB & ... & 0.98(6) & ... & 4.96(8) & ... & 3.039(7) & 4830(160) & 0.0(3) && 0.46(2) & ... & ... & & 0.32 & 103.0 & ... & 9.9(4.0) & Cmp eclipsed \\ 
10015516 & RC & ... & 1.75(9) & ... & 9.6(2) & ... & 2.716(7) & 4830(130) & -0.4(2) && 1.33(6) & 1.52(3) & 6909(177) & & 0.00 & 67.7 & 66(6) & 0.8(0.2) & Ca H\&K em \\ 
10074700 & ... & ... & 0.84(4) & ... & 3.53(5) & ... & 3.267(6) & 5070(100) & -0.4(1) && 0.6(3) & 0.56(1) & 4020(520) & & 0.29 & 365.6 & ... & 13.9(3.9) & Faint \\ 
\hline
\end{tabular}}
\tablefoot{The parameters $M$, $R$, $\log g$, $T\ind{eff}$, and [Fe/H] refer to stellar masses, radii, surface gravities, effective temperatures, and metallicities, respectively. The subscripts ``dyn'' and ``seis'' stand for dynamical modeling and asteroseismic scaling relations, respectively. In the comment column, em. stands for spectral ``emission''; Ca H \& K indicate the calcium lines at 3968.5 and 3933.7 \AA, respectively, and H$_\alpha$ the hydrogen Balmer line at 6562.8 \AA. The comments ``RG eclipsed'' and ``Cmp eclipsed'' indicate that only one type of eclipse is visible in the light curve, either the RG is eclipsed (RG) or the companion (Cmp).
For SB1 systems, the parameters of the companion stars are deduced by combining asteroseismic masses and radii of the RG with the mass function obtained from light curve and radial velocity modeling. For RC stars, the age uncertainty may be underestimated (see discussion in Sect.~\ref{subsec:age_stellar_model}). We could not determine the mass, radius, and age of the giant component of KIC~4473933 because this binary is an SB1 (see Fig.~\ref{fig:RV_eclipses_SB1}) and its power spectrum has no detectable oscillations (see Fig.~\ref{fig:psd_non_osc}). The values of $T\ind{eff}$, [Fe/H], $P\ind{orb}$, and $e$ are reported to provide a view of each system at a glance. The exact values with uncertainties are shown in Tables \ref{tab:atmosphericParam} and \ref{tab:seismicParam}.}
\end{sidewaystable*}

\subsubsection{SB2 systems with a pulsating giant}
This group is interesting because it can be used to test the accuracy of the seismic scaling relations (Sect.~\ref{subsec:testScalingLaws2}). These four systems are KIC~4054905, KIC~4663623, KIC~7293054, and KIC~9153621.

KIC~4054905 is composed of two nearly identical-mass stars ($M\approx0.95~\Msun$) on a 275-day eccentric orbit ($e= 0.37$). The primary component is a giant with a radius of 8~$\Rsun$ and its companion is a subgiant with a radius of 3~$\Rsun$. We only detect the oscillations of the largest component. The mixed-mode analysis reveals that the giant component belongs to the RC, which means that it has already traveled through the tip of the RGB, where its radius was on the order of 180~$\Rsun$. Knowing the eccentricity and that the semimajor axis of the orbit is 1~AU, the distance in between the two components at the periastron is 137~$\Rsun$. This means that this system has likely experienced mass exchanges during the evolutionary phase where its primary component was at the tip of the RGB. This star is one of the very few eccentric binary systems that contain a helium core-burning star. While the existence of such binary systems was disputed by \citet{VerbuntPhinney1995}, \citet{Becketal2018} show that such systems do exist. Such systems are also relevant to determine the efficiency of the equilibrium tide.

KIC~4663623 is a system composed of two $\approx1.5~\Msun$ stars on a 358-day eccentric orbit ($e=0.4$). The giant component has a radius of 10~$\Rsun$ whereas its companion lies close to the MS with a radius of 1.8~$\Rsun$. The $l=1$ modes are severely, although not totally depleted, which makes the mixed-mode analysis inconclusive. While \citet{Gaulmeetal2014} conclude it was a RGB, our analysis suggests it is a RC, which is supported by the measurements of \citet{Kallingeretal2018}. The evolutionary tracks that we ran to determine the age of the system are compatible with the RG on the RGB and the companion on the MS, but the error bars are very large. The 180-$\Rsun$ distance in between the two components at periastron suggests that the stars exchanged mass earlier in their evolution if the giant is on the RC.

The system KIC~7293054 is composed of a 1.6 $\Msun$ RG and a 1.4 $\Msun$ companion orbiting in about 672 days on a very eccentric orbit ($e=0.8$). While the asteroseismic analysis indicates an 11.6 $\Rsun$ for the giant, it is impossible to determine the radius of the companion because this binary has no secondary eclipse (Fig.~\ref{fig:RV_eclipses_SB2}). Moreover, since we found no mixed modes in the power spectrum of the giant, it is difficult to accurately determine its evolutionary state. The high eccentricity of this system makes it an interesting test case for testing tidal theory. While the orbital semimajor axis is 500~$\Rsun$, the distance between the two components at periastron is less than 80~$\Rsun$.

KIC~9153621 is composed of a giant of mass 1.1~$\Msun$ and radius 10~$\Rsun$ and a MS star of mass 0.9~$\Msun$ and radius 1~$\Rsun$.
The high uncertainty on the primary dynamical mass is due to the poor constraint on $K_2$.
The relation between these two quantities is given in Eq.~\eqref{eq:determine_m1_dyn}.
The absence of mixed modes in the power spectrum does not allow us to determine the evolutionary state of the primary component.
Because of its large orbital eccentricity ($e=0.7$) and semimajor axis of 1~AU, the distance at periastron (65~$\Rsun$) is small enough that mass exchange either has happened or will happen during the RGB ascension of the primary component.

\subsubsection{SB2 systems with a non-oscillating giant}

These five EBs (KIC~5193386, KIC~6307537, KIC~7133286, KIC~8435232, and KIC~11235323) have similar characteristics.
They constitute the systems with the shortest orbital periods of the sample and their values range from 20 days to 50 days.
The masses and radii of the primary components range from 1.03~$\Msun$ to 1.38~$\Msun$ and from 3.6~$\Rsun$ to 12.9~$\Rsun$, respectively.
The mass ratios $M_2 / M_1$ are very homogeneous and range from 0.82 to 0.96.
Finally, their orbits are all circular and their giant components are active.
These properties are further discussed in Sect.~\ref{subsub:nonOscRG}.

\subsubsection{SB1 systems with a hot MS companion}

These two systems (KIC~4473933 and KIC~10015516) are challenging because they are SB1s despite luminosity ratios of 9\% and 11\%, respectively.
They are both composed of an active RG and a MS star with $\TeffComp \simeq$ 7\,000~K.
The non-detection of these two companions, despite favorable fractional light ratios, is most probably due to the broadening of their spectral lines resulting from their rapid rotation.
We note that, in the two systems, the rotation period of the giant component is equal to the tidal pseudo-synchronization period predicted by \citet{Hut1981}.
Despite these similarities, their orbital configurations are slightly different, which we detail in this section.

KIC~4473933 is the only non-oscillating RG that does not belong to an SB2 among the 36 known RGs in EBs. Both the photometric and spectroscopic data allow us to affirm that KIC~4473933 is composed of a giant and a hot MS star. First, since the deepest eclipse has a flat bottom and the shallowest has a round bottom, the component with the biggest radius is the coldest, which indicates that it is an evolved star that has left the MS. This is confirmed by the analysis of its disentangled optical spectrum, which showed that its surface gravity and effective temperature were those of a RG (see Table~\ref{tab:atmosphericParam}). As mentioned above, this giant is active and its rotation is pseudo-synchronized as predicted by \citet{Hut1981}, which indicates that this binary has experienced tidal interaction. We note that the orbital period of this system, $\Porb \simeq$ 104~d, is significantly longer than those of the other systems with no detectable oscillations. We also note that this system has a significant nonzero eccentricity ($e$ = 0.279), which is further discussed in paragraph~\ref{subsub:nonOscRG}.

The system KIC~10015516 is particularly interesting because it is composed of an oscillating RG and a pulsating $\gamma$~Doradus ($\gamma$~Dor) star. Although analyzing the pulsations of the companion is out of the scope of the present work, we estimate the mass and radius of the $\gamma$~Dor by combining the seismic mass and radius of the giant with the orbital parameters of the system. The resulting parameters are fully compatible with a $\gamma$~Dor variable star (see Table~\ref{tab:massesAndRadii}). Another interesting feature of this system is that its giant is active and its rotation is synchronized to the orbit. It is the only binary in the sample presented in this work whose primary component is active and has detectable oscillations. We show in Sect.~\ref{sec:tides} that this system is a precious test case for tidal evolution theories.

\begin{figure}
\centering
\includegraphics[width=\hsize]{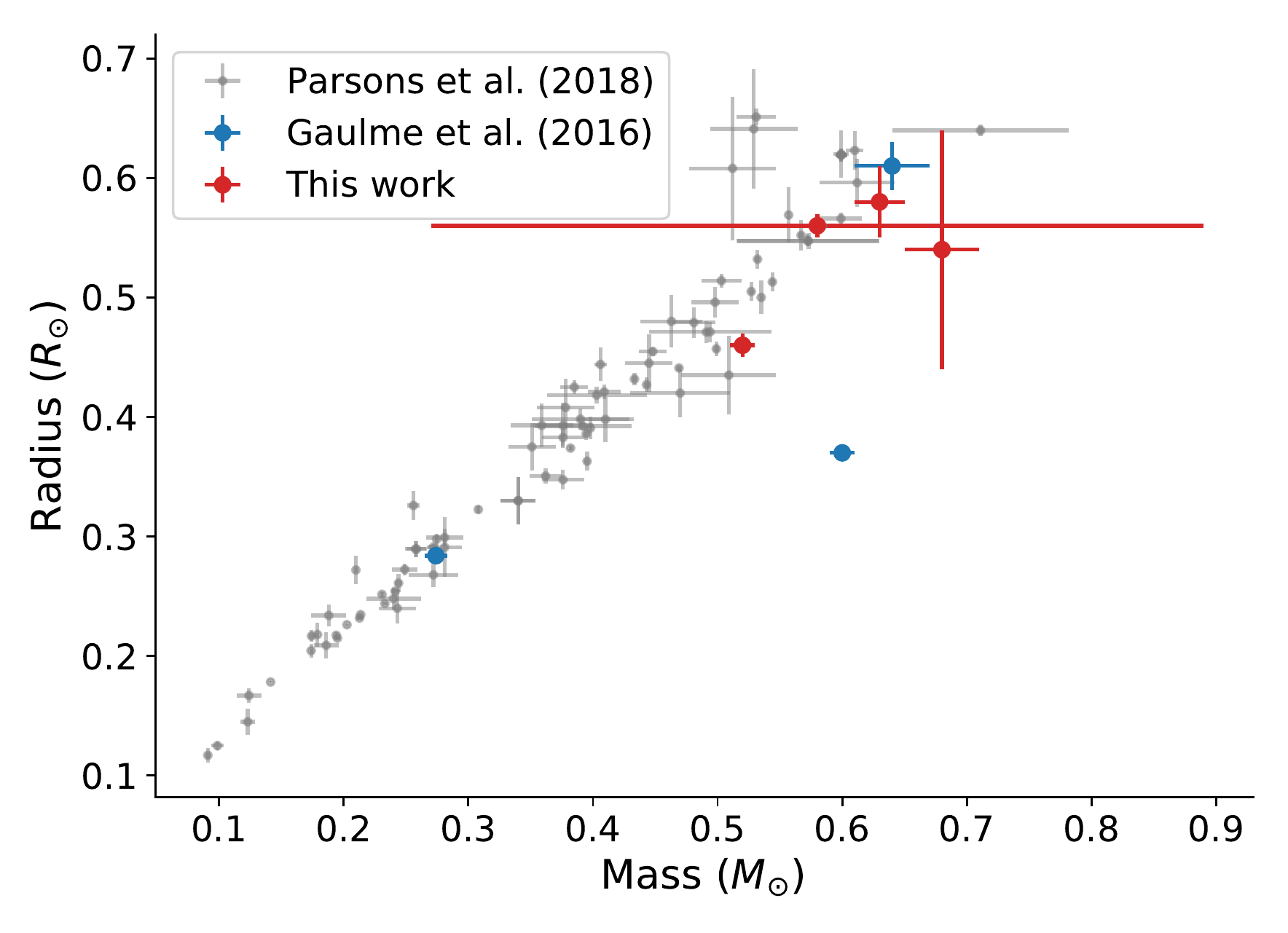}
\caption{Radii of M and K dwarfs as a function of their masses. The gray dots correspond to the measurements obtained by \citet{Parsonsetal2018} by compiling catalogs and observing new targets. Among the values presented by these authors, only those with a relative uncertainty less than 10\% are shown.
We note that the uncertainty on the radius (resp. mass) of the companion of KIC~4360072 (resp. KIC~10074700) is greater than 10\%.
We however decided to keep them on the plot to allow for comparison with the most up-to-date scatterplot.
}
\label{fig:mDwarfRelationMR}
\end{figure}

\subsubsection{SB1 systems with a red-dwarf companion}
\label{subsub:nature_systems_sb1_red_companion}

The six binaries of this group (KIC~4360072, KIC~5866138, KIC~6757558, KIC~7768447, KIC~9904059, and KIC~10074700) are composed of an oscillating RG and an MS star less massive than 0.7~$\Msun$. As noted by, for example, \citet{Parsonsetal2018}, characterizing such low-mass stars is important since they are priority targets for the search for habitable planets. These authors also noted that the mass and radius of these objects were generally poorly constrained because of their low luminosity. Since the number of known M or K dwarfs for which the mass and radius are measured within an uncertainty less than 10\% is about 40, any new star is a precious addition.

We characterize the dwarf companions from the seismic properties of the giant and the orbital parameters from JKTEBOP.
Since KIC~6757558 and KIC~9904059 have no secondary eclipse, the radius of the dwarf component cannot be determined.
We also note that we could not measure any radial velocity for KIC~10074700 because of its low luminosity.
Consequently, the secondary star in this system was only characterized using photometric data.

Concerning KIC~4360072, the very high effective temperature of the companion is surprising given its mass and radius ($\TeffComp$ = 5730 K, $M_2 = 0.68\,\Msun$, $R_2 = 0.54\,\Rsun$, see Table~\ref{tab:massesAndRadii}).
The most probable cause for this discrepancy is the surface activity of the giant in this system, which can be seen in Fig.~\ref{fig:RV_eclipses_SB1}.
Because of the very long orbital period of this system (1084 d), only one primary eclipse and two secondary eclipses were observed by \textit{Kepler}.
As a result, the surface activity signal was not averaged out in the phase-folded light curve, which might have biased the dynamical analysis.
A more detailed investigation is required to better characterize the companion of this system, which is out of the scope of the present work. 

We compare our results with the known mass--radius relationship for this type of objects.
In Fig.~\ref{fig:mDwarfRelationMR}, we plot the radius of the 40 stars published by \citet{Parsonsetal2018} as a function of their masses and add the dwarf companions of our sample.
It appears that our stars are in the middle of the already known distribution.
The large uncertainty on the mass of the secondary component of KIC~10074700 is a consequence of the absence of radial velocity measurements for this system.
We also plotted the three dwarf companions of the sample of \citet{Gaulmeetal2016}.
Whereas two of these are in good agreement with the mass--radius relationship, the secondary of KIC~5179609, with $M = 0.60~\Msun$ and $R = 0.37~\Rsun$, lies significantly below this value.
This most likely results from an underestimation of the ratio of the radii.
Since this EB displays very shallow secondary eclipses, modeling these secondary eclipses is challenging because they are difficult to disentangle from the other signals present in the light curve such as the variability caused by the oscillations.
This makes errors on their depth and width more probable and, as a consequence, can cause significant errors on the radius and luminosity ratios of the system.

\begin{figure*}
\includegraphics[width=17cm]{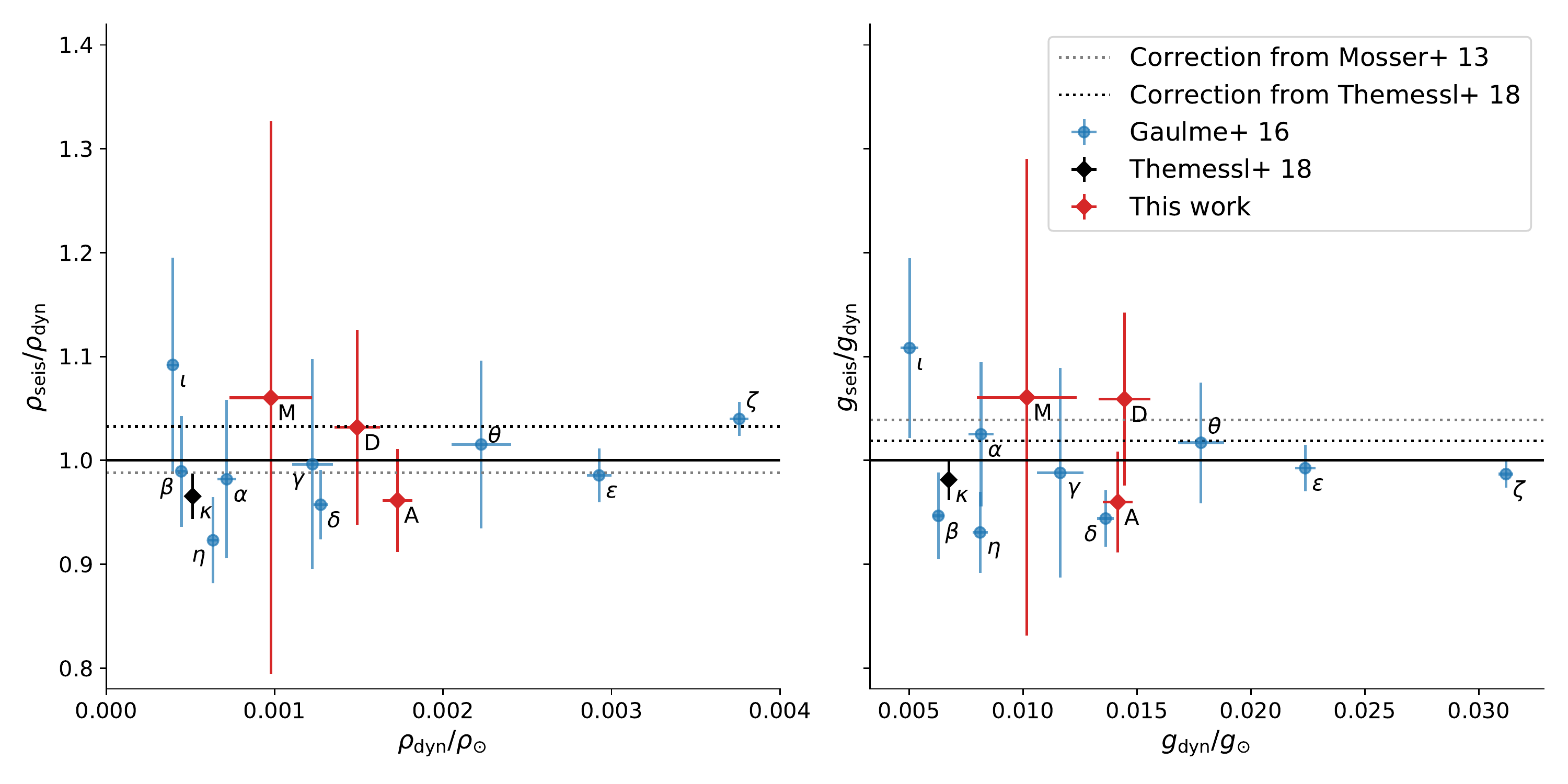}
\caption{Comparison between the asteroseismic estimates and dynamical measurements of the mean density and surface gravity. Left panel: The ratio of the seismic mean density, $\rhoSis$, divided by the dynamical mean density, $\rhoDyn$, as a function of the dynamical mean density. Right panel: The ratio of the seismic surface gravity, $\gSis$, divided by the dynamical surface gravity, $\gDyn$, as a function of the dynamical surface gravity. Each dot corresponds to an SB2 EB with an oscillating RG component. The blue dots represent those published by \citet{Gaulmeetal2016}, the black dot, the star discovered by \citet{Themessletal2018} and the red dots, those published in this paper, to which we added KIC~4663623, the star from \citet{Gaulmeetal2016} that we reanalyzed in this study. The correspondence between the letters and the KIC indices are given in Tables.~\ref{tab:atmosphericParam} and~\ref{tab:benchmarksFromLiterature}. KIC~7293054 is not shown in this figure because its radius could not be determined dynamically. The seismic estimates for the points shown were obtained with the correction described in subsection~\ref{subsec:testScalingLaws2}. The dotted gray and black lines represent the means obtained with the reference values of \citet{Mosseretal2013} and \citet{Themessletal2018}, respectively. The black plain line indicates a ratio of 1. }
\label{fig:compDynSisRhoGravLetters}
\end{figure*}

\subsection{Testing the accuracy of the asteroseismic scaling relations}
\label{subsec:testScalingLaws2}

One of the main interests of studying solar-like oscillators in EBs is to check whether the stellar masses and radii obtained with asteroseismology agree with the dynamical quantites.
Since the latter method relies on basic physical principles, it can be considered a robust and unbiased method to determine the mass and radius of a star.
The dynamical and seismic analyses have already been compared in previous works, in particular by \citet{Frandsenetal2013}, \citet{Gaulmeetal2016}, \citet{Brogaardetal2018}, and \citet{Themessletal2018}.
Based on a sample of ten RGs, \citet{Gaulmeetal2016} found that the asteroseismic scaling relations tend to overestimate the masses and radii by about 15\% and 5\%, respectively.
Such a discrepancy is significant and may cause large biases when studying the evolution of stellar populations.

A subsample of three of these stars were studied a second time by \citet{Brogaardetal2018}, who obtained new radial velocity measurements and new effective temperatures and metallicities. Beyond small discrepancies, these authors confirmed the mass and radius overestimation reported by \citet{Gaulmeetal2016} when employing the asteroseismic scaling relations. In addition, for the first time, they tested the online software PARAM \citep[][]{Rodriguesetal2017}\footnote{\url{http://stev.oapd.inaf.it/cgi-bin/param}}, which uses grids of stellar evolution models to retrieve RG masses and radii from $\numax$, $\Dnu$, $\Teff$, and [Fe/H]. They observed a good agreement between the PARAM results and the dynamical measurements. Finally, \citet{Themessletal2018} led a similar study with the ten systems studied by \citeauthor{Gaulmeetal2016} and a new system that had not been published before. For the new system and two others, \citet{Themessletal2018} measured new radial velocities and determined the dynamical mass and radius. To circumvent the mass and radius overestimation, they proposed an adjustment of reference values ($\DnuRef$, $\numaxRef$, $\TeffRef$) which provides a better agreement between the asteroseismic and dynamical stellar properties.

We performed a similar experiment with all the binaries that have been published and the new systems presented in this work. In other words, the total sample now comprises 14 systems, including the following:
1) the ten binaries that were already published in 2016 \citep{Frandsenetal2013,Rawlsetal2016,Gaulmeetal2016,Brogaardetal2018}; 2) KIC~5640750 \citep{Themessletal2018}; and 3) KIC~4054905, KIC~7293054, and KIC~9153621 from the present study.

First of all, our results are consistent with the conclusions of \citet{Gaulmeetal2016}.
Indeed, when we apply the same asteroseismic scaling relations, that is, when we use the approach and solar reference values of \citet{Mosseretal2013}, seismic masses are overestimated by approximately 15\,\% $\pm$ 10\,\%, and radii by 5\% $\pm$ 3\,\%.
The correction of \citet{Mosseretal2013} consists of multiplying the observed value of $\Dnu$ by a correcting factor (equal to 1.038 for RGs) and using solar reference values provided in their paper ($\Dnu_{\odot,\mathrm{M13}}$ = 138.8 $\mu$Hz, $\nu_{\mathrm{max},\odot,\mathrm{M13}}$ = 3104 $\mu$Hz)\footnote{
Appropriately applying the correction of \citet{Mosseretal2013} actually involves a recalibration of the solar global seismic parameters $\Dnu_{\odot}$ and $\nu_{\mathrm{max},\odot}$ to ensure these reference values are consistent with the used seismic analysis pipeline.
We decided however not to proceed to this calibration because our pipeline is very similar to that of \citet{Mosseretal2013}.
}.
Secondly, when we apply the latest corrections of the scaling relations proposed by \citet{Themessletal2018}, the seismic masses and radii match the dynamical values within the error bars.
We observe a slight underestimation of 1\% for both masses and radii on average, with a dispersion of 9\% and 3\%, respectively.
We thus confirm that their proposed tuned reference values reduce the biases in mass and radius
\footnote{
It should be noted that a suitable application of the \citet{Themessletal2018} correction would require us to tune the solar reference values so as to cancel the average deviation between seismic and dynamical masses and radii in the 14-benchmark star sample.
In this work, we directly used the values they reported for two reasons.
First, the point of this section is to show that a proper calibration of the reference values has to be done using the stellar mean density and surface gravity instead of mass and radius.
Second, since directly using their reference values already gave a satisfying result, recalibrating these values was not necessary to show the improvement they yielded.
}.

\begin{figure*}
\includegraphics[width=17cm]{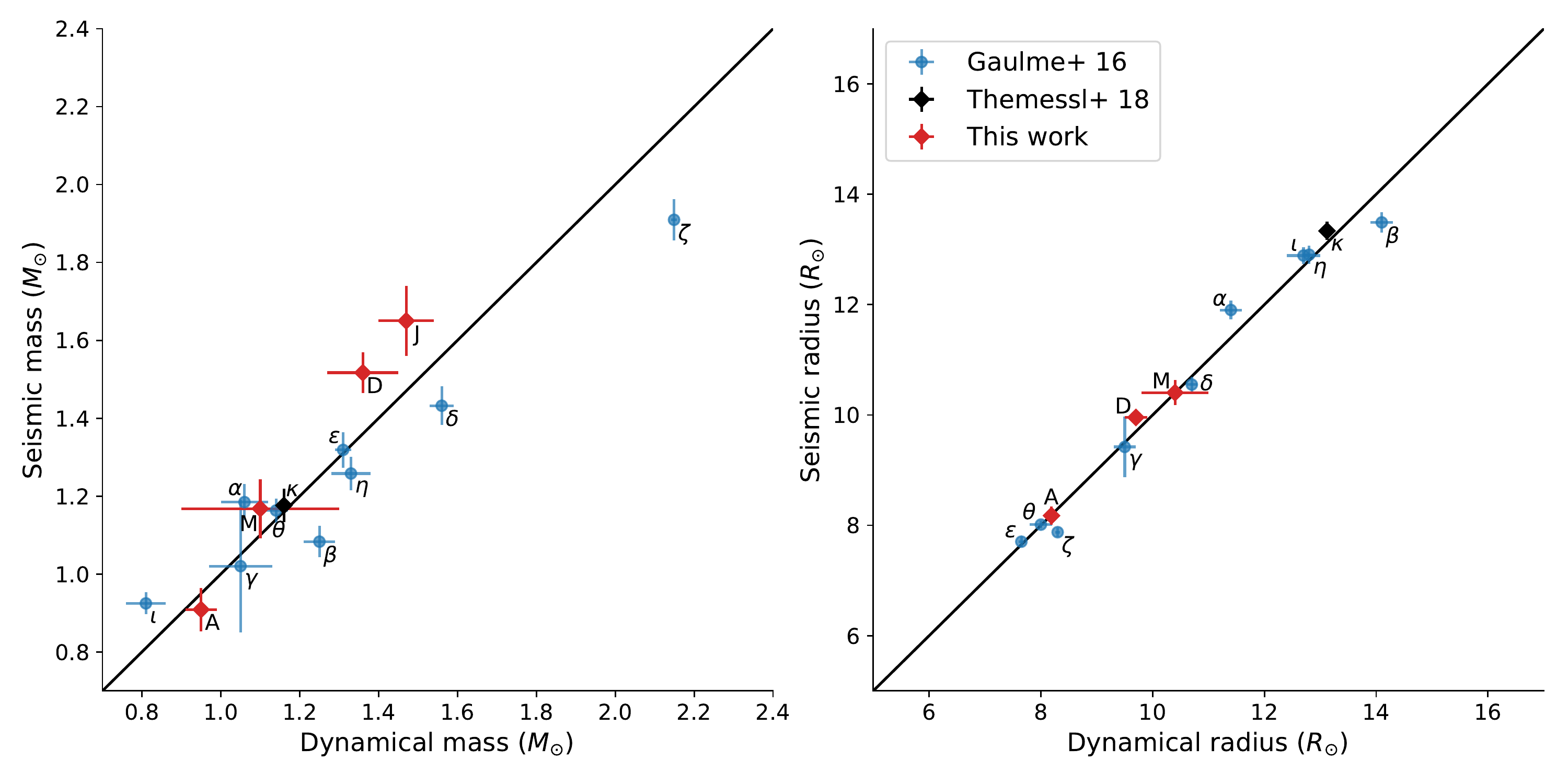}
\caption{Comparison between the asteroseismic estimates of the stellar masses and radii and their dynamical measurements. The seismic estimates for the points shown were obtained using the unbiased correction described in subsection~\ref{subsec:testScalingLaws2}. Left panel: Comparison in mass. Right panel: Comparison in radius. See Tables.~\ref{tab:atmosphericParam} and~\ref{tab:benchmarksFromLiterature} for the correspondence between the letters and the KIC indices.}
\label{fig:compDynSisMassRadLetters}
\end{figure*}

However, a proper comparison between the dynamical method and the seismic scaling relations should be done in terms of mean density and surface gravity instead of stellar mass and radius \citep[e.g.,][]{Kallingeretal2018}.
Indeed, the scaling relations connect the global seismic parameters to the stellar mean density $\bar{\rho}$ and the surface gravity $g$.
When we compare the dynamical measurements of these quantities ($\gDyn$, $\rhoDyn$) to their seismic estimates ($\gSis$, $\rhoSis$), we observe that the reference values used by \citet{Gaulmeetal2016} lead to overestimating the surface gravity by 4\% $\pm$ 6\% and underestimating the mean density by 2\% $\pm$ 5\%.
We note that, although both density and gravity agree within the error bars, the masses and radii do not.
This is because of the opposite signs of the biases in $g$ and $\bar{\rho}$, and the strong dependence of stellar mass as a function of gravity and density ($M \propto g^3 / \bar{\rho}^2$).
In contrast, using the reference values of \citet{Themessletal2018} leads to overestimating both the surface gravity and the mean density by 2\% $\pm$ 6\% and 3\% $\pm$ 5\%, respectively, which leads to an unbiased result on $M$ and $R$.

In this work, we propose to calibrate the seismic reference values $\DnuRef$, $\numaxRef$, and $\TeffRef$ directly from mean density and surface gravity. The principle is to choose these references so that the average value over the sample of 14 RGs of the seismic-to-dynamical density and gravity ratios, $<\rhoSis / \rhoDyn>$ and $<\gSis / \gDyn>$, are both equal to 1. The value of $\DnuRef$ is determined by the constraint on $<\rhoSis / \rhoDyn>$, while the constraint on $<\gSis / \gDyn>$ fixes the value of the product $\numaxRef \sqrt{\TeffRef}$. Consequently, there is one degree of freedom in the choice of $\numaxRef$ and $\TeffRef$. We decide to fix $\TeffRef$ = 5771.8~K, which is a recent solar effective temperature estimate used by \citet{Themessletal2018}. In these conditions, the new seismic reference values are given as
\begin{align}
\DnuRef   &= 132.9~\mu\mbox{Hz}, \label{eq:DnuRefNew} \\
\numaxRef &= 3227~\mu\mbox{Hz}. \label{eq:numaxRefNew}
\end{align}
With these new reference values, the dispersion of the mismatch between seismic and dynamical values is 5\% for both the mean density and surface gravity (Fig.~\ref{fig:compDynSisRhoGravLetters}).
With these values of $\DnuRef$ and $\numaxRef$, the stellar masses and radii too are unbiased (the overstimation is precisely $1.9 \times 10^{-3}$ for the mass and $5.6 \times 10^{-5}$ for the radius, see Fig.~\ref{fig:compDynSisMassRadLetters}).
These results are comparable to those of \citet{Kallingeretal2018}, who developed a correction based on the mean density and surface gravity.
The dispersion is 9\% for the mass and 3\% for the radius, following the correction from \citet{Themessletal2018}.
This plot confirms that these new reference values allow us to correct the biases of the seismic scaling relations.
Ensuring that seismic estimates of masses and radii are unbiased is important because such quantities can be used to determine other properties of stars and stellar populations \citep[see for instance][]{Miglioetal2012,SilvaAguirreetal2018,Sharmaetal2020}.
The recent studies of \citet{Halletal2019} and \citet{Zinnetal2019} also focused on the accuracy of the scaling relations.
In Table~\ref{tab:massesAndRadii}, we used the global seismic parameters of Table~\ref{tab:seismicParam} and the reference values of Eqs.~\eqref{eq:DnuRefNew} and~\eqref{eq:numaxRefNew} to estimate the mass and radius of each of the pulsating stars in our sample.
With these estimates, we can characterize all the systems including the SB1s for which only a mass ratio can be determined from the orbital dynamics.

We note that the reference values of Eqs.~\eqref{eq:DnuRefNew} and~\eqref{eq:numaxRefNew} are not strictly universal. They are specific to our seismic pipeline and should in principle be recomputed if another analysis software is used. With a different pipeline, the observed $\numax$ and $\Dnu$ may be slightly different, which would bias the outcome. In addition, we note that these reference values are obtained with a reference effective temperature of $\TeffRef$ = 5771.8~K. Another choice of reference effective temperature (e.g., 5777~K as for \citealt{Gaulmeetal2016}) would lead to different values of $\DnuRef$ and $\numaxRef$. Finally, this correction is valid in the range of mass and evolutionary state covered by the sample used in this paper. Our approach does not allow us to extrapolate outside this parameter space.

Developing a more generalizable correction would require us to take into account more accurately the stellar structure and the influence of the evolutionary state on the global seismic parameter. With this in mind, we compared the results of the method proposed by \citet{Sharmaetal2016} with those of the dynamical method\footnote{
The routine developed by \citet{Sharmaetal2016} is publicly available at \url{http://www.physics.usyd.edu.au/k2gap/Asfgrid/)}.
}.
This correction is based on searching in a grid of stellar models the synthetic star that best matches the observed asymptotic $\Dnu$, $\numax$, metallicity, evolutionary state, effective temperature, and surface gravity.
In our sample, the method of \citet{Sharmaetal2016} yielded an average overestimation of 10\% on mass and 3\% on radius, which is a slight improvement compared to the method used by \citet{Gaulmeetal2016}.
These results suggest that more accurate mass and radius estimates can be obtained if more parameters (e.g., metalicity and evolutionary state) are taken into account.
This is supported by \citet{Pinsonneaultetal2018}, who showed that RGB and RC stars may have different reference values.

The dependence of the seismic reference values on the evolutionary state implies that empirical corrections that treat RGB and RC stars equally, for example, those from \citet{Mosseretal2013}, \citet{Themessletal2018}, and the present paper, introduce systematic uncertainties at the few percent level on mass and radius.
This is an additional reason why future corrections to the seismic scaling relations will benefit from taking the evolutionary state into account.
However, testing such corrections with precision requires knowledge of the evolutionary state of the benchmark stars, which is not always the case.
For example, in this paper, the absence of mixed modes in the spectra of KIC 7293054 and KIC 9153621 prevented us from determining the evolutionary state (see Sect.~\ref{subsub:mixed_modes}).
Moreover, a larger sample of stars will be required to draw significant conclusions regarding the influence of the evolutionary state on the seismic-relation calibration.
These issues are out of the scope of the present paper.

As a final comparison, we used the PARAM online tool \citep{Rodriguesetal2017}, which was successfully tested by \citet{Brogaardetal2018}\footnote{
  In addition, \citet{Brogaardetal2018} also find good agreement between asteroseismic and dynamical estimates of mass and radius after application of standard scaling relations with $\Dnu$ corrections from \citet{Rodriguesetal2017}.
} on three systems, to estimate the masses and radii of all 14 of our systems.
The principle, similar to that of \citet{Whiteetal2011ApJ743} and \citet{Sharmaetal2016}, consists of fitting a stellar evolution model to the available data set.
The basic set is the pair ($\Dnu$, $\numax$), which can be complemented by the effective temperature, metallicity, period spacing of the mixed modes, and absolute luminosity.
We observed a strong discrepancy in the results of the new target KIC~4054905, one of the few systems that belong to the RC, when we opted to provide $\DPi$ as a constraint.
The PARAM code estimates the mass at $1.03\pm 0.05 M_\odot$ without $\DPi$ and $2.06\pm 0.02 M_\odot$ with $\DPi$, whereas the dynamical value is $M_{\mathrm{dyn}} = 0.95 \pm 0.04 M_\odot$.
An explanation to this specific case could be the fact that the RG belongs to the RC and may have experienced mass transfer with its companion at the tip of the RGB, which could have altered its structure and evolution, thus making the output of the stellar evolution models irrelevant.
Other sources of discrepancy between the predicted and observed values of $\DPi$ were studied by \citet{Constantinoetal2015}.
Their work constitutes a useful reference to investigate this question in detail.
Regarding the other systems, including a prior on the evolutionary state in the PARAM optimization allows us to get masses and radii overestimated only by 5\% and 0.5\%, which is very encouraging.
Meanwhile a simple empirical approach that is model free, as we propose, remains competitive for deriving accurate RG parameters. 

\subsection{Oscillation suppression and surface activity in close-in systems}
\label{subsub:nonOscRG}

\begin{figure}
\includegraphics[width=8.7cm]{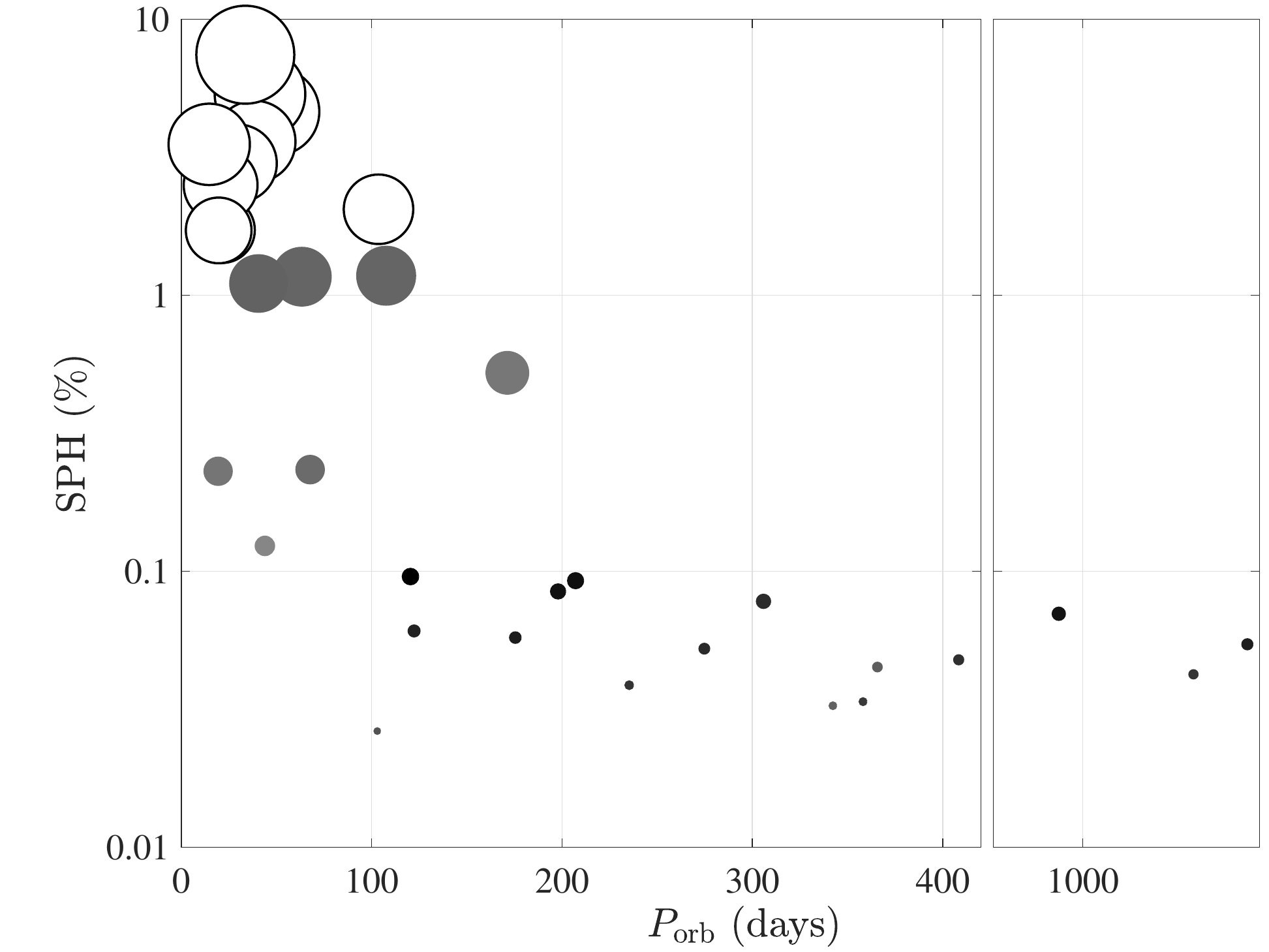}

\includegraphics[width=8.7cm]{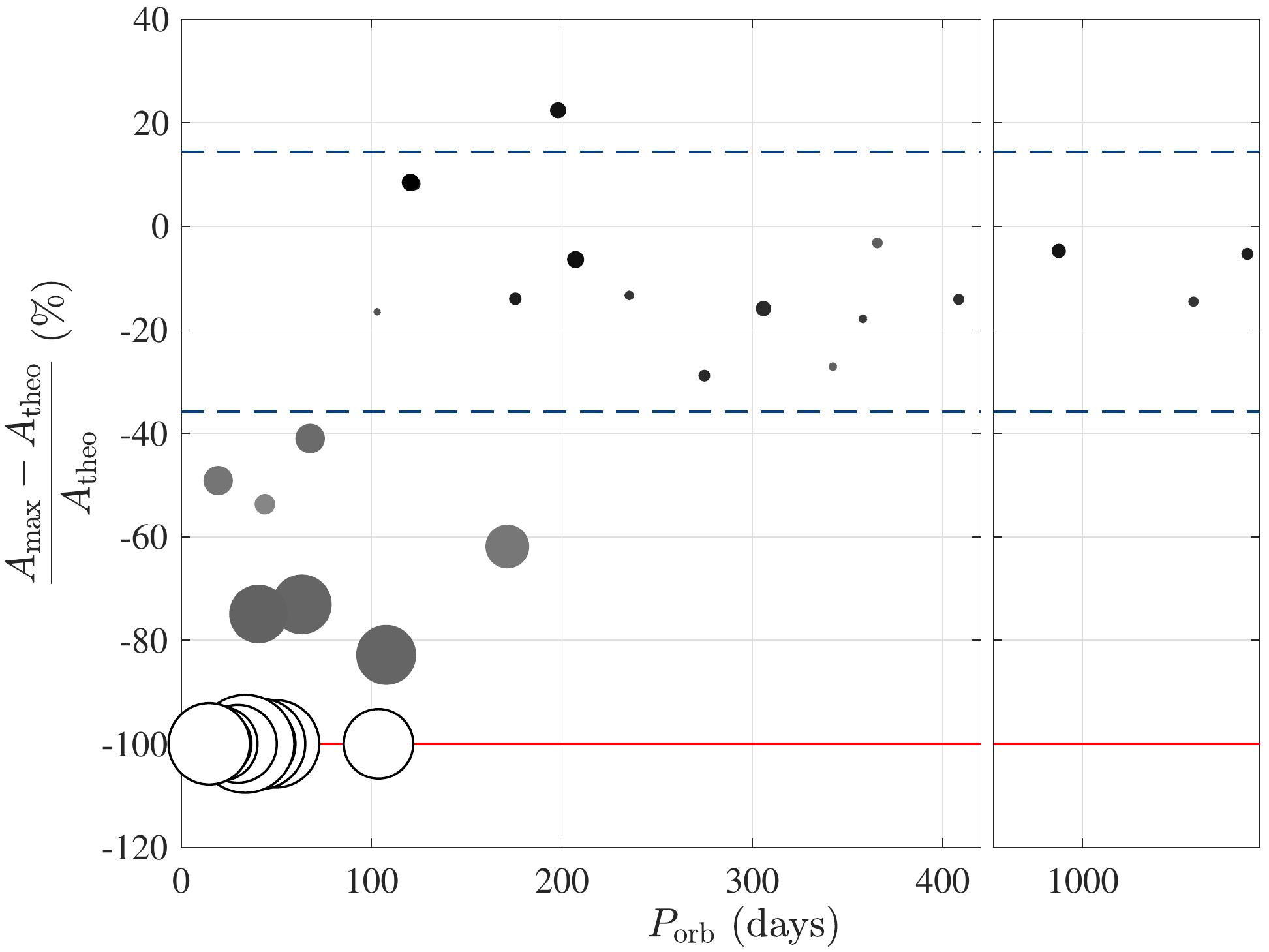}

\includegraphics[width=8.7cm]{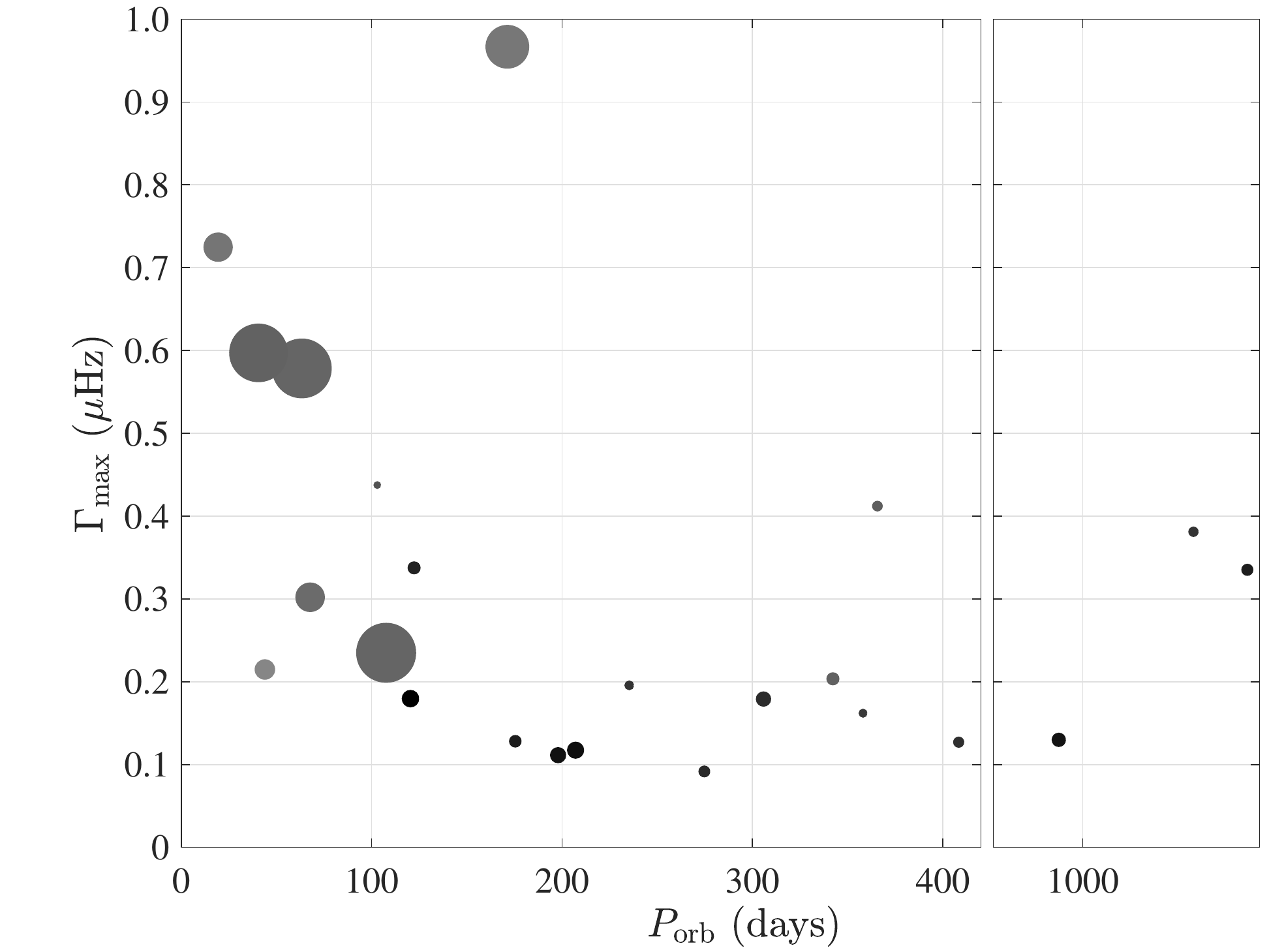}
\caption{
Properties of rotational modulations and radial ($l=0$) modes near the frequency at maximum amplitude $\numax$.
Top panel: The photometric magnetic activity level proxy $\Sph$ (\%) as a function of orbital period (days).
The $\Sph$ values are given in Tables~\ref{tab:benchmarksFromLiterature} and~\ref{tab:massesAndRadii}.
Middle panel: The relative difference in between expected and measured oscillation amplitudes (\%) as a function of orbital period $P\ind{orb}$ (days). The red line indicates 0, i.e., stars whose oscillations are not detected.
The two dashed blue lines represent the region in which relative mode amplitude lies for systems with orbital periods longer than 180 days within two sigma.
Bottom panel: The mode width $\Gamma\ind{max}$ ($\mu$Hz) as a function of orbital period $P\ind{orb}$ (d).
In all panels the size of each symbol represents the amplitude of stellar variability $\Sph$ and the gray scale indicates the pulsation mode amplitude (white-no modes; black-large amp.).
}
\label{fig:mode_amp_width}
\end{figure}

In their sample of 19~binaries, \citet{Gaulmeetal2014} found that four giants were active and had no detectable pulsations.
These four systems displayed strong pseudoperiodic modulations due to surface activity (hereafter called rotational modulations) and all had orbital periods shorter than 50~days, sums of fractional radii $(R_1 + R_2) / a$ larger than 15\,\%, and companions hotter than the giant component ($T_2>T_1$).
Shortly after that publication, some concerns about the existence of such a mode suppression were raised in the community for the following reasons.
Firstly, the oscillation signal could have been diluted because of the contribution of the companion to the light of the binary system.
The relatively small fraction of the luminosity of the companion ($L_2 / L_1 < 15\,\%$) makes that option extremely unlikely.
This was further confirmed by \citet{Rawlsetal2016}, who characterized the oscillations of two pulsating RGs that form a 171-day orbital period eclipsing binary system with a luminosity ratio close to 1; and more recently by the numerical simulations of \citet{Sekaranetal2019}, who showed that in RG binaries with a luminosity ratio on the order of 10\,\%, the determination of the seismic parameters of the more luminous component was not biased by light dilution.
Secondly, the oscillation frequencies could be out of the range covered by \textit{Kepler}.
Despite mass and radius estimates of the companions based on orbital parameters and asteroseismic properties of the giant, such an option needed to be checked.
\citet{Gaulmeetal2016} confirmed the original conclusion of \citet{Gaulmeetal2014} based on radial velocity measurements: given their masses and radii, all large companions were RGs and all should have been displaying detectable oscillations.

\begin{table}
\caption{Predicted $\Dnu$ and $\numax$ for the non-oscillating RGs in SB2 systems}
\label{tab:predictDnuNumax}
\centering
\begin{tabular}{lll}
\hline\hline
KIC & $\Dnu$ ($\mu$Hz) & $\numax$ ($\mu$Hz) \\
\hline
 5193386 & 16.51 $\pm$ 0.29  & 233.6 $\pm$ 7.7 \\
 6307537 & 16.18 $\pm$ 0.52  & 218   $\pm$ 14  \\
 7133286 & 4.902 $\pm$ 0.087 & 43.5  $\pm$ 1.6 \\
 8435232 & 3.162 $\pm$ 0.080 & 25.5  $\pm$ 1.2 \\
11235323 & 20.05 $\pm$ 0.12  & 272.9 $\pm$ 6.4 \\
\hline
\end{tabular}
\tablefoot{
Since no uncertainty on the effective temperature of the primary component of KIC~5193386 was available, we computed the error bar on its $\numax$ parameter using $\sigma_{\Teff}$ = 100~K, which is the typical order of magnitude of the uncertainty for this parameter.
We note that the APOGEE DR14 rather suggests an uncertainty of 70~K \citep{Abolfathietal2018,Pinsonneaultetal2018}.
However, given the low luminosity of this system (visible magnitude of 14.7, while the typical value for the rest of the sample is 13), its high luminosity ratio ($L_2/L_1 = 0.33$), and the uncertainties for the other binaries studied in this work (typically 150~K), we suspect this value to be underestimated.
For the other stars, we used the uncertainty provided in Table~\ref{tab:atmosphericParam}.
}
\end{table}

The present work adds seven more systems with clear rotational modulations, including six with totally suppressed oscillations and one with partially suppressed oscillations (KIC 10015516).
We computed the expected global seismic parameters of the five non-oscillating giants that belong to SB2 systems from their dynamical mass and radius, and $\Teff$ (Table~\ref{tab:predictDnuNumax}) by inverting the asteroseismic scaling relations.
These estimates confirm that the oscillations frequencies of the non-oscillating RGs in EBs lie in the frequency range accessible by \textit{Kepler}.
The sixth system with an active and non-oscillating RG is KIC~4473933.
This binary is different because it is not circular ($e$ = 0.279) and its 103.6 d orbital period is significantly longer than that of the other systems with a non-oscillating giant.
The rotation period of its RG component is close to the pseudo-synchronization period predicted by \citet[][see also \citealt{Lurieetal2017,Zimmermanetal2017}]{Hut1981}, which suggests that this binary has already undergone significant tidal synchronization
\footnote{
It should be noted that the ab initio modeling of the dissipation of the equilibrium tide of \citet{Remusetal2012} can lead to a dependence of the tidal torque on the rotation and orbital periods similar to those of the Hut model (see also Sect.~\ref{sec:tides}).
}
.

\begin{figure}
\includegraphics[width=8.7cm]{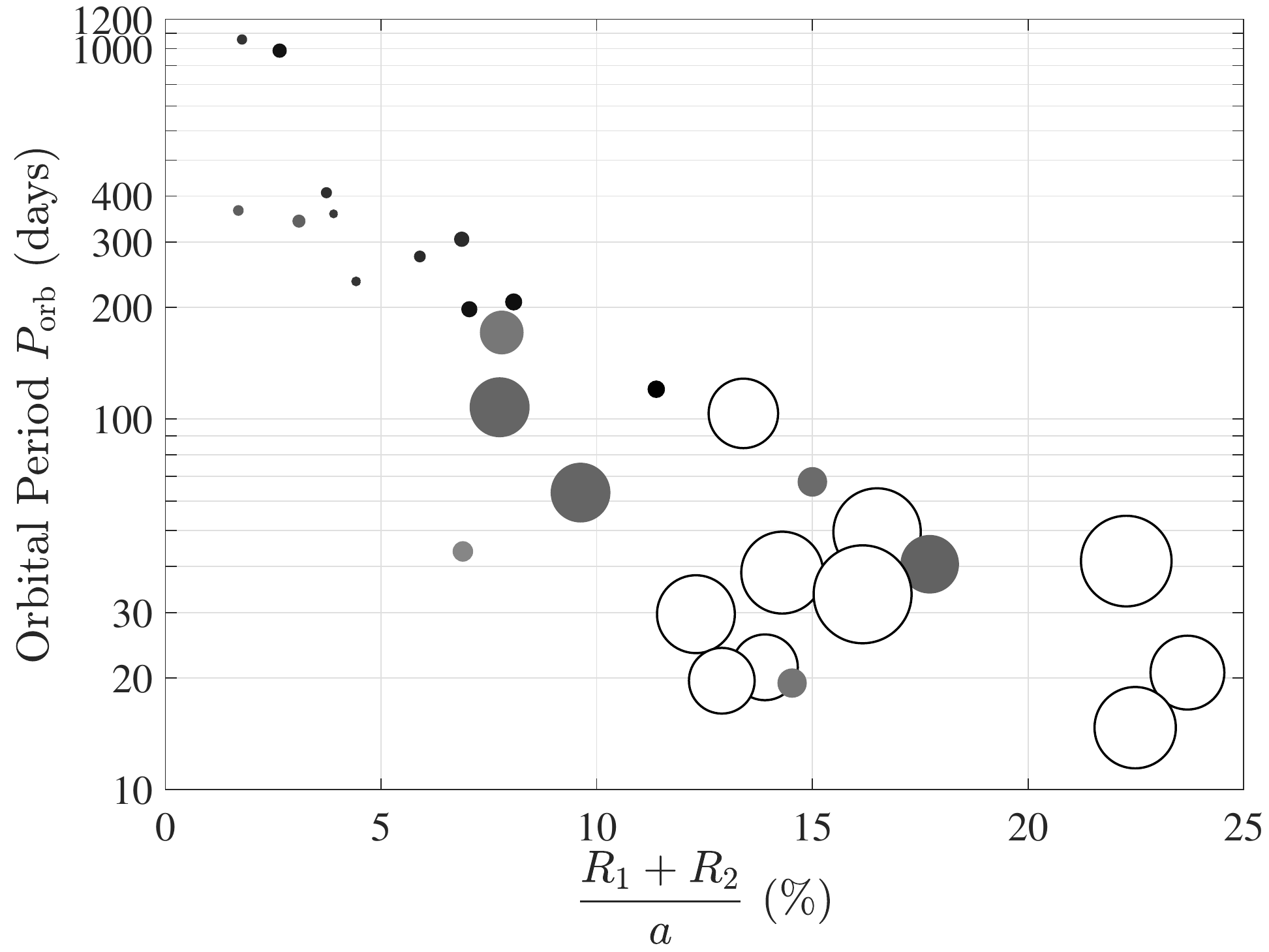}

\includegraphics[width=8.7cm]{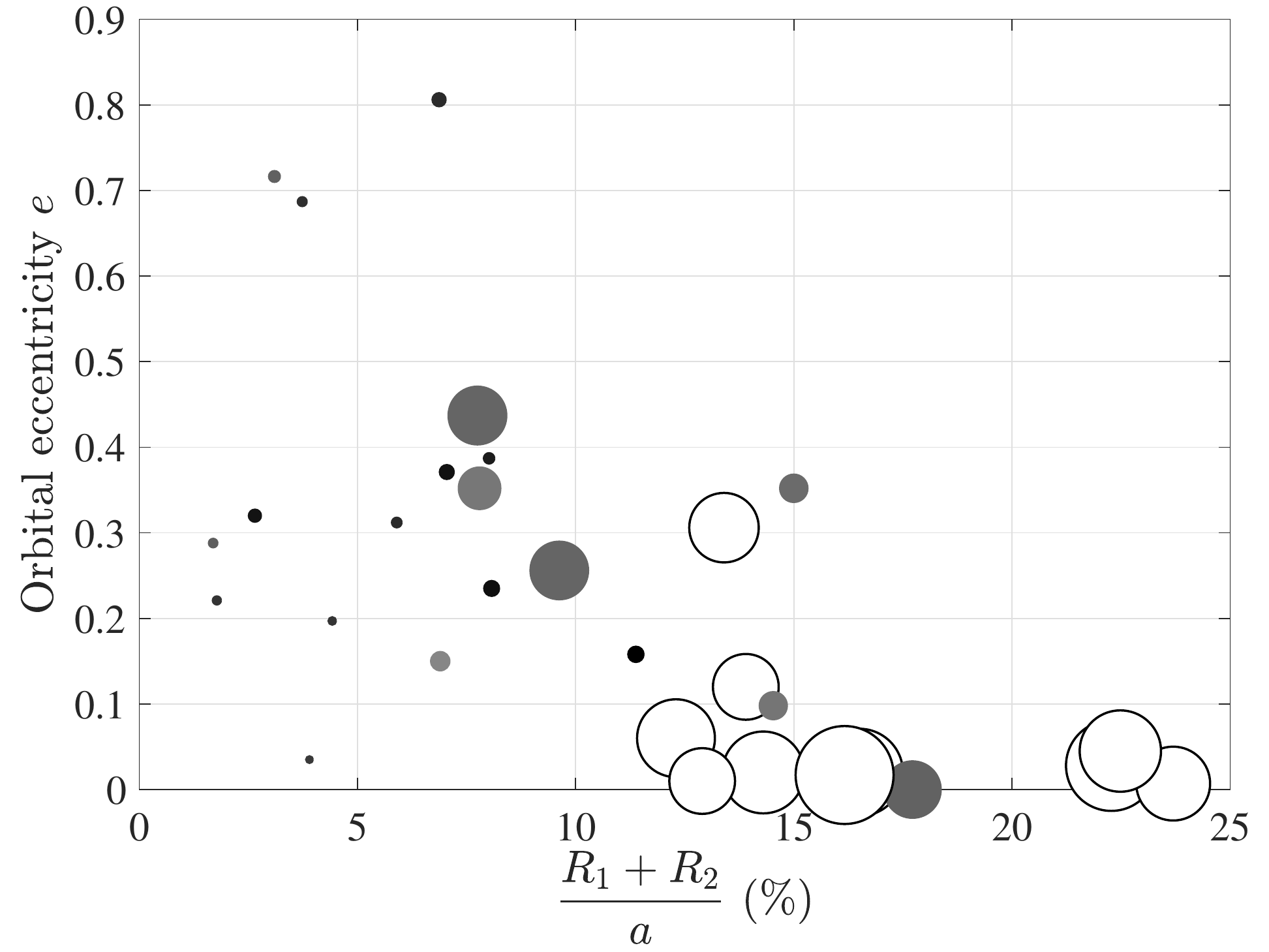}
\caption{Orbital configurations of the systems. Top: Orbital period vs. relative radii for the 35 systems from \citep{Gaulmeetal2014} and this paper. The size of each symbol represents the amplitude of stellar variability and the gray scale indicates the pulsation mode amplitude (white -- no modes; black -- large amp.) Bottom: The same but for the eccentricity of the binary system. This figure is an updated version of Fig. 6 from \citet{Gaulmeetal2014}.}
\label{fig:mode_Porb_e_sumR}
\end{figure}

\begin{figure}
\includegraphics[width=8.7cm]{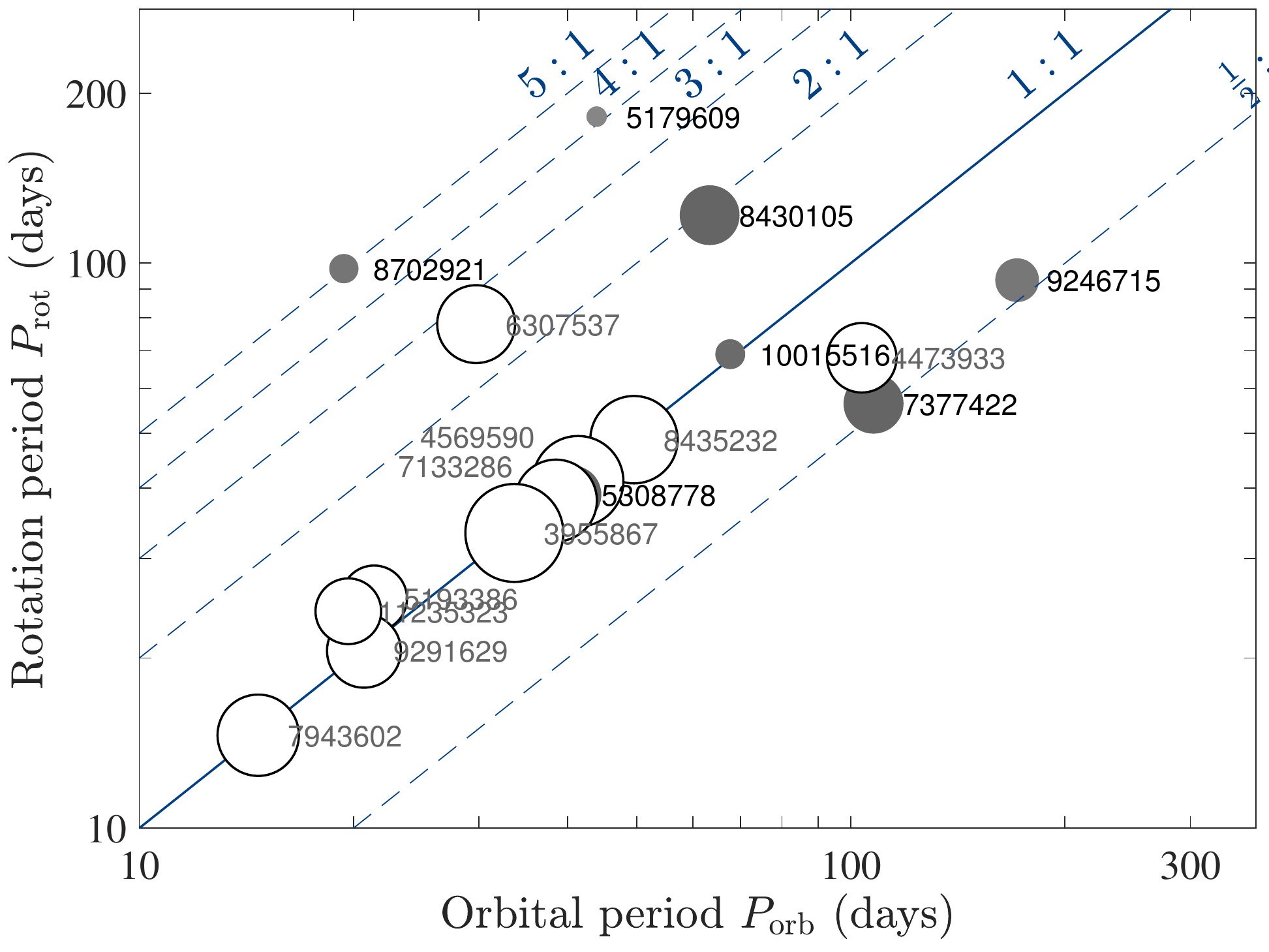}
\caption{Orbital period vs. the rotation period for the 17 systems (out of the set 35 RG/EBs studied by \citet{Gaulmeetal2014} and this paper), where significant surface activity is detected. The size of each symbol represents the amplitude of stellar variability and the gray scale indicates the pulsation-mode amplitude (white—no modes; black—large amp). The parallel lines indicate isolevels of the ratio $P\ind{rot}/P\ind{orb}$. This figure is an updated version of Fig. 5 from \citet{Gaulmeetal2014}.}
\label{fig:Prot_Porb}
\end{figure}

Beyond an almost doubled number of systems of RGs in EBs with respect to \citet{Gaulmeetal2014}, there are two main improvements regarding the study of rotational modulations and oscillation properties in close binary systems. Firstly, instead of measuring the peak-to-peak amplitude of stellar variability from the time series, we compute the photometric magnetic activity index $\Sph$ as defined by \citet{Mathur_2014}.
It is the standard deviation of the light curve over five times the rotation period $P\ind{rot}$ when surface activity is detected. In the absence of photometric modulation, we compute the mean of the running standard deviation of the light curve in five-day ranges. These stars correspond to the smallest, darkest symbols in Figs.~\ref{fig:mode_amp_width} and~\ref{fig:mode_Porb_e_sumR}. The values of $\Sph$ used in this section are reported in Tables~\ref{tab:seismicParam} and~\ref{tab:benchmarksFromLiterature}. Secondly, we properly measure the amplitude of radial modes by employing a peak-bagging method (Sect. \ref{sect:peak_bagging}).
In the former work, the height of the largest peak of the oscillation power spectra was used as a proxy of mode amplitude. It was actually inaccurate as modes of the same amplitude may have different heights and widths \footnote{For a single-sided PSD, the amplitude of a given mode is defined as $A = \sqrt{HW \pi/2}$ \citep{Appourchaux_2015}, where $H$ and $W$ are its height and width.}. The current approach is better as it entails fitting each oscillation mode with a Lorentzian function. That way, we are able to look for unusual mode widths, which are inversely proportional to the mode lifetime: the wider a mode, the shorter its lifetime.
We also reanalyzed the sample of 19 RG/EBs characterized by \citet{Gaulmeetal2014} to ensure consistency between the studies.
Our goal is to emphasize the link in RG binaries between oscillation suppression, magnetic activity, and tidal interactions.
We combine the results from the dynamical modeling, rotational modulations, and peak-bagging in Figs. \ref{fig:mode_amp_width}, \ref{fig:mode_Porb_e_sumR}, and \ref{fig:Prot_Porb}.
Our conclusions are summarized at the end of this section.

Figure~\ref{fig:mode_amp_width} (top panel) shows the photometric magnetic activity level proxy $\Sph$, that is, the amplitude of the photometric signal caused by the presence of spots on the RG surfaces, as a function of the orbital period.
The correlation is clear: for orbital periods longer than 180 days, $\Sph$ never exceeds 0.1\%.
At shorter orbital periods, $\Sph$ becomes larger, reaching 10\% of the relative stellar flux.
This analysis is consistent with the conclusions of \citet{Mathur_2019} who studied the correlation of surface activity and oscillation detection for a sample of 2,000 MS stars observed by \textit{Kepler}.
These authors report that stars with $\Sph$ larger than 0.2\,\% have almost no chance of having oscillations detected (98.3\% probability).
We detect oscillations with $\Sph \approx 1\,\%$ but our sample consists of RGs, which may explain the difference with \citet{Mathur_2019}.

High $\Sph$ values are correlated with suppressed oscillations.
This can be seen in Fig. \ref{fig:mode_amp_width} (middle panel), in which we compare the amplitude of the closest radial mode to $\nu\ind{max}$ with the empirically expected maximum radial mode amplitude predicted by \citet{Corsaro_2013}. This estimated mode amplitude is essentially a scaling law based on the amplitude of solar oscillations, which is a function of $\numax$, $\Dnu$,  $T\ind{eff}$ and the solar mode amplitude. It arises that the relative amplitudes of oscillations are fairly uniform (dispersion standard deviation of $12\,\%$) for systems with $P\ind{orb} \geq 180$ d. We note a small offset: the average mode amplitude of RGs should be 0 instead of $-14\,\%$ in the RGs that are not suppressed. It could either mean that even for longer orbits modes are partially suppressed or that the peak-bagging algorithms used by Corsaro to draw the amplitude scaling law and by Gaulme provide slightly different results. We favor the second option because an offset of that order of magnitude was reported by \citet{Gaulme_2016b} when both Corsaro code and Gaulme measured the amplitude of solar oscillations in SOHO photometric time series. In any case, all RGs in systems with orbits shorter than 100 days display significant mode suppression. In between 100 and 180 days, both situations (suppressed and not) are met.

In shorter-period systems, partially suppressed oscillations also have a shorter lifetime.
This is visible in Fig.~\ref{fig:mode_amp_width} (bottom panel), where we plot the mode width as a function of $P\ind{orb}$ where it clearly appears that modes with a suppressed amplitude are also wider, that is, with shorter lifetimes. The mean width of the largest radial modes is about 0.15~$\mu$Hz for systems with $P\ind{orb} \geq 180$ d, but can reach about 1~$\mu$Hz in the most extreme case. This shorter mode lifetime suggests that the surface activity prevents the pressure waves from forming long-lived standing modes.

Figures \ref{fig:mode_Porb_e_sumR} and \ref{fig:Prot_Porb} are updates of Figs. 5 and 6 from \citet{Gaulmeetal2014}, where we confirm that oscillations and surface activity are both correlated with the orbital configuration of the systems, that is, to tidal interactions.
All systems with relative distance $(R_1+R_2)/a \leq 7\,\%$ show no significant rotational modulations, regular oscillation amplitudes, and random eccentricities.
Among the 35 confirmed RGs in EBs that have been studied with radial velocity measurements by \citet{Frandsenetal2013}, \citet{Gaulmeetal2016}, and \citet{Themessletal2018}, a total of 18 meet this low surface activity threshold.
When $7\,\%\leq (R_1+R_2)/a \leq 15\,\%$ rotational modulations get higher ($\Sph > 0.1\,\%$) and $e\leq 0.4$.
For all systems with $(R_1+R_2)/a \geq 15\,\%$ (that is, orbits shorter than about 50 days)\ orbits are circular, rotational modulations are maximum ($1 \leq \Sph \leq 10\,\% $), and only one RG displays (weak) oscillations out of six.
Figure \ref{fig:Prot_Porb} shows that all active systems with partially or totally suppressed oscillations are either synchronized and circularized, or pseudo-synchronized.

In this section, we first showed that, for RGs in binary systems, the photometric magnetic activity level proxy, $\Sph$, was linked to the orbital period: the shorter the period, the higher the value of $\Sph$.
We then used $\Sph$ and the mode-amplitude predictor of \citet{Corsaro_2013} to confirm on a larger sample the result of \citet{Gaulmeetal2014,Gaulmeetal2016}, who showed that active RGs in binaries had suppressed or partially suppressed oscillations.
We also demonstrated a correlation between orbital and rotation period, as shown in Fig. \ref{fig:Prot_Porb}, which suggests that the most active giants in our sample were spun up by tidal interactions.
We finally showed evidence for tidal circularization in our sample: close-in systems with $(R_1+R_2)/a$ > 15\,\% all have an orbital eccentricity smaller than 0.1.
Taken together, these elements support the scenario proposed by \citet{Gaulmeetal2014} to explain the suppression of oscillations in the shorter-period systems of their sample:
in these binaries, tidal interactions act to accelerate the rotation of the giant;
as a consequence, this component becomes more active, which results in its oscillations being suppressed.

Recently, \citet{Gaulmeetal2020} went further in analyzing the link between tidal interactions and surface activity.
In particular, they studied a sample of 4\,500 RGs among which 370 displayed rotational modulations and showed that RGs members of binary systems had higher surface activity levels than single RGs with the same rotation periods.
This suggests that tidal interactions enhance the magnetic activity of RGs.
Investigating these observations will be the goal of a future work.
More discussion about tidal circularization is described in the next section.

\subsection{Observational constraints on tidal circularization}
\label{sec:tides}

\begin{figure*}
\centering
\includegraphics[width=15cm]{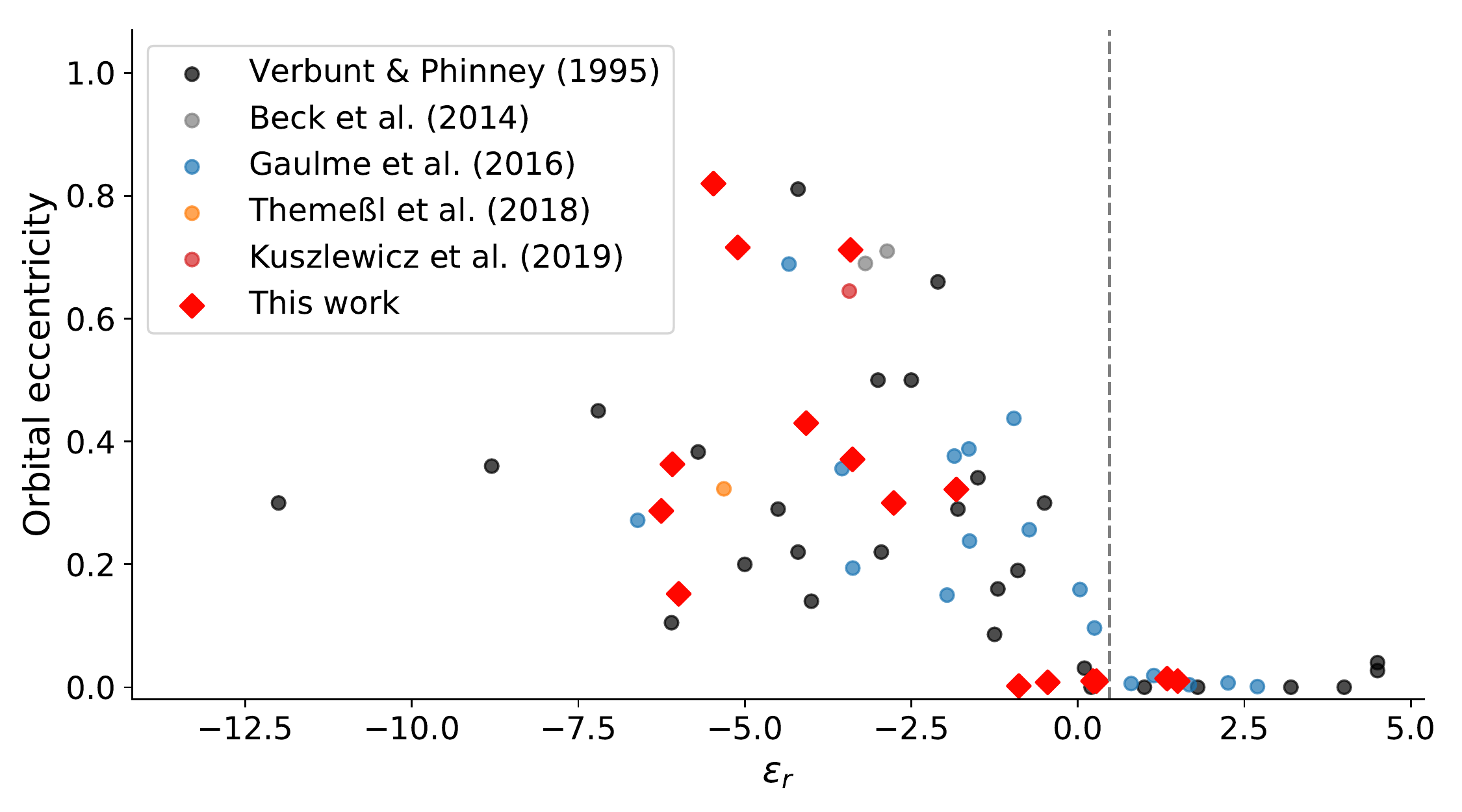}
\caption{Orbital eccentricity as a function of the circularization proxy, $\epsR = \log_{10} (- \Dlne / f)$. The vertical dashed line represents the critical value $\epsCrit = \log_{10} (3)$ predicted by \citet{Zahn1989} and confirmed by \citet{VerbuntPhinney1995} and \citet{Becketal2018}.}
\label{fig:tidalTheoryPaper}
\end{figure*}

Several teams have exploited the \textit{Kepler} data to study the tidal circularization of binary systems. The two main observational works dedicated to test theoretical predictions were those of \citet{VanEylenetal2016} and \citet{Becketal2018}. \citeauthor{VanEylenetal2016} investigated the influence of the spectral type of the components of binary systems on their circularization timescale, denoted $\tcirc$. These authors first computed the orbital period and $e \cos \omega$ for 945 stars in the KEBC. They then used the photometric effective temperatures given in the catalog of \citet{Armstrongetal2014} to divide their sample into three groups: the systems composed of two hot stars, those composed of two cool stars, and those composed of a hot star and a cool star. The boundary between hot and cool was set at $\Teff$ = 6250~K, which approximately corresponds to the limit separating late-type MS stars with a convective envelope from earlier-type MS stars with a radiative envelope. These authors found that, among the hot--hot binaries, the fraction of eccentric short-period systems was higher than in the other two groups. This result suggests that the tidal dissipation is more efficient in late-type stars, as predicted by \citet{Zahn1975} and \citet{Zahn1977}. However, as noted by the authors, this feature in the eccentricity distribution could be caused by other effects, which is why the interpretation above should be reinvestigated by more detailed studies on smaller samples.

\citet{Becketal2018} focused on a sample of 59~binary systems with at least one RG component, which had been characterized by \citet{VerbuntPhinney1995}, \citet{Becketal2014,Becketal2018AA}, and \citet{Gaulmeetal2014,Gaulmeetal2016}. \citet{Becketal2018} tested the equilibrium tide\footnote{The equilibrium tide is the large-scale flow induced by the hydrostatic adjustement of a star induced by the presence of a companion \citep{Zahn1966a,Remusetal2012}. This flow is completed by the so-called dynamical tide \citep{Zahn1975}, which is constituted by oscillation eigenmodes that are excited by the tidal force, that is, tidal gravity waves in stably stratified radiation zones \citep[e.g.,][]{Zahn1975} and tidal inertial waves in convective regions \citep{OgilvieLin2004}.} model of \citet{Zahn1966a,Zahn1966b,Zahn1977,Zahn1989} by investigating if the eccentricity variations predicted by these theoretical calculations allowed them to explain the orbital eccentricity distribution in their sample. \citet{Becketal2018} found a good agreement between theory and observations, which we explain in more detail in the next paragraphs. The approaches of \citet{VanEylenetal2016} and \citet{Becketal2018} are complementary because they implemented observational tests of the tidal theoretical results through two independent methods. In this subsection, we explain why the 16~binaries characterized in this work\footnote{That is, all our sample except from KIC~4473933.} complete the sample of 59~systems of \citet{Becketal2018}.

\citet{Becketal2018} followed the approach of \citet{VerbuntPhinney1995}, which relies on linking the orbital eccentricity variations undergone by binary systems to their current orbital and structural parameters. After adapting the equations of \citet{Zahn1977} to the case of detached binaries with a RG, \citet{VerbuntPhinney1995} obtained the following relation:
\begin{equation}
\frac{\Dlne}{f} = - 1.3\times 10^{4} \left(\frac{M_1}{\Msun}\right)^{-11/3} \left(\frac{R_1}{\Rsun}\right)^{6.51} q(1+q)^{-5/3} \left(\frac{\Porb}{\mathrm{day}}\right)^{-16/3},
\end{equation}
where $\Dlne = \ln (e_{\mathrm{today}}) - \ln (e_{\mathrm{ZAMS}})$ is the variation of the logarithmic eccentricity undergone by the binary, $f$ is a dimensionless factor that accounts for the effects of the turbulent friction applied by the convection on the equilibrium tide, and $q = M_2 / M_1$ is the mass ratio of the system. The crucial parameter in this formula is the value of the multiplicative constant $f$. This factor is predicted by \citet{Zahn1989} to be on the order of 1, which implies that systems for which $(- \Dlne / f) > 3$ are circularized and that the only circular systems with $(- \Dlne / f) < 3$ formed circular.

\citet{Becketal2018} computed $\epsR = \log_{10} (-\Dlne / f)$ for each binary in their sample. When these authors plotted the orbital eccentricities of their binaries as a function of $\epsR$, they found that the value $\epsCrit = \log_{10} (3)$ was a relevant boundary between circularized and non-circularized systems. Thus, they confirmed the results that \citet{VerbuntPhinney1995} obtained on a smaller sample. In addition, \citet{Becketal2018} demonstrated that the dissipation of tidal inertial waves in the deep convective envelopes of RG can be neglected \citep{OgilvieLin2007,Mathis2015,Galletetal2017AA604} because these stars are close to the configuration of a full convective sphere where this mechanism becomes less efficient \citep{Wu2005}. They also showed that this $\epsCrit$ marked the separating line between oscillating and non-oscillating systems, indicating that the non-oscillating binaries are those with short circularization timescales.

To extend the sample presented by \citet{Becketal2018}, we compute $\epsR$ for the 16 new systems as well as for the two recent RG binaries published by \citet{Themessletal2018} and \citet{Kuszlewiczetal2019}. The updated picture is shown in Fig.~\ref{fig:tidalTheoryPaper}, where the orbital eccentricity is plotted as a function of $\epsR$ for the whole sample. Among the 18 new binaries, the eccentric systems all verify $\epsR < \epsCrit$, as predicted by the theory of the equilibrium tide. However, only two out of six new circular binaries have $\epsR > \epsCrit$, whereas almost all the circular systems considered by \citet{Becketal2018} verified this condition.

Among our sample, the four circular binaries with $\epsR < \epsCrit$ are KIC~5193386, KIC~6307537, KIC~11235323, and KIC~10015516.
The first three systems have an active, non-oscillating RG, as explained in paragraph~\ref{subsub:nonOscRG}.
The latter, KIC~10015516, is composed of an active RG and a hot MS companion with $\gamma$~Dor oscillations (Sect.~\ref{subsec:natureSystems}).
Several scenarios could explain the position of these four binaries in Fig.~\ref{fig:tidalTheoryPaper}.
First, these stars could have formed circular.
This hypothesis can not be discarded with the data available to date.
The second possibility is that they have undergone some circularization due to the dissipation of the dynamical tide during the MS phase of their giant component.
The dissipation of tidal gravity waves resulting from their breaking during the subgiant and RG phases could be another mechanism to study \citep{BarkerOgilvie2010,Weinbergetal2017,Sunetal2018}.
Finally, these systems could also have been circularized during the pre-MS phase of the primary component.
Two theories could explain this pre-MS circularization: first, the dissipation of the equilibrium tide during this phase because of the important thickness of the convective envelope \citep[see][]{ZahnBouchet1989} and second, the dissipation of tidal inertial waves over this period \citep[see for instance][]{Mathis2015,Galletetal2017AA604}.
Testing these hypotheses requires secular evolution models of binary stars, similar to those developed for star-planet systems developed by \citet{Benbakouraetal2019}, which is beyond the scope of this paper.
In this context, the four systems mentioned in this paragraph are the most interesting cases to test evolution scenarios involving multiple tidal mechanisms.

Figure~\ref{fig:tidalTheoryPaper} also shows a dearth of binaries with eccentricities between 0.5 < $e$ < 0.6.
This was first discovered by \citet{Becketal2018} in their RG binary analysis and is further discussed by Beck et al. (submitted).
To this point, no theoretical explanation has been suggested.

\section{Conclusions and prospects}
\label{sec:conclusion}

In this paper, we reported the identification of 16 new EBs hosting a RG observed by the \textit{Kepler} main mission. Thus, this paper brings the number of confirmed stars that belong to EBs from the original \textit{Kepler} mission from 20 to 36 \citep[][and this paper]{Hekkeretal2010,Gaulmeetal2013,Gaulmeetal2014,Themessletal2018,Kuszlewiczetal2019}. We presented the result of ground-based high-resolution spectroscopic observations of these 16 systems plus KIC~4663623, which was observed with a poor sampling by \citet{Gaulmeetal2016}. For each system, we obtained 15 radial velocity measurements on average that we combined with the \textit{Kepler} light curve to perform a photometric, seismic, and spectroscopic analysis. From this study, we divided the sample into four groups: four SB2 with a pulsating giant, five SB2 with a non-oscillating giant, two SB1 with a hot MS companion, and six SB1 with a red-dwarf companion.

The first group, composed of KIC~4054905, KIC~4663623, KIC~7293054, and KIC~9153621, is useful to test the asteroseismic scaling relations\footnote{
Among these four systems, one (KIC~4663623) had already been studied by \citet{Gaulmeetal2016}. The other three binaries are novel detections.
}.
Our study brings three new systems to the 11 already known EBs that allow for such a test.
In Sect.~\ref{sec:results}, we compared the dynamical masses and radii of these 14 benchmarks to their seismic estimates and confirmed the discrepancy that was reported in previous works \citep{Huber2015,Gaulmeetal2016,Brogaardetal2018,Themessletal2018}.
We determined a set of empirical reference values for $\numax$ and $\Dnu$ for RG stars that allowed us to compute unbiased seismic estimates of stellar masses and radii on our sample.
The values provided in Eqs.~\eqref{eq:DnuRefNew} and~\eqref{eq:numaxRefNew} are specific to our seismic pipeline but can easily be recomputed for any other seismic analysis code.

We are aware that a sample of 14 stars is still small in a statistical sense for providing robust empirical calibrations and we suggest looking for more of these systems.
Three RG/EB candidates, which were not studied in this paper, have been identified in the targets of the original \textit{Kepler} mission by \citet{GaulmeGuzik2019} during a systematic search for pulsating stars in all of the known \textit{Kepler} data.
In addition, there are promising data sets from the K2 \citep{Howelletal2014} and TESS \citep{Rickeretal2014} missions, even though the frequency resolution of TESS is significantly worse.
However, early results about RG seismology with TESS are encouraging \citep{SilvaAguirreetal2019}.
Alternative types of systems should be explored as well. It is possible to measure the mass of stars belonging to ``heartbeat'' binaries \citep[e.g.,][]{Welsh_2011}, hierarchical triple systems \citep[e.g.,][]{Borkovits_2016}, and visual binaries \citep[e.g.,][]{Marcadon_2018} with the help of complementary high-resolution spectroscopy.

Regarding radial velocity measurements, we note that gravitational redshift and convective blueshift should be taken into account ab initio in the dynamical modeling of the binary systems. With JKTEBOP, the best method to account for these two phenomena is to allow the systemic radial velocity of the companion to differ from to that of the giant (see Sect.~\ref{subsec:dynamical}). Since their effects can both yield a difference on the order of 500\,m~s$^{-1}$, they are responsible for an uncertainty of 1\% to 3\% on the dynamical masses, which would be suppressed if they were not treated as free parameters. Whereas the gravitational redshift can easily be computed from the mass and radius of a star, the convective blueshift is still poorly understood. Useful references concerning this phenomenon are the papers of \citet{Gray2016}, \citet{MeunierLagrangeMbembaKabuikuetal2017,MeunierMignonLagrange2017}, and \citet{Daietal2019}.

Our work confirms the inverse relation between surface magnetic activity and amplitude of the modes. The stars shown the largest surface magnetism do not show visible modes in the PSD.

The systems analyzed in this work are also precious test cases for understanding how binarity impacts the internal structure of stars and how tidal interaction shapes the systems. In these evolved binaries, the angular momentum exchange associated with tidal interaction arises simultaneously to the structural evolution of their components. Therefore, they lie beyond the theoretical framework of \citet{Hut1981}, which has been used as a reference by observational studies of tidal evolution \citep[e.g.,][]{Lurieetal2017,Zimmermanetal2017}. This work adds six non-oscillating giants to the four already published by \citet{Gaulmeetal2014,Gaulmeetal2016}. Such stars are very helpful to understand the tidal pseudo-synchronization of the components of binary systems. Moreover, the whole sample we considered represents a significant contribution to the study led by \citet{Becketal2018} on tidal circularization of evolved binaries. It confirms the importance of the dissipation of the equilibrium tide and highlights in some cases the necessity to explore other tidal dissipation mechanisms.

Consequently, the present work shows the necessity of secular evolution models of binary systems taking into account the structural and rotational evolution of the components simultaneously to the tidal angular momentum exchanges. Significant theoretical advances have been done by \citet{Siessetal2013} and \citet{Davisetal2014} for intermediate-mass and massive stars by \citet{Flemingetal2019} for MS solar-like stars and by \citet{Benbakouraetal2019} in the case of star-planet systems. Understanding the tidal interactions that occur in binary systems is of great importance to explain the stellar rotational distributions and their link to stellar dynamics, structure, and evolution \citep{Mathis2019}.

\begin{acknowledgements}
The authors thank the referee for their careful reading and constructive comments which allowed us to improve this article.
We also thank B. Mosser and T. Appourchaux for verifying the mixed-mode analysis of KIC 4663623.
This paper includes data collected by the \textit{Kepler} mission. Funding for the \textit{Kepler} mission is provided by the NASA Science Mission directorate.
Some of the data presented in this paper were obtained from the Mikulski Archive for Space Telescopes (MAST).
STScI is operated by the Association of Universities for Research in Astronomy, Inc., under NASA contract NAS5-26555.
This work is also based on observations obtained with the Apache Point Observatory 3.5-meter telescope, which is owned and operated by the Astrophysical Research Consortium.
Part of our spectroscopic observations were done at the Observatoire de Haute-Provence.
We acknowledge the technical team for their onsite support as well as the ``Programme National de Physique Stellaire'' (PNPS) of CNRS/INSU for their financial support.
M.B. acknowledges support from the ``Deutscher Akademischer Austauschdienst'' (DAAD) and the Université Paris Diderot.
P.G. was supported by the German space agency (Deutsches Zentrum für Luft- und Raumfahrt) under PLATO data grant 50OO1501.
P.G. and J.J. acknowledge NASA grant NNX17AF74G for partial support. M.B., R.A.G., and St.M. have been funded by the PLATO CNES grant.
St.M. acknowledges support from the ERC SPIRE 647383 grant.
Sav.M. acknowledges support by the National Aeronautics and Space Administration under Grant NNX15AF13G, by the National Science Foundation grant AST-1411685 and the Ramon y Cajal fellowship number RYC-2015-17697.
The research leading to these results has (partially) received funding from the European Research Council (ERC) under the European Union's Horizon 2020 research and innovation programme (grant agreement N$^\circ$670519: MAMSIE), from the KU~Leuven Research Council (grant C16/18/005: PARADISE), from the Research Foundation Flanders (FWO) under grant agreement G0H5416N (ERC Runner Up Project), as well as from the BELgian federal Science Policy Office (BELSPO) through PRODEX grant PLATO.
Funding for the Sloan Digital Sky Survey IV has been provided by the Alfred P. Sloan Foundation, the U.S. Department of Energy Office of Science, and the Participating Institutions.
SDSS acknowledges support and resources from the Center for High-Performance Computing at the University of Utah. The SDSS web site is \url{www.sdss.org}.
SDSS is managed by the Astrophysical Research Consortium for the Participating Institutions of the SDSS Collaboration including the Brazilian Participation Group, the Carnegie Institution for Science, Carnegie Mellon University, the Chilean Participation Group, the French Participation Group, Harvard-Smithsonian Center for Astrophysics, Instituto de Astrofísica de Canarias, The Johns Hopkins University, Kavli Institute for the Physics and Mathematics of the Universe (IPMU) / University of Tokyo, the Korean Participation Group, Lawrence Berkeley National Laboratory, Leibniz Institut für Astrophysik Potsdam (AIP), Max-Planck-Institut für Astronomie (MPIA Heidelberg), Max-Planck-Institut für Astrophysik (MPA Garching), Max-Planck-Institut für Extraterrestrische Physik (MPE), National Astronomical Observatories of China, New Mexico State University, New York University, University of Notre Dame, Observatório Nacional / MCTI, The Ohio State University, Pennsylvania State University, Shanghai Astronomical Observatory, United Kingdom Participation Group, Universidad Nacional Autónoma de México, University of Arizona, University of Colorado Boulder, University of Oxford, University of Portsmouth, University of Utah, University of Virginia, University of Washington, University of Wisconsin, Vanderbilt University, and Yale University.
\end{acknowledgements}

\bibliographystyle{aa}
\bibliography{biblio}

\appendix

\section{Power spectra, background fits, and échelle diagrams}

\begin{figure*}
\centering
\begin{tabular}{cc}
\includegraphics[width=.42\linewidth]{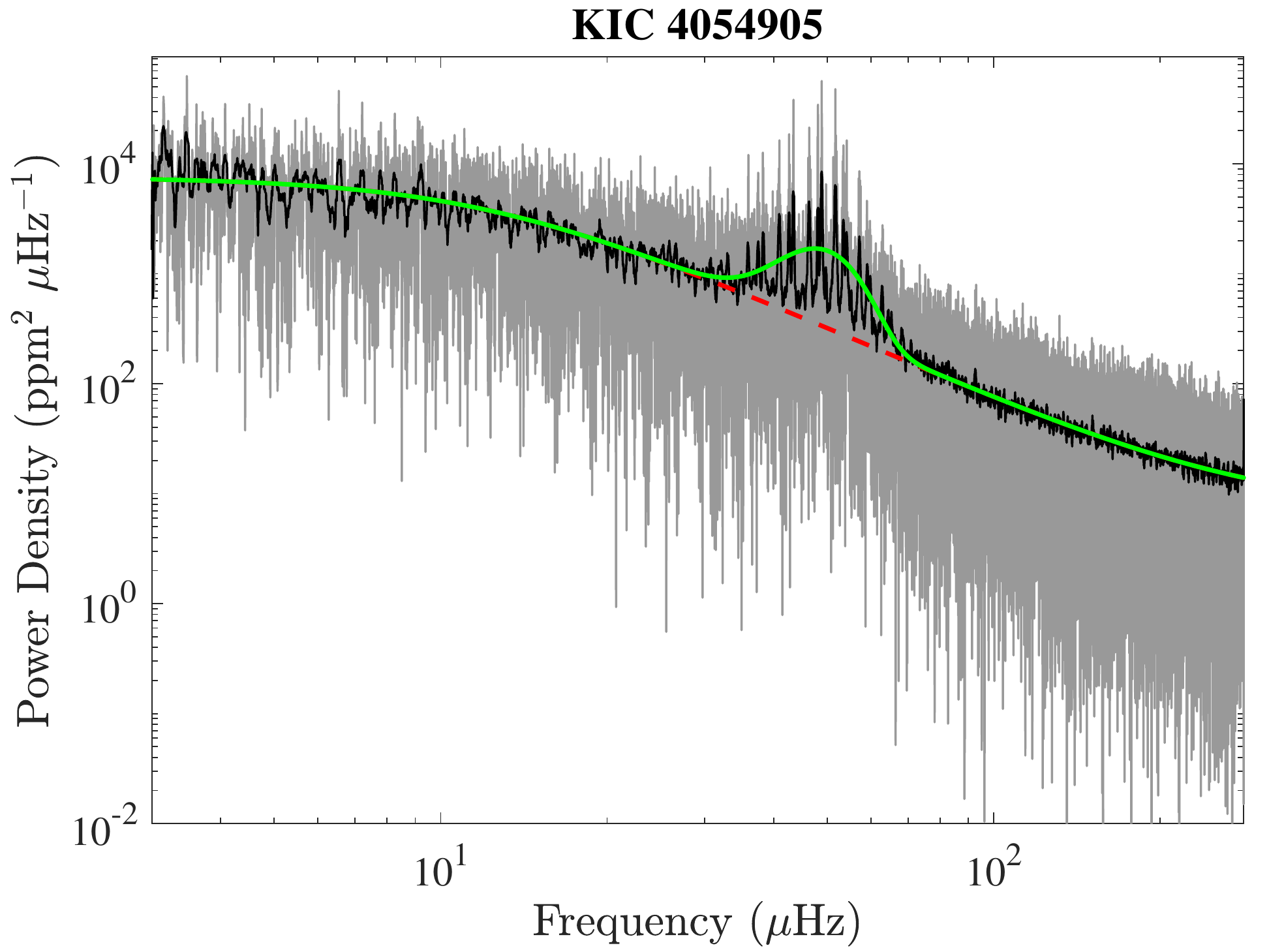}  &
\includegraphics[width=.42\linewidth]{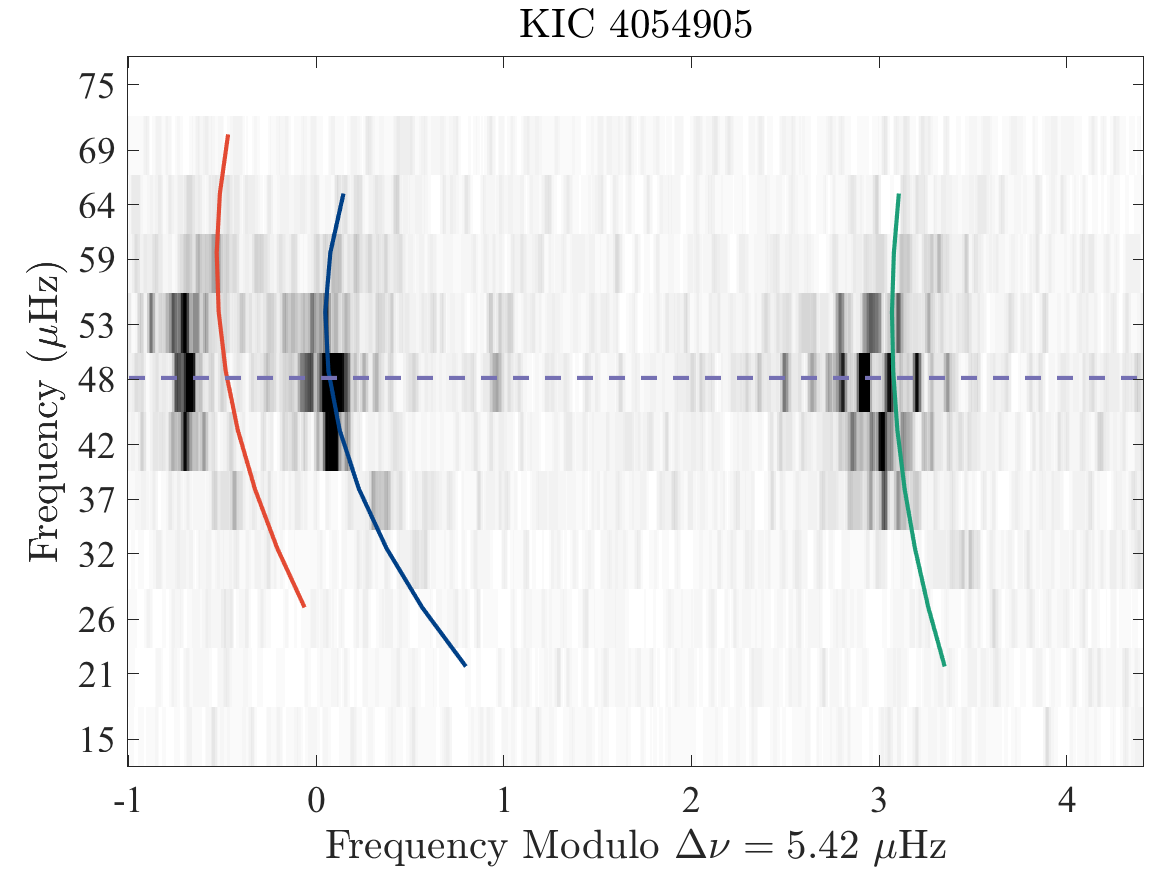}  \\
\includegraphics[width=.42\linewidth]{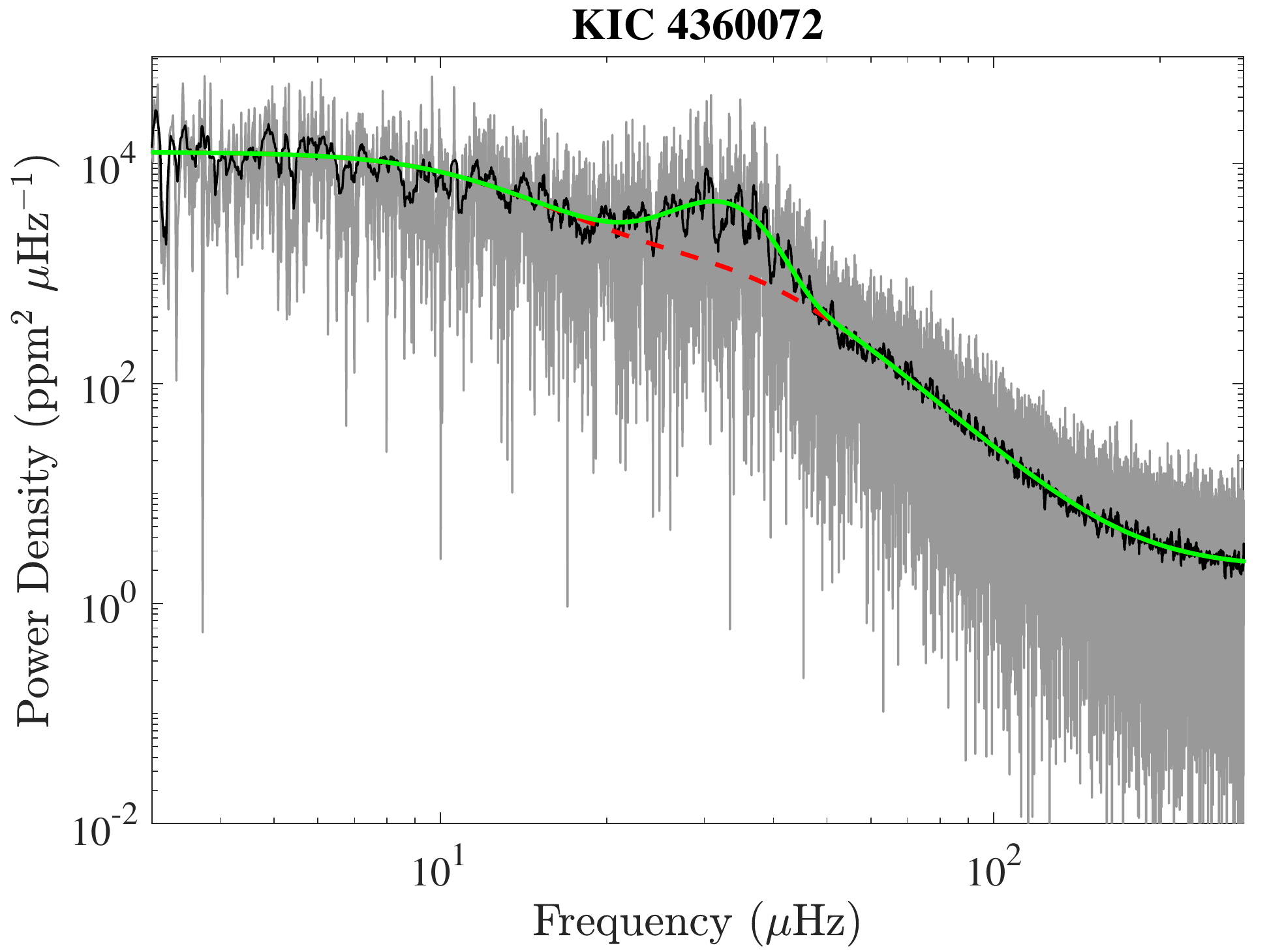}  &
\includegraphics[width=.42\linewidth]{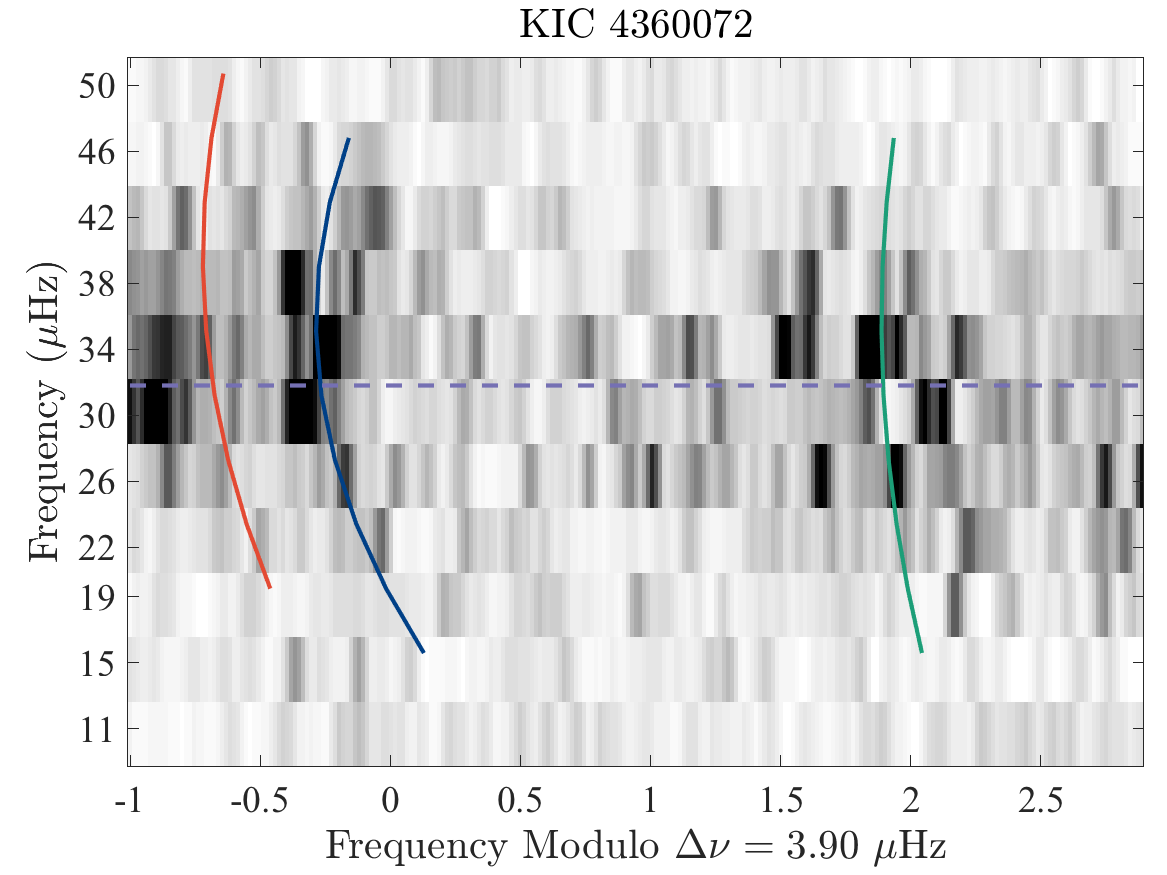}  \\
\includegraphics[width=.42\linewidth]{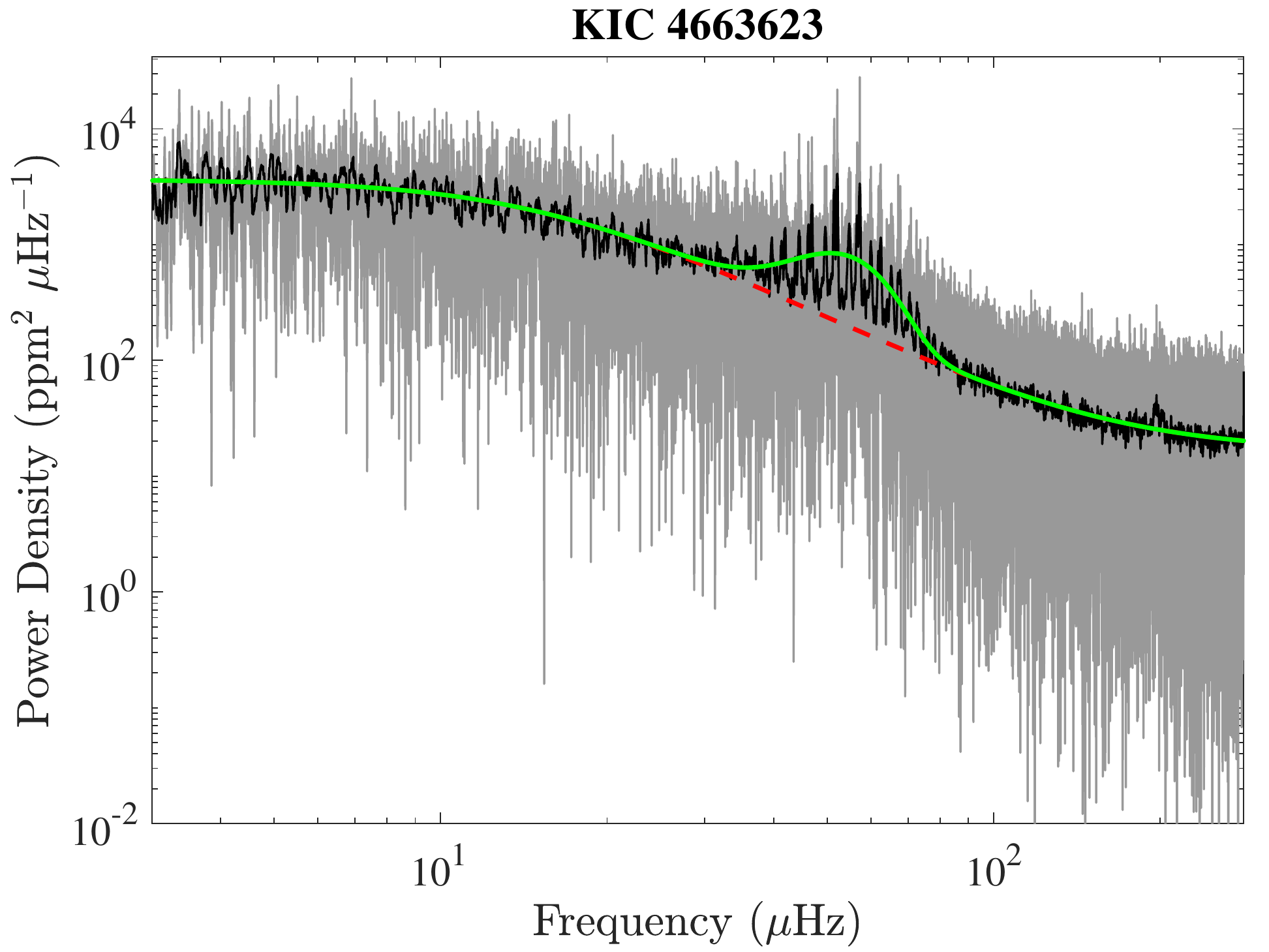}  &
\includegraphics[width=.42\linewidth]{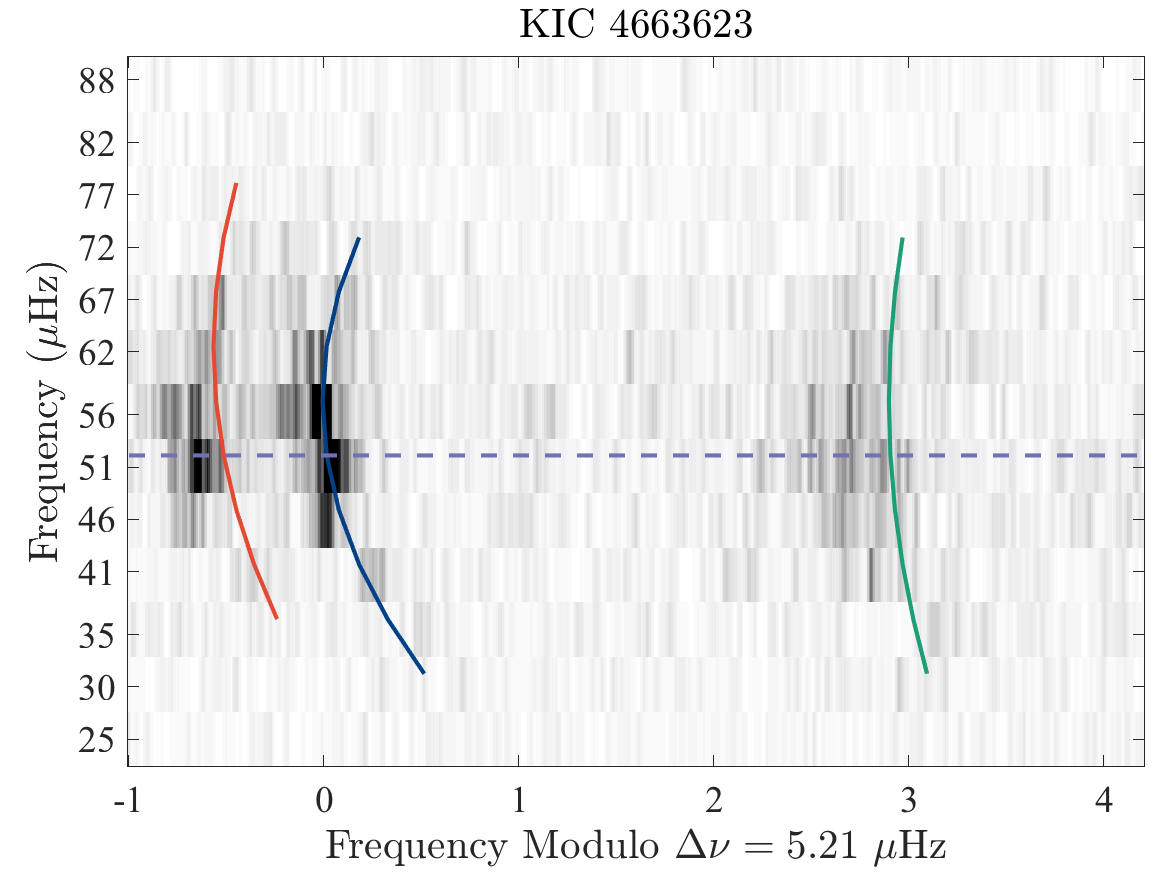}  \\
\includegraphics[width=.42\linewidth]{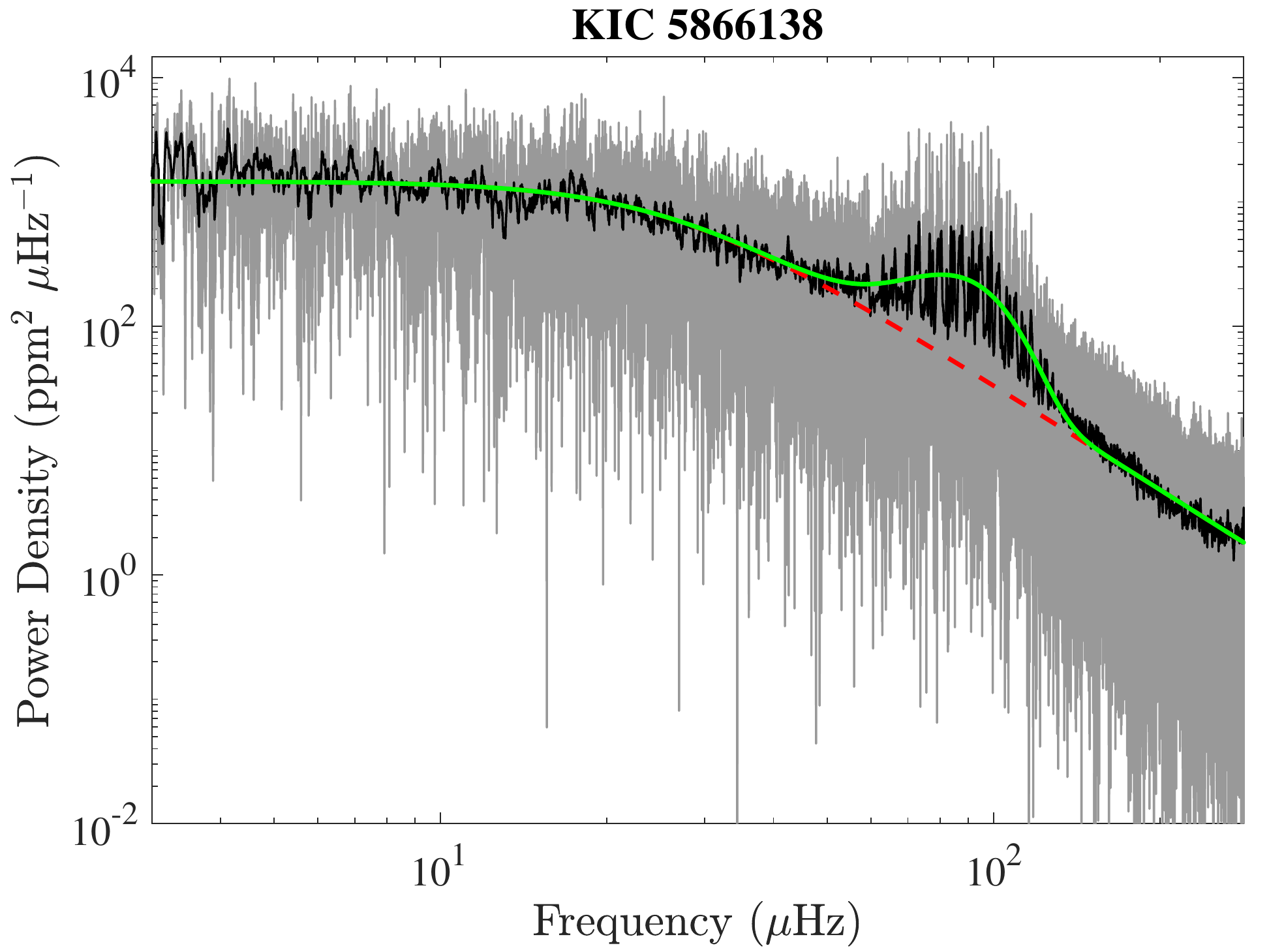}  &
\includegraphics[width=.42\linewidth]{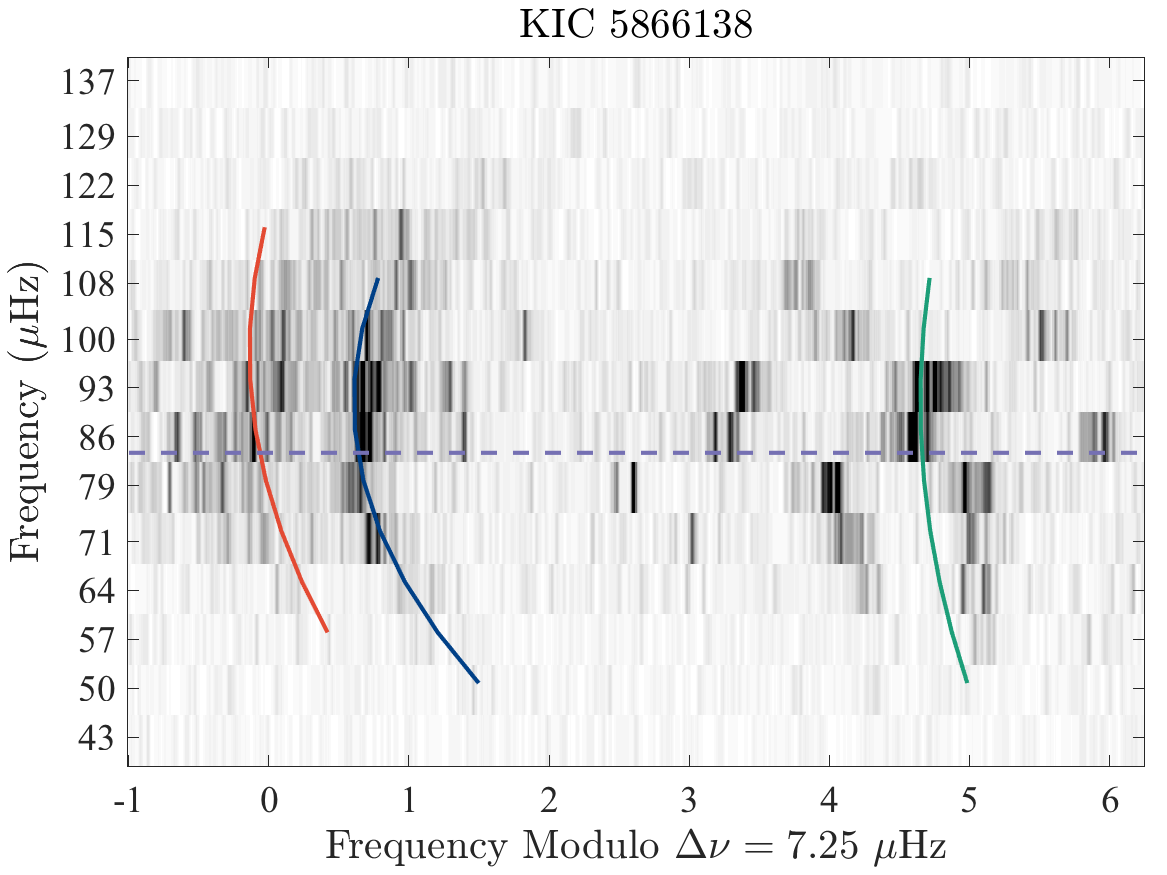}  \\
\end{tabular}
\caption{
Power spectrum fit of the oscillating stars of the sample.
Left: Background fitting. The gray line represents the power spectral density and the black line corresponds to its smoothed version.
The sum of the two super-Lorentzian functions and the white noise model is shown in red and the total background, with the Gaussian envelope of the modes, in green.
See Sect.~\ref{sect:globa_param} for a description of the method. Right: An \'echelle diagram.
The blue, green, and red lines correspond to the ridges of the radial, dipolar, and quadrupolar modes, respectively. These curves were computed from Eqs. \eqref{eq:echelle_mosser_l0}, \eqref{eq:echelle_mosser_l1}, and \eqref{eq:echelle_mosser_l2}. The dashed horizontal line indicates the value of $\numax$.
The value of $\Dnu$ is reported under the $x$-axis.
}
\label{fig:seismic_fit_1}
\end{figure*}

\begin{figure*}
\centering
\begin{tabular}{cc}
\includegraphics[width=.42\linewidth]{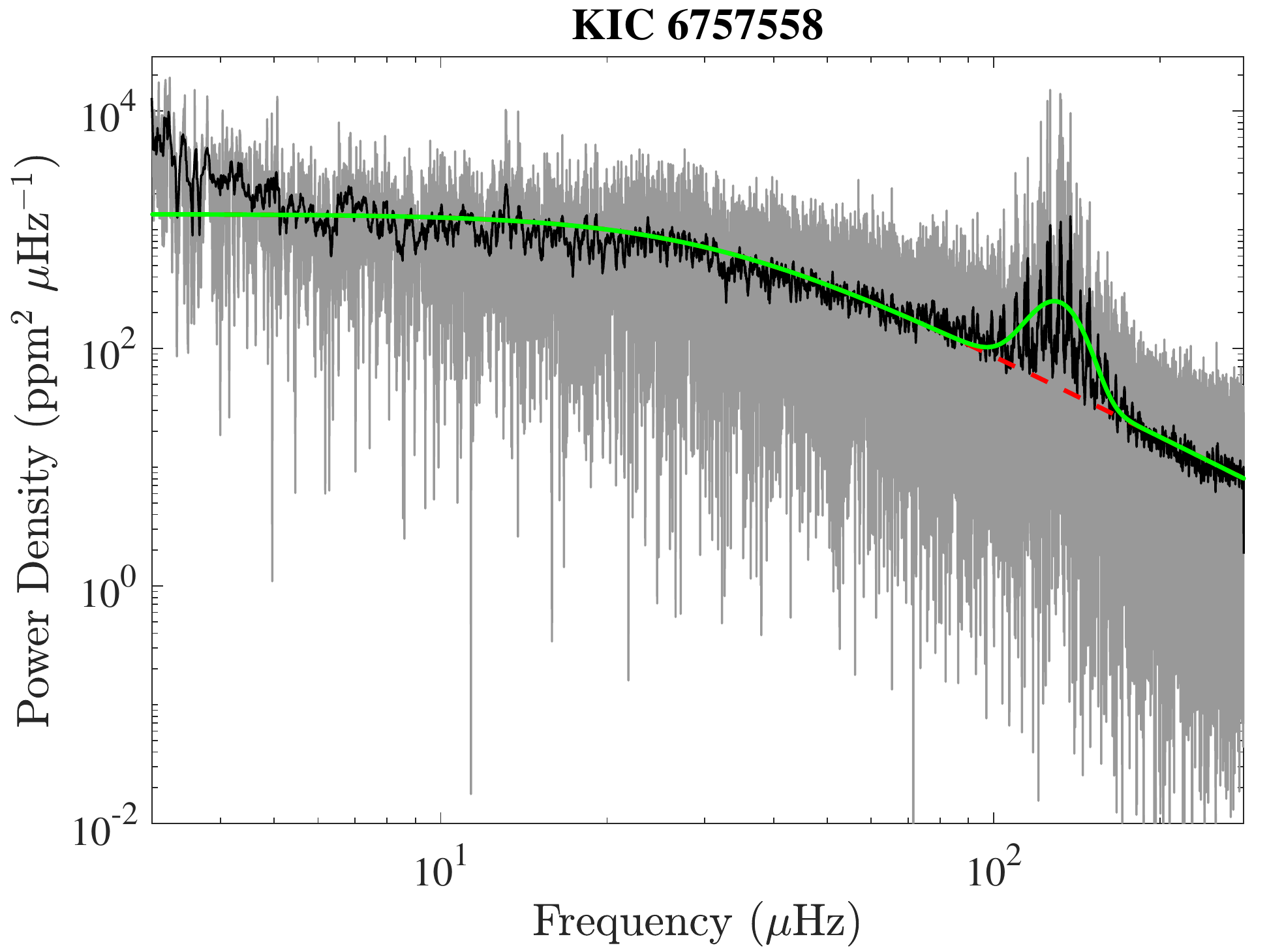}  &
\includegraphics[width=.42\linewidth]{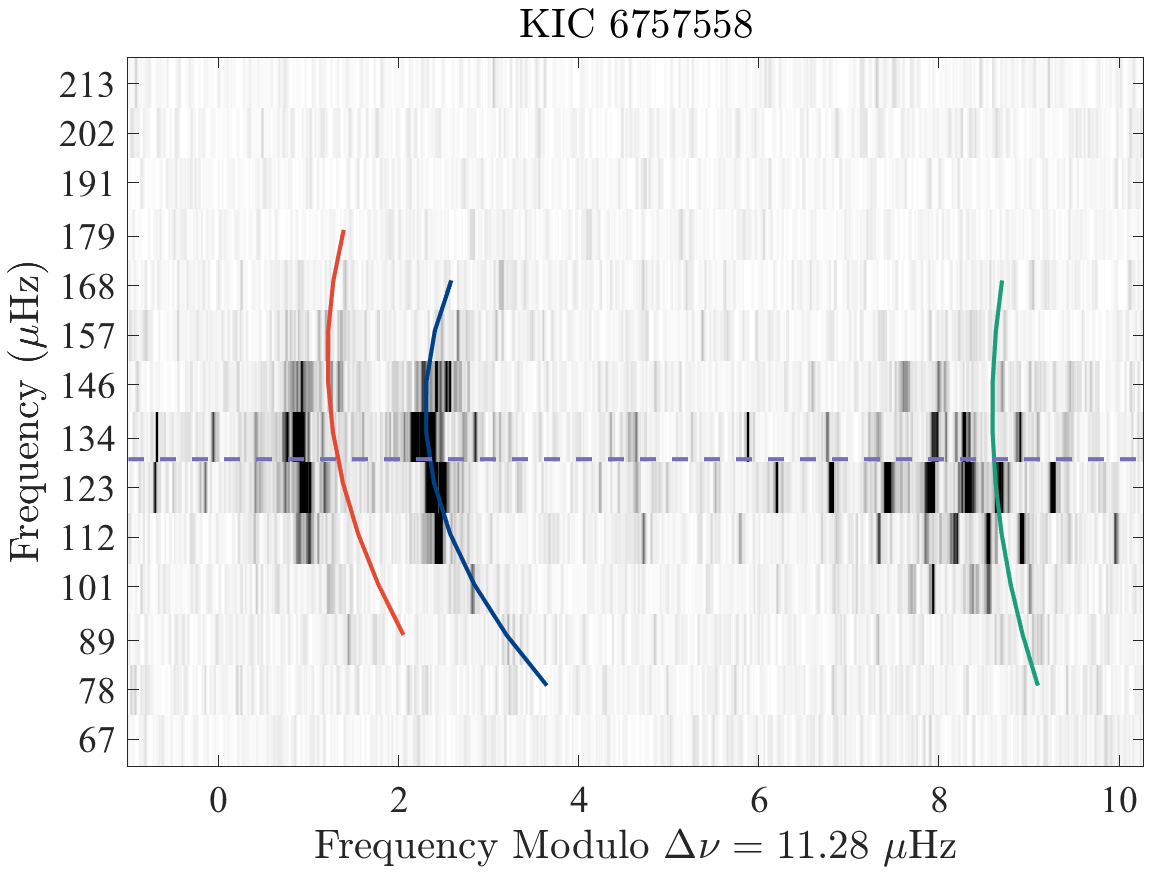}  \\
\includegraphics[width=.42\linewidth]{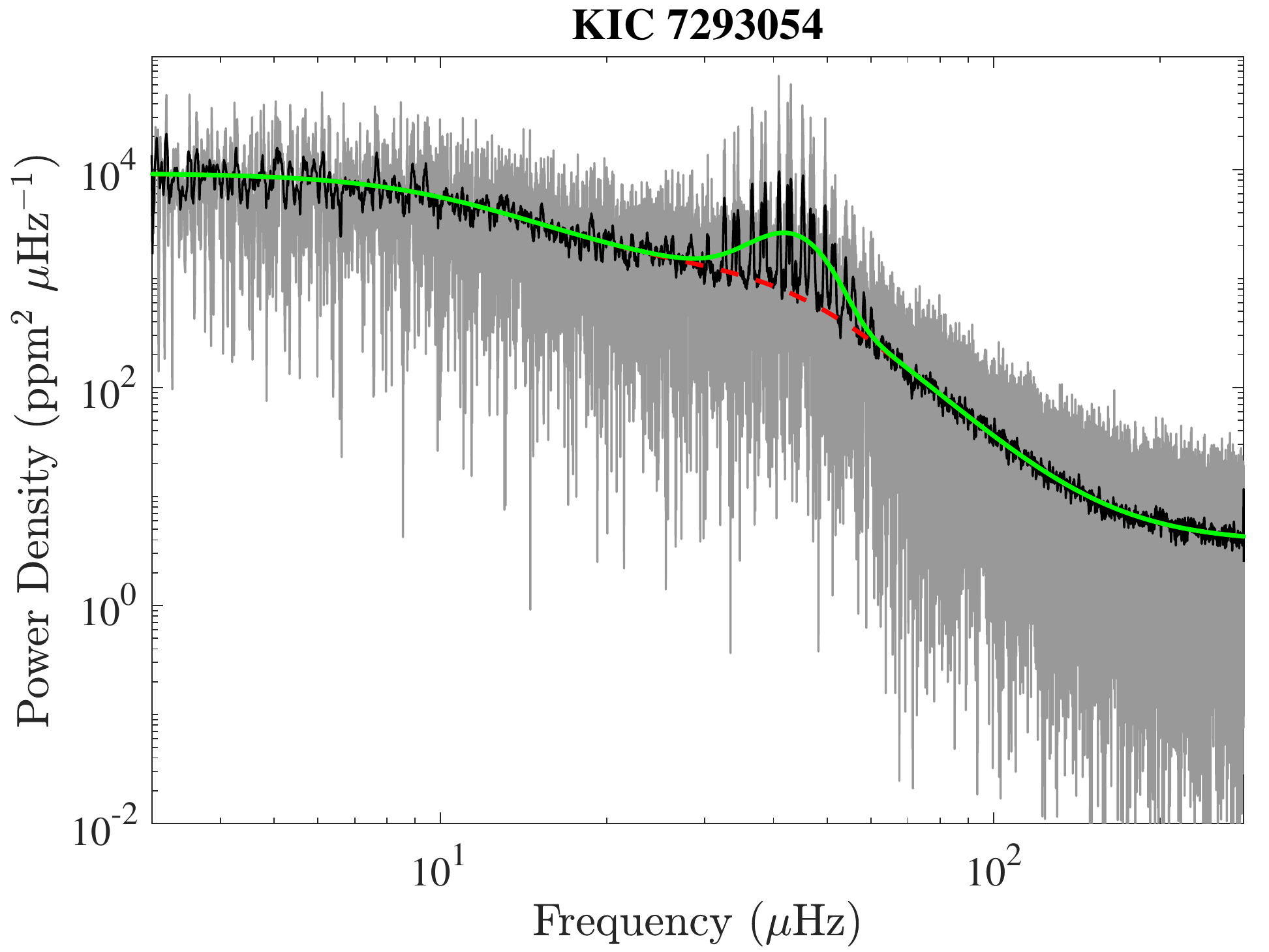}  &
\includegraphics[width=.42\linewidth]{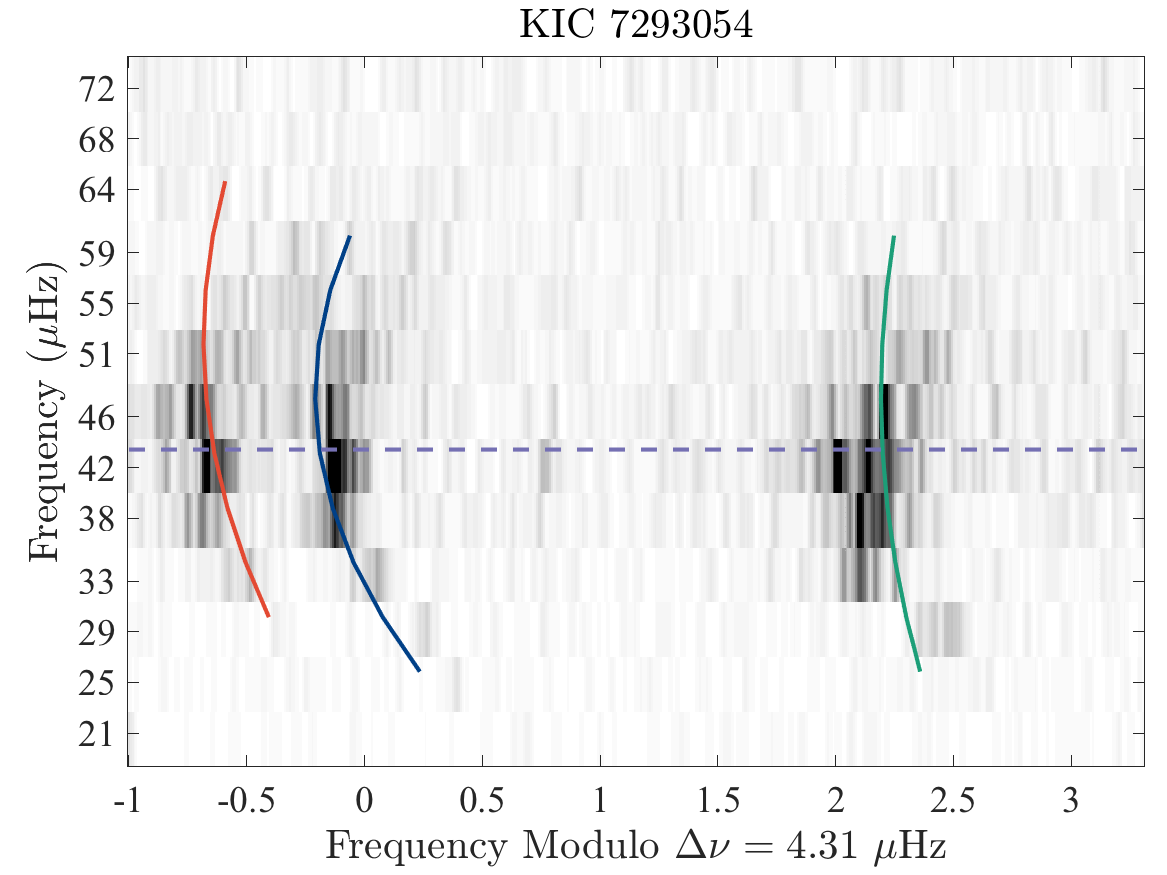}  \\
\includegraphics[width=.42\linewidth]{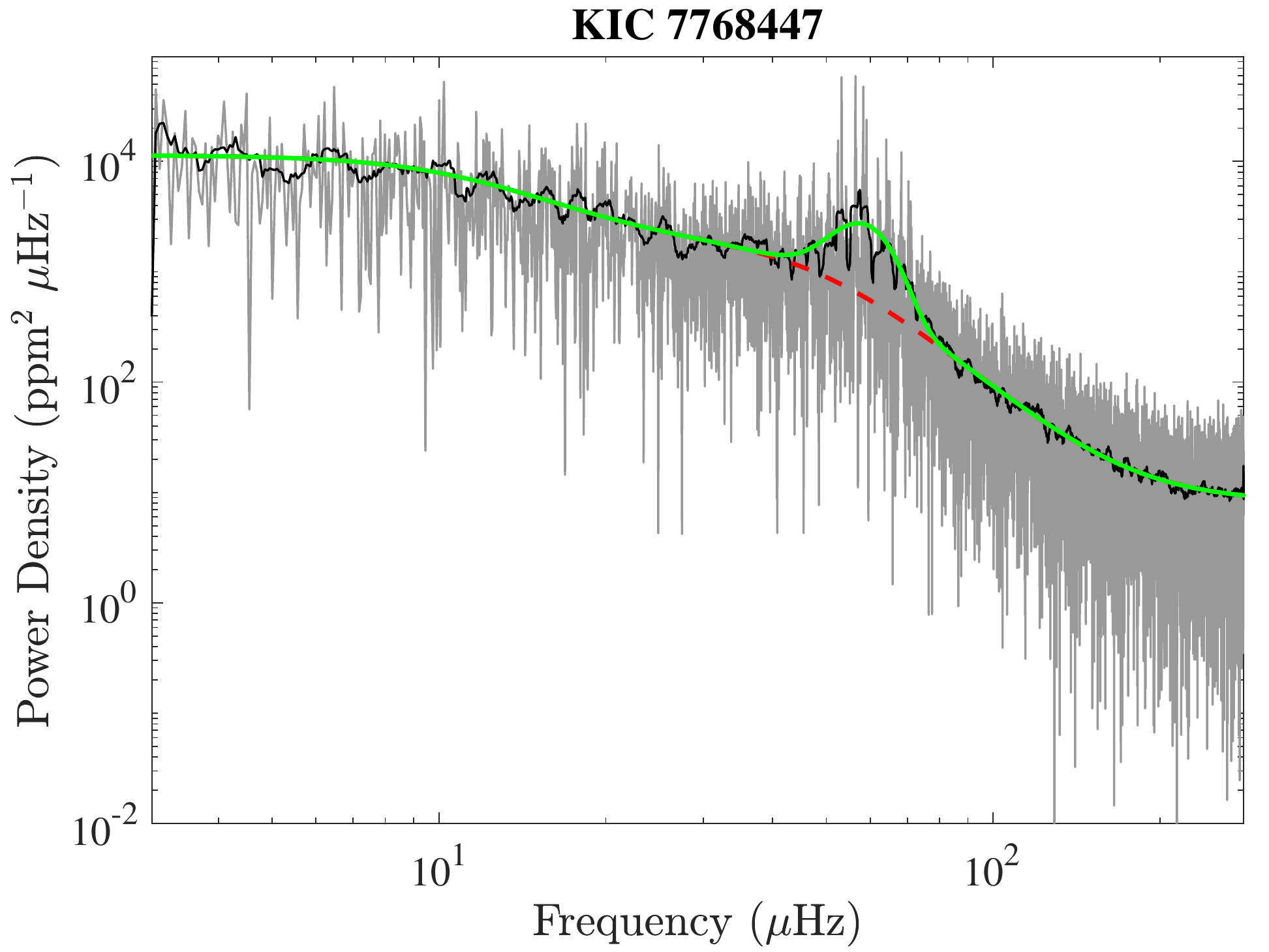}  &
\includegraphics[width=.42\linewidth]{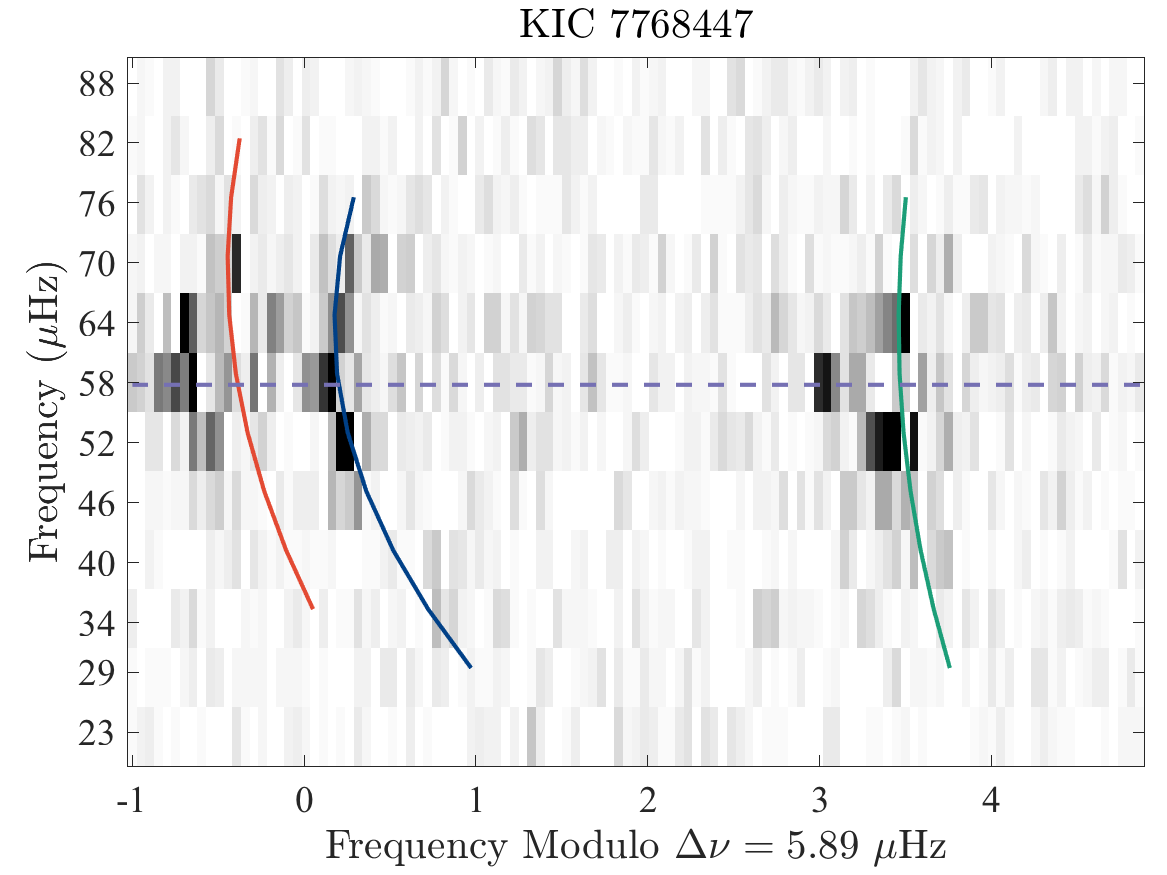}  \\
\includegraphics[width=.42\linewidth]{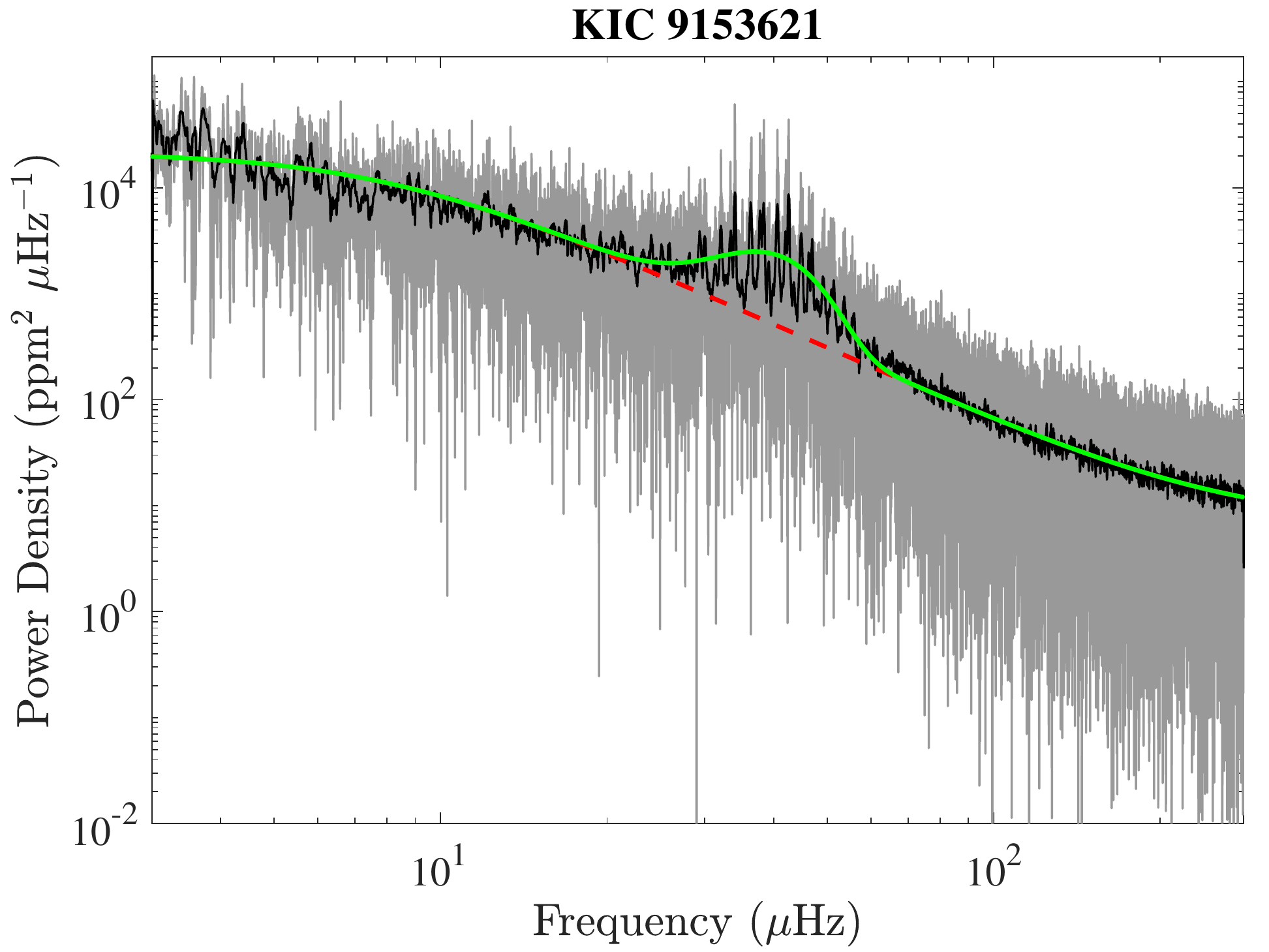}  &
\includegraphics[width=.42\linewidth]{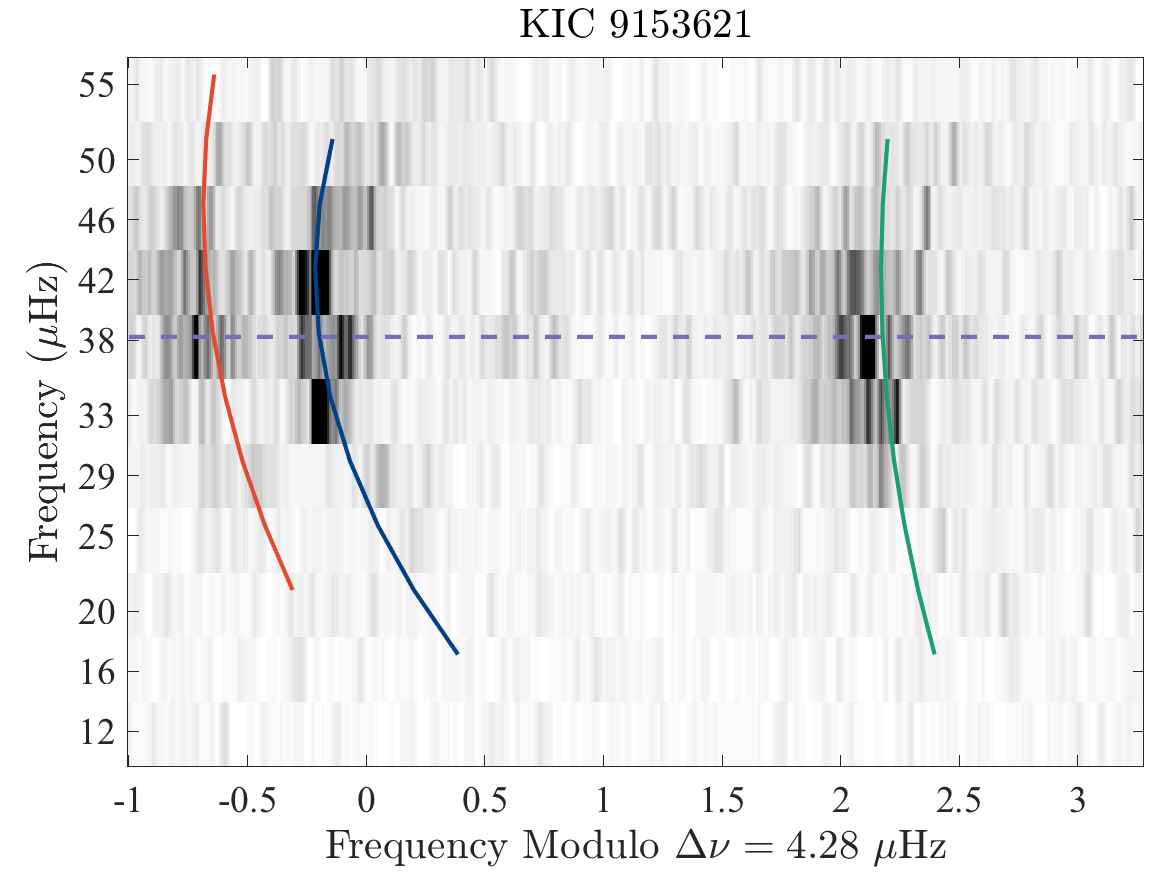}  \\
\end{tabular}
\caption{Same as Fig. \ref{fig:seismic_fit_1}.}
\label{fig:seismic_fit_2}
\end{figure*}

\begin{figure*}
\centering
\begin{tabular}{cc}
\includegraphics[width=.42\linewidth]{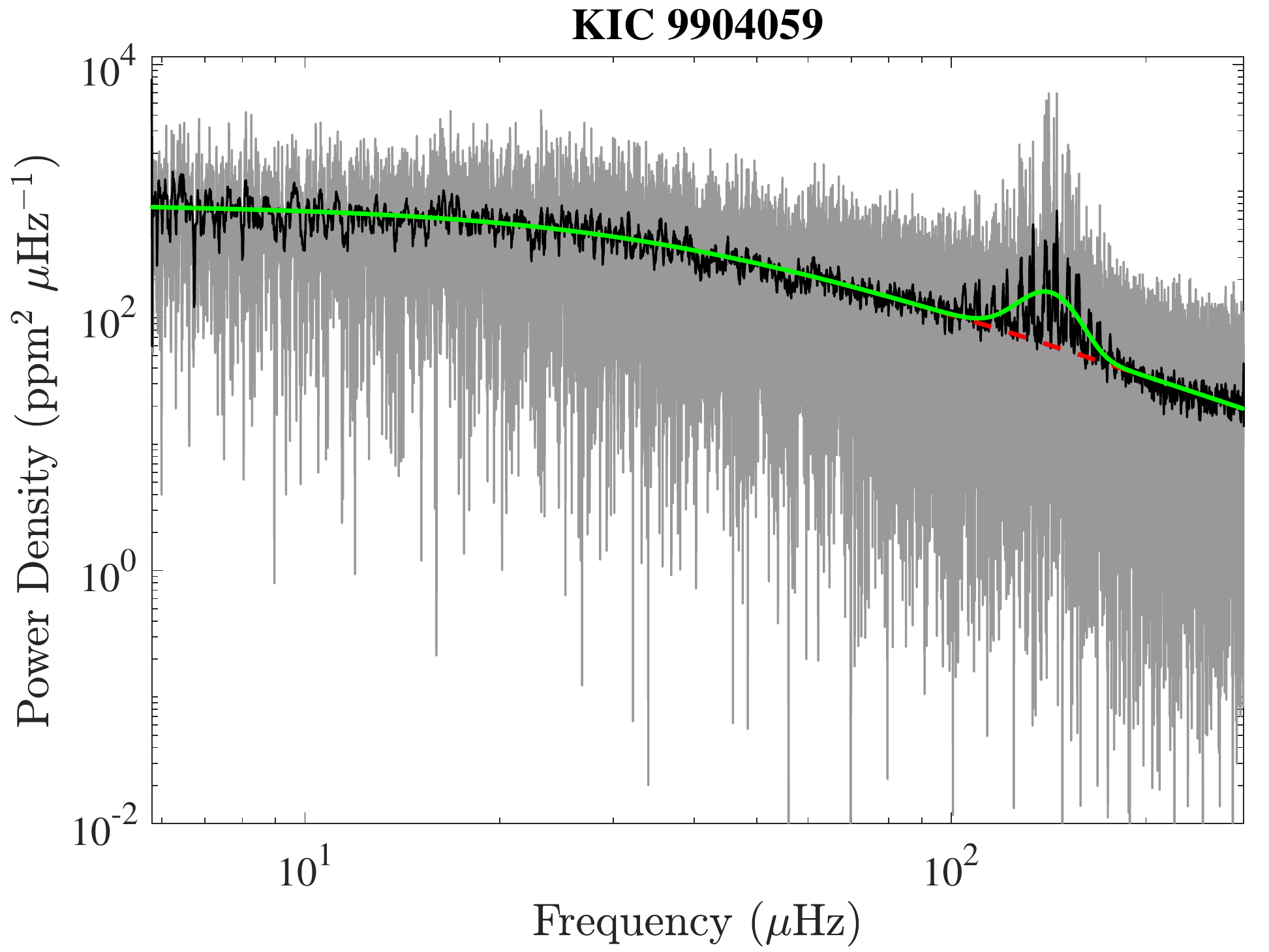}  &
\includegraphics[width=.42\linewidth]{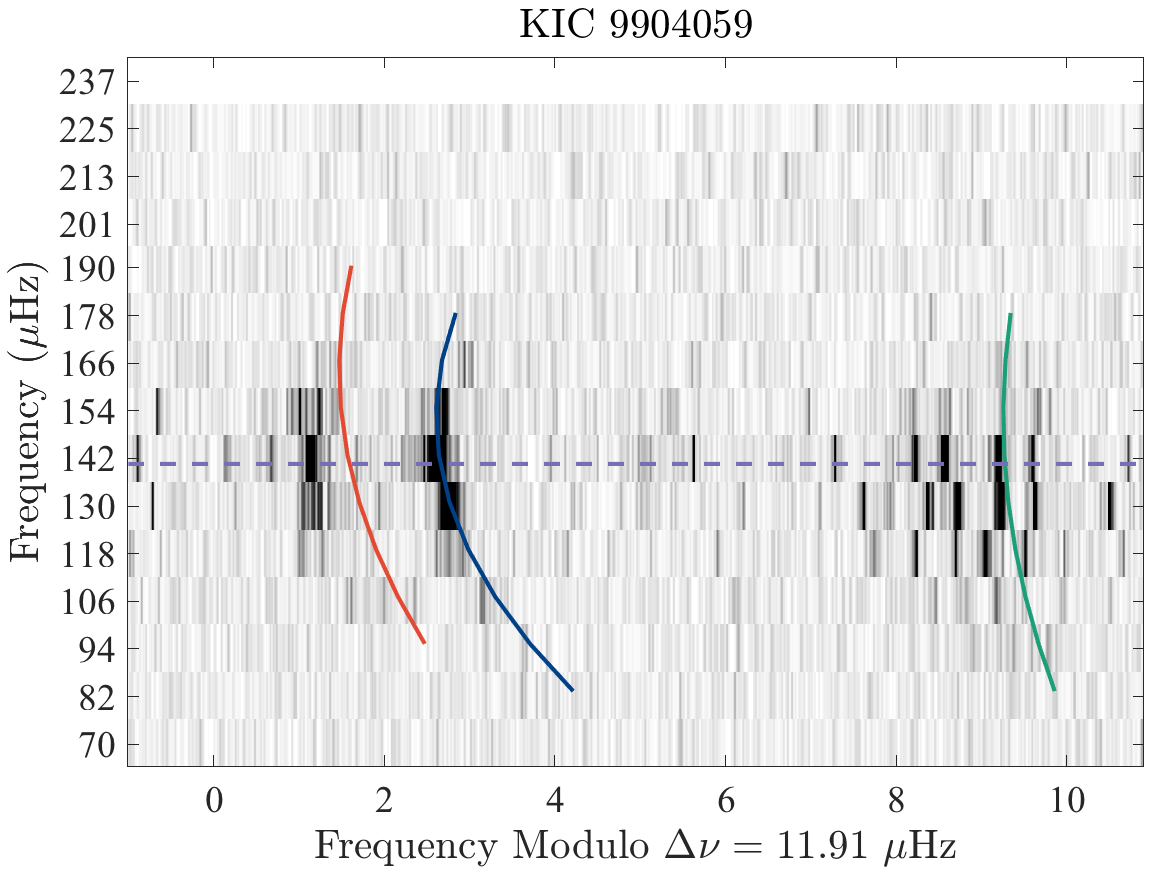}  \\
\includegraphics[width=.42\linewidth]{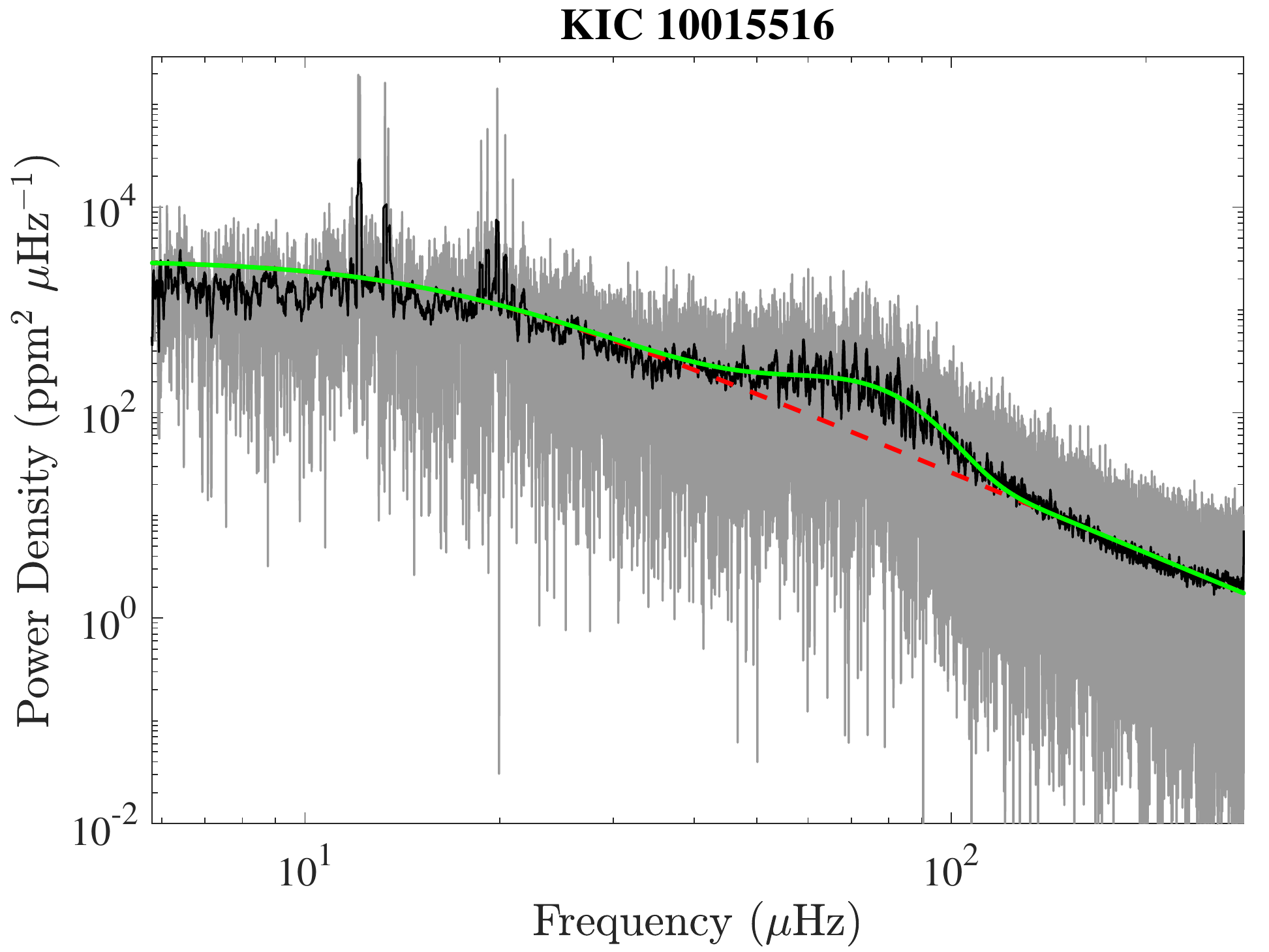} &
\includegraphics[width=.42\linewidth]{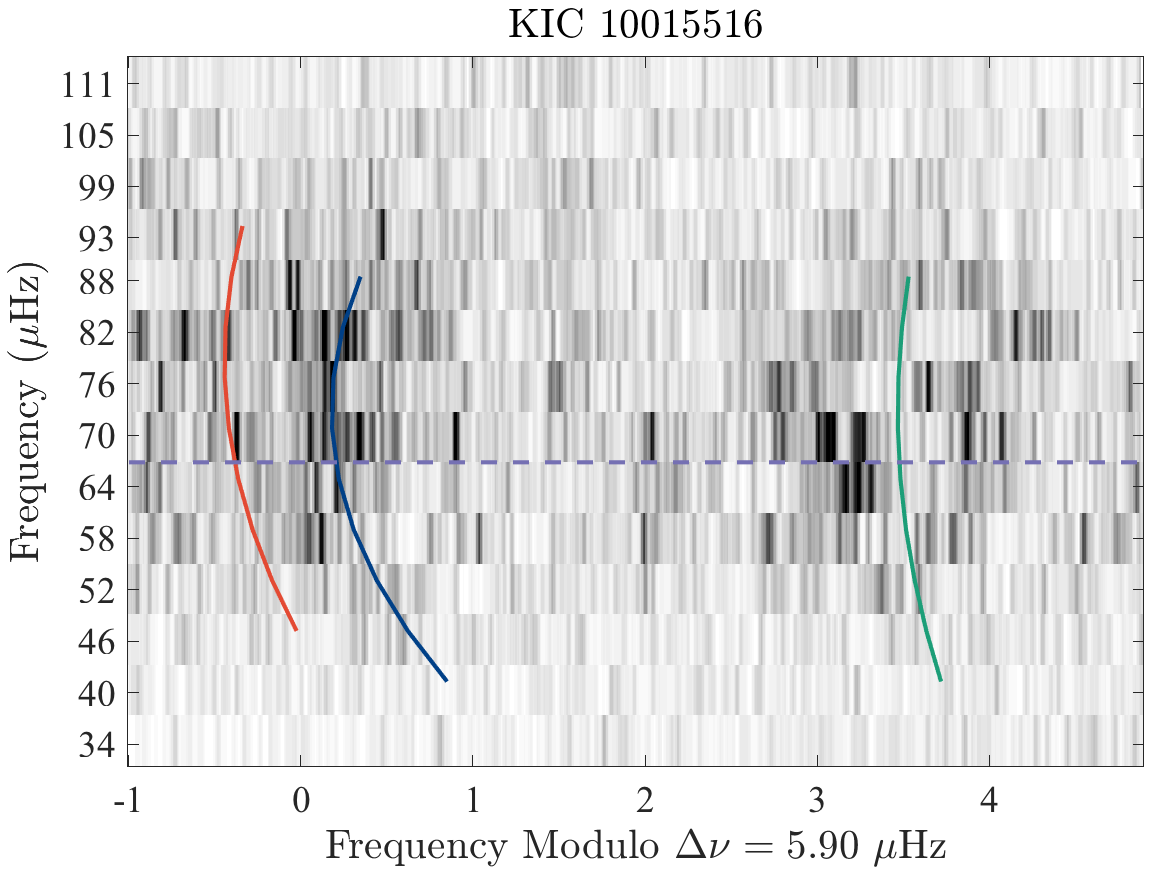} \\
\includegraphics[width=.42\linewidth]{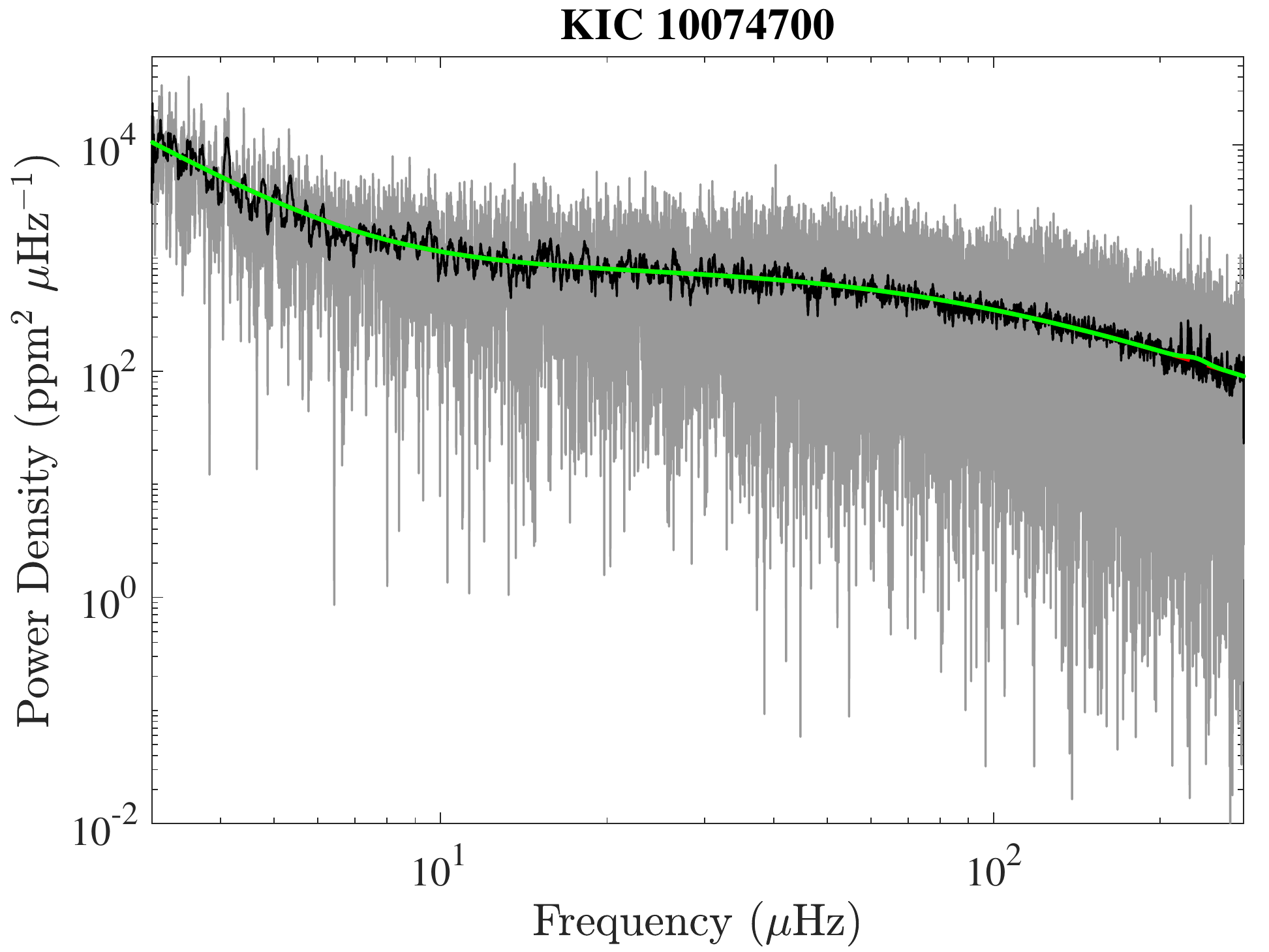} &
\includegraphics[width=.42\linewidth]{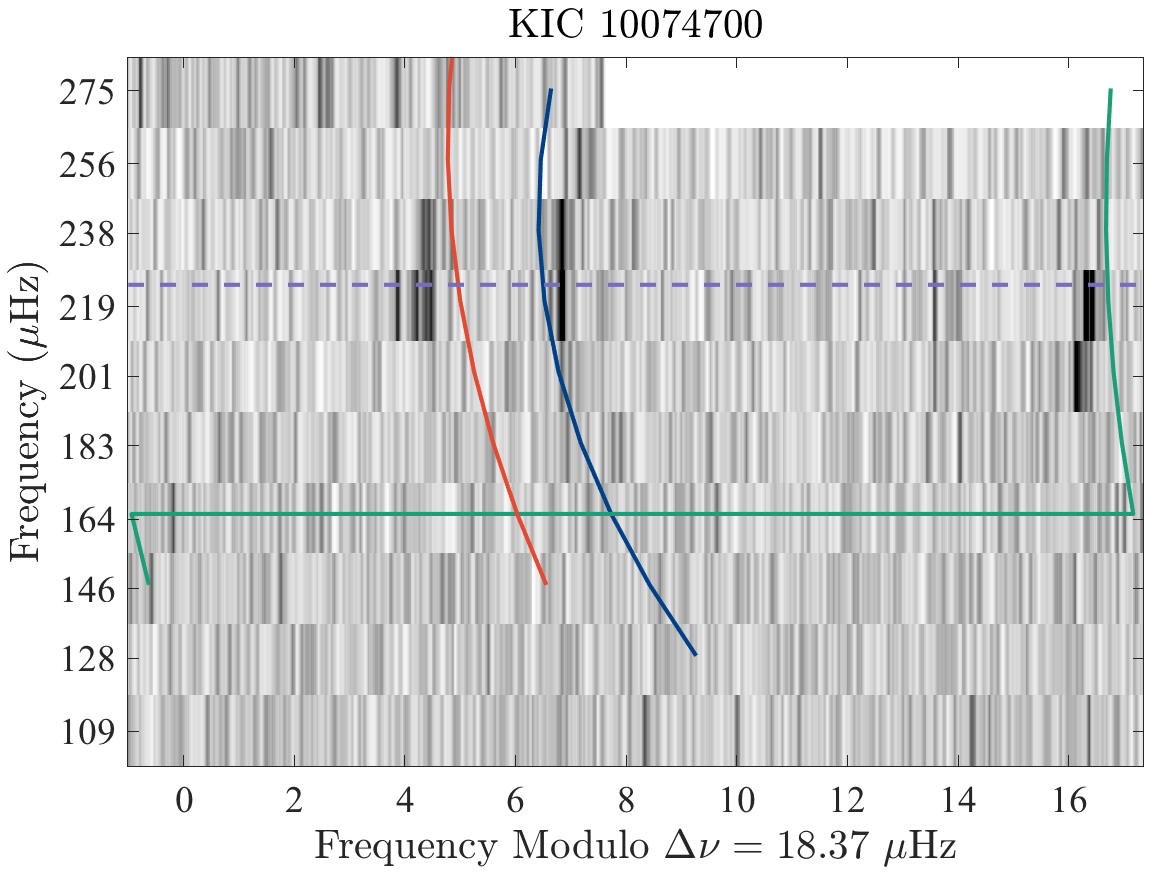} \\
\end{tabular}
\caption{Same as Fig. \ref{fig:seismic_fit_1}.}
\label{fig:seismic_fit_3}
\end{figure*}

\begin{figure*}
\centering
\begin{tabular}{cc}
\includegraphics[width=.42\linewidth]{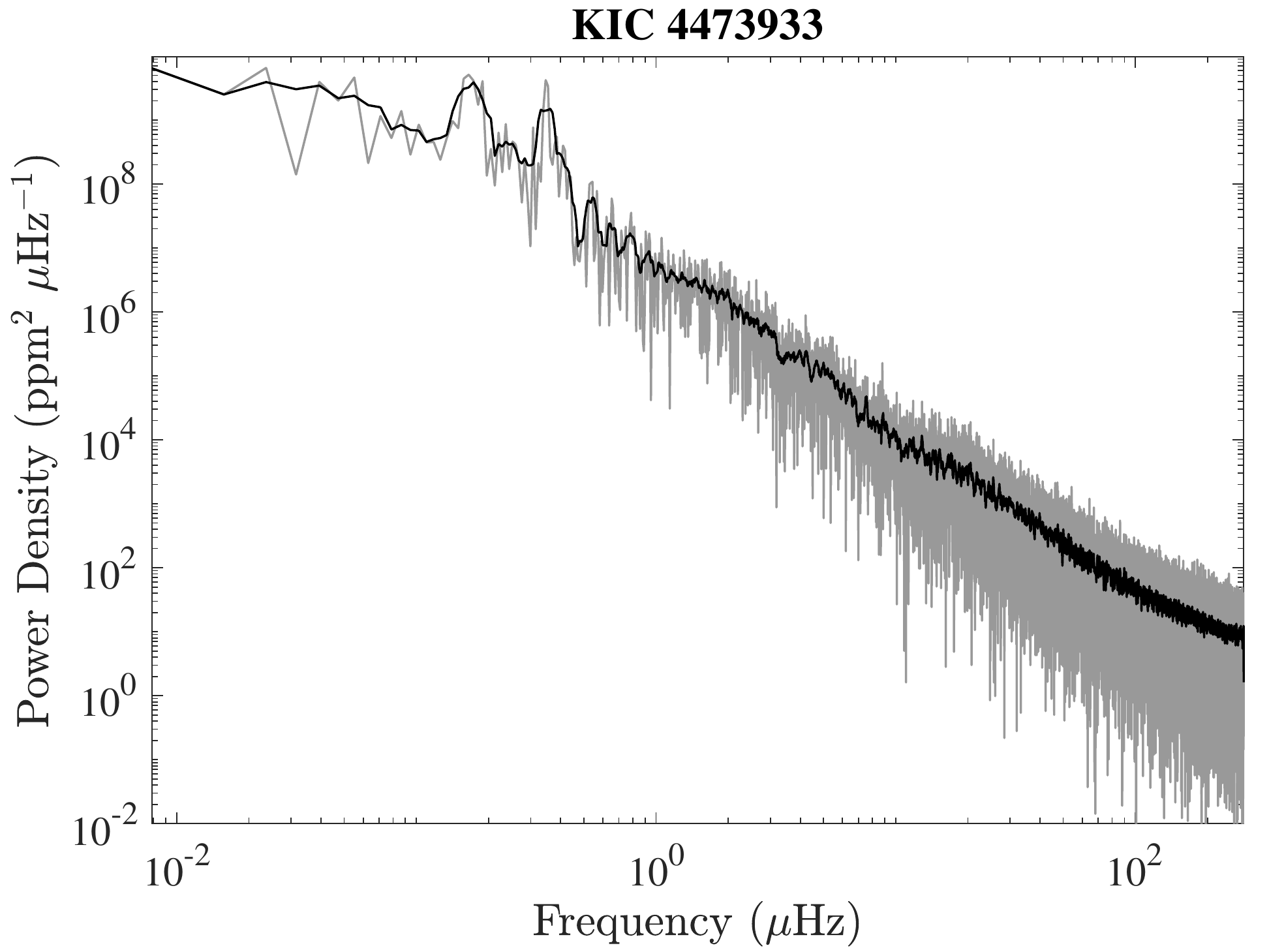}  &
\includegraphics[width=.42\linewidth]{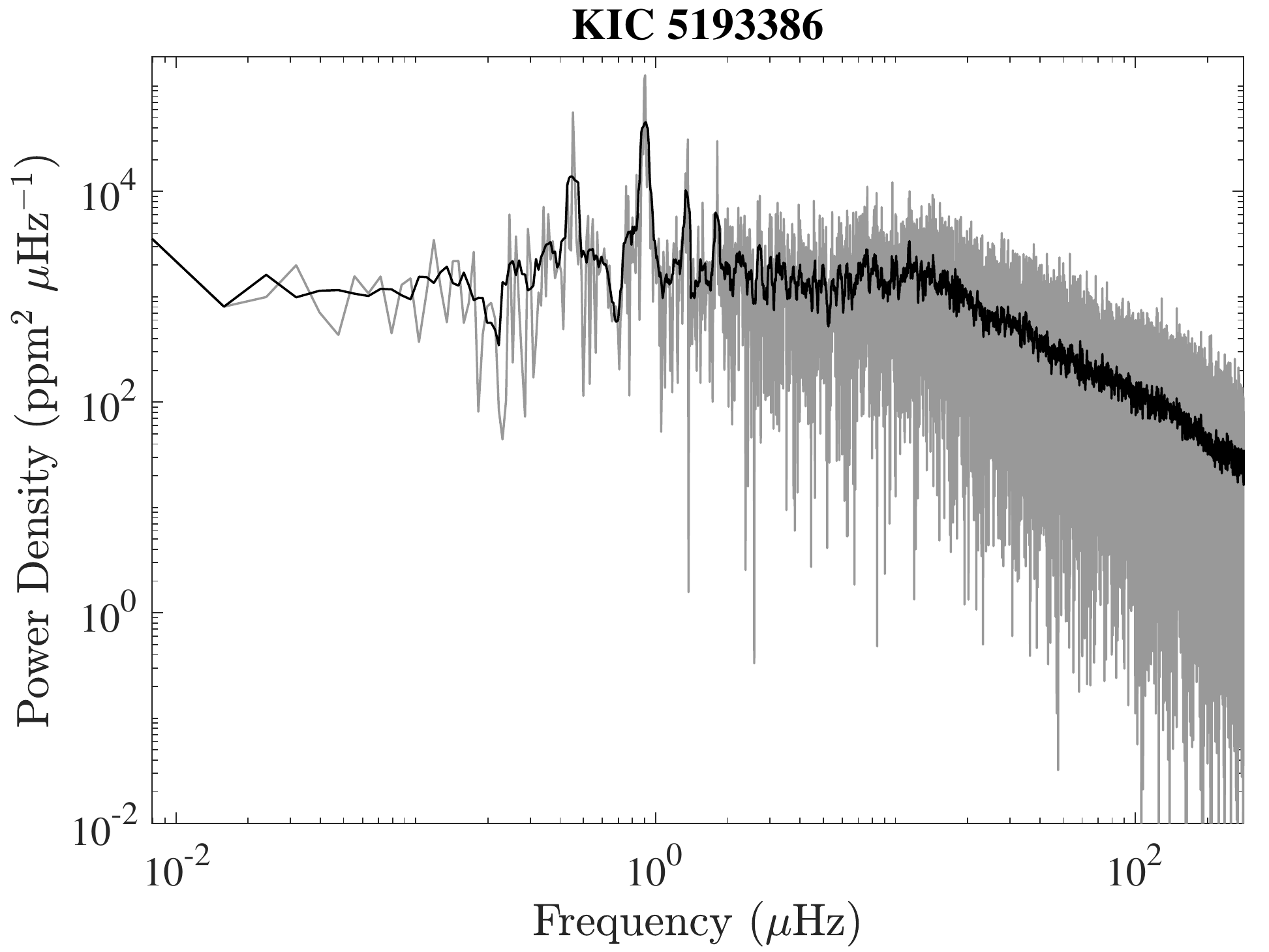}  \\
\includegraphics[width=.42\linewidth]{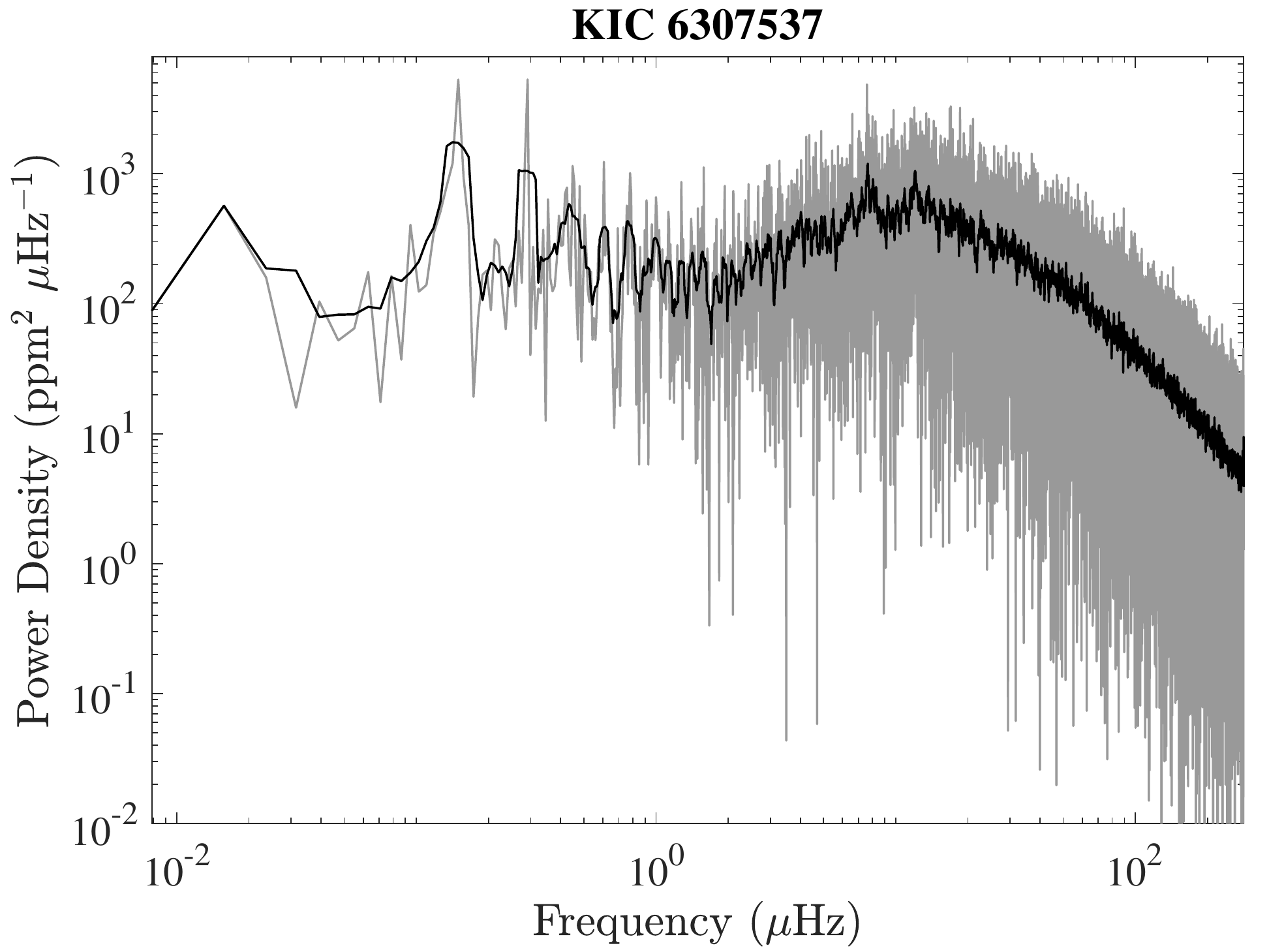}  &
\includegraphics[width=.42\linewidth]{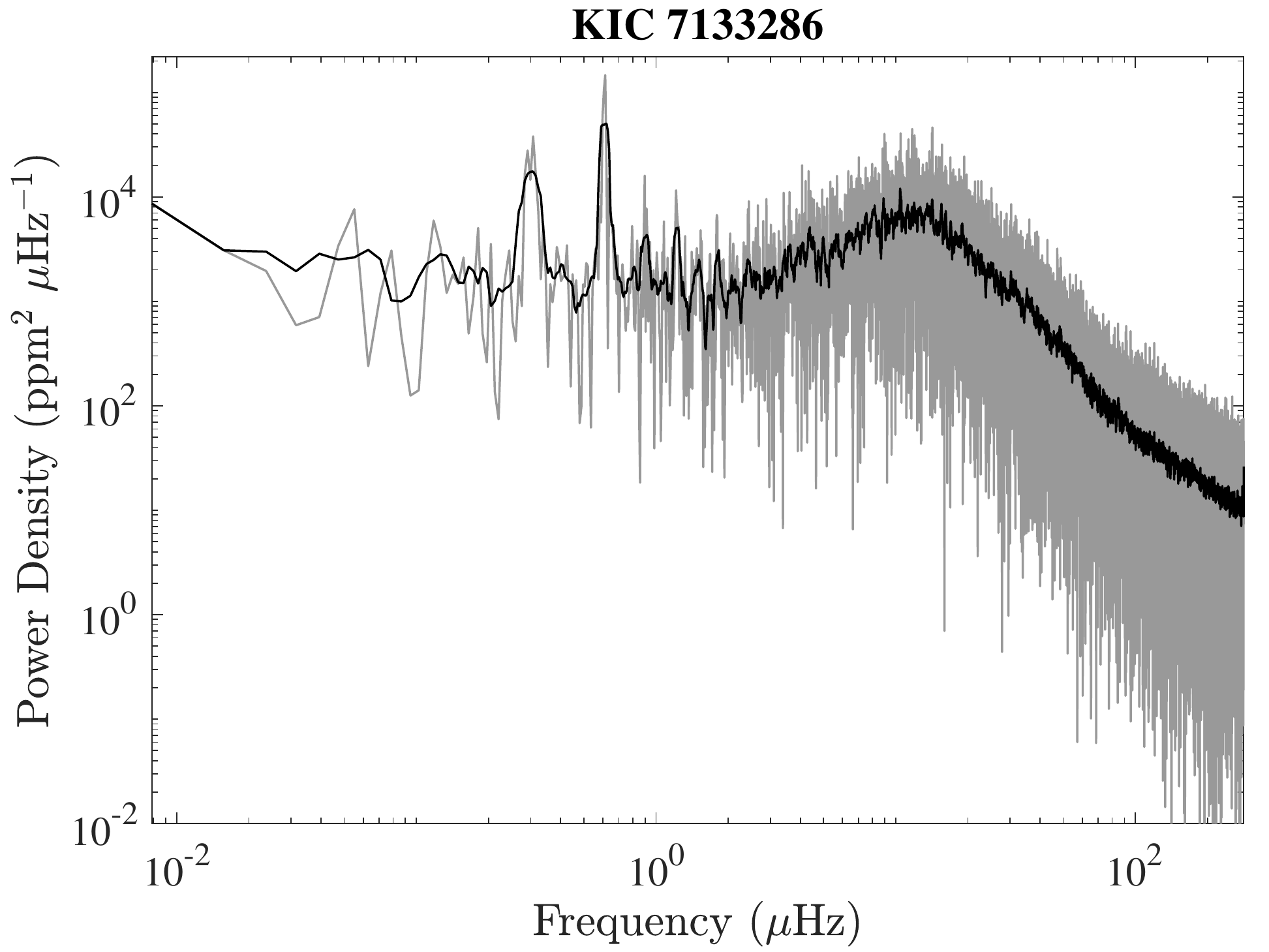}  \\
\includegraphics[width=.42\linewidth]{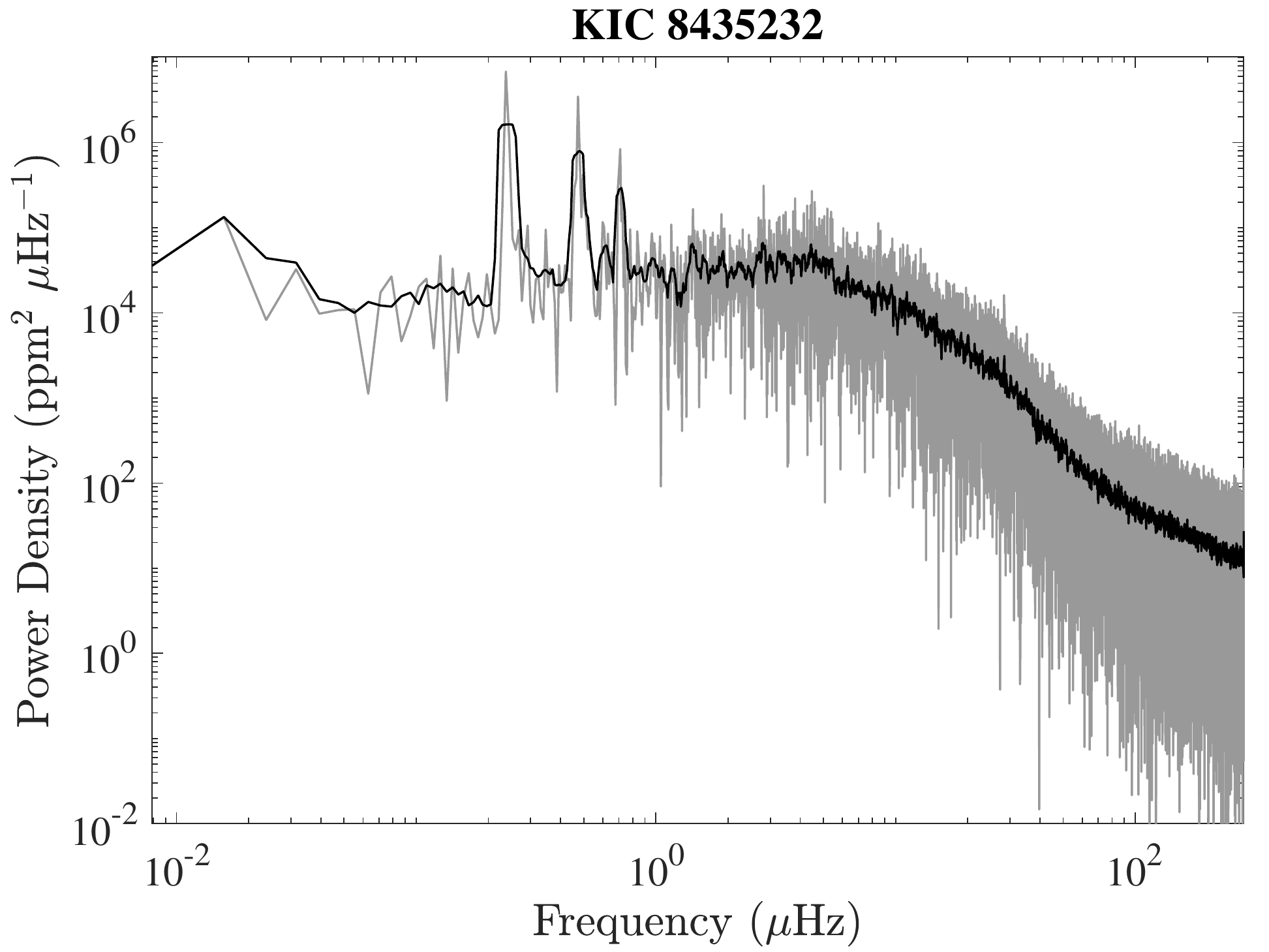}  &
\includegraphics[width=.42\linewidth]{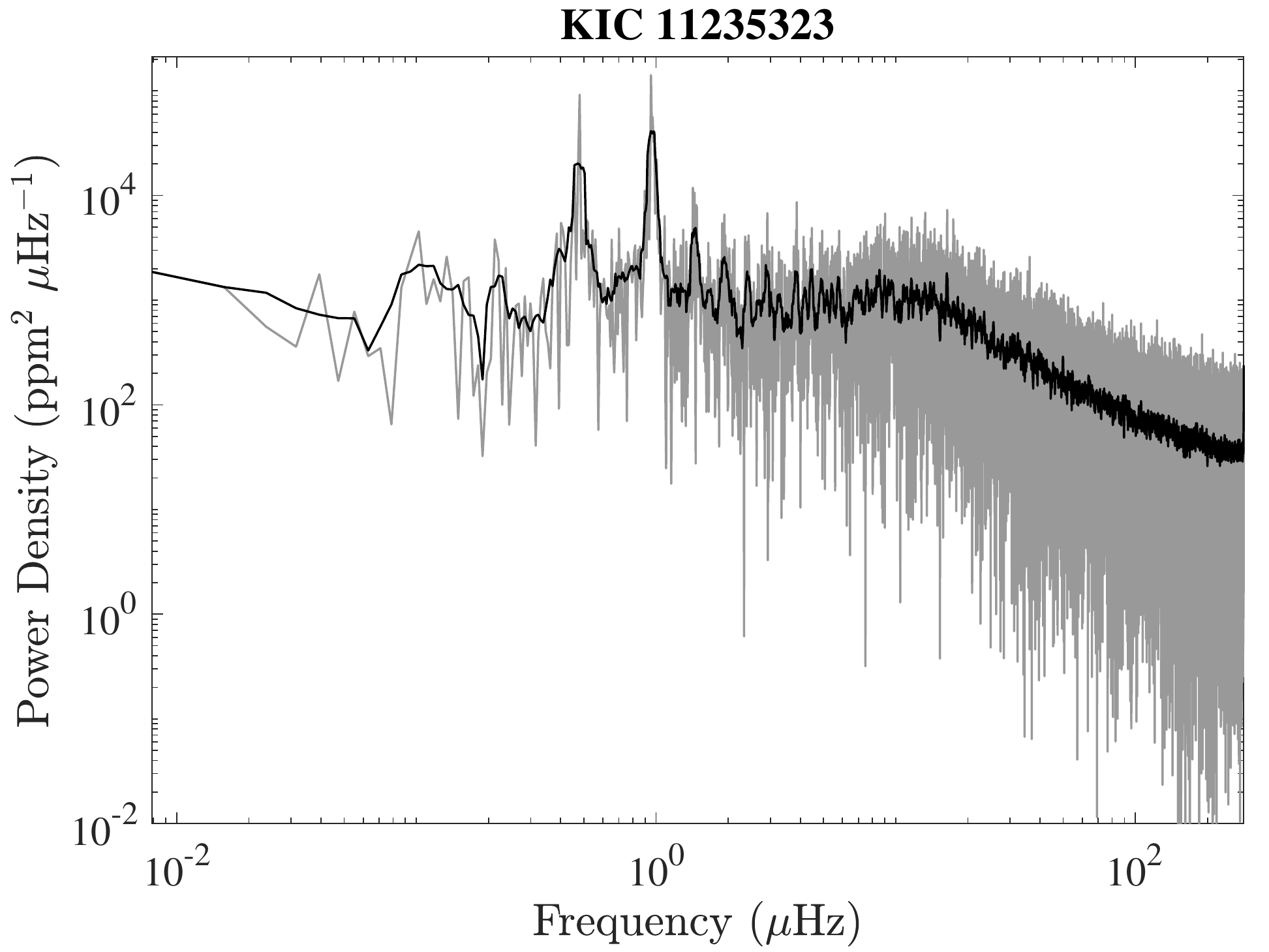} \\
\end{tabular}
\caption{
Power spectrum fit of the non-oscillating stars of the sample.
In each panel, the gray line represents the power spectral density and the black line corresponds to its smoothed version.
A description of the method is given in Sect. 3.1.1.
}
\label{fig:psd_non_osc}
\end{figure*}

\section{Table with the dynamical masses and radii from the literature}

\begin{sidewaystable*}
\caption{Benchmark EBs from the literature. The acronyms F13, G16, B18, and T18 refer to \citet{Frandsenetal2013}, \citet{Gaulmeetal2016}, \citet{Brogaardetal2018}, and \citet{Themessletal2018}, respectively.}
\label{tab:benchmarksFromLiterature}
\centering
\begin{tabular}{ccccccccccc}
\hline\hline
Label & KIC & $\Mdyn$  & $\Rdyn$  & $\Teff$ & $\log g$ & [Fe/H] & $\numax$ & $\DnuObs$ & $\Sph$ & References \\
      &      & $\Msun$ & $\Rsun$  &  K      & dex      & dex    &  $\mu$Hz & $\mu$Hz   &  \% &  \\
\hline
  $\alpha$ & 5786154  & 1.06(6)  & 11.4(2)  & 4747  $\pm$ 100 & 2.6(2)   & -0.06(6) & 29.75  $\pm$ 0.16 & 3.523 $\pm$ 0.014 & 0.08 & G16 \\
   $\beta$ & 7037405  & 1.25(4)  & 14.1(2)  & 4500  $\pm$ 80  & 2.5(2)   & -0.34(1) & 21.75  $\pm$ 0.14 & 2.792 $\pm$ 0.012 & 0.09 & G16, B18 \\
  $\gamma$ & 7377422  & 1.05(8)  & 9.5(2)   & 4938  $\pm$ 110 & 3.1(2)   & -0.33(6) & 40.1   $\pm$ 2.1  & 4.643 $\pm$ 0.052 & 1.18 & G16 \\
  $\delta$ & 8410637  & 1.56(3)  & 10.7(1)  & 4699  $\pm$ 91  & 2.7(1)   &  0.16(3) & 46.00  $\pm$ 0.19 & 4.641 $\pm$ 0.017 & 0.05 & F13, G16, T18 \\
$\epsilon$ & 8430105  & 1.31(2)  & 7.65(5)  & 5042  $\pm$ 68  & 3.04(9)  & -0.49(4) & 76.70  $\pm$ 0.57 & 7.138 $\pm$ 0.031 & 1.17 & G16 \\
   $\zeta$ & 9246715  & 2.149(7) & 8.30(4)  & 5030  $\pm$ 45  & 3.0(2)   &  0.05(2) & 106.40 $\pm$ 0.80 & 8.310 $\pm$ 0.020 & 0.52 & G16 \\
    $\eta$ & 9540226  & 1.33(5)  & 12.8(1)  & 4680  $\pm$ 80  & 2.2(1)   & -0.33(4) & 27.07  $\pm$ 0.15 & 3.216 $\pm$ 0.013 & 0.06 & G16, B18, T18 \\
  $\theta$ & 9970396  & 1.14(3)  & 8.0(2)   & 4860  $\pm$ 80  & 3.1(1)   & -0.23(3) & 63.70  $\pm$ 0.16 & 6.320 $\pm$ 0.010 & 0.04 & G16, B18 \\
   $\iota$ & 10001167 & 0.81(5)  & 12.7(3)  & 4700  $\pm$ 66  & 2.6(1)   & -0.69(4) & 19.90  $\pm$ 0.09 & 2.762 $\pm$ 0.012 & 0.10 & G16 \\
  $\kappa$ & 5640750  & 1.16(1)  & 13.12(9) & 4525  $\pm$ 75  & 2.266(6) & -0.29(9) & 24.1   $\pm$ 0.2  & 2.960 $\pm$ 0.006 & 0.07 & T18 \\
\hline
\end{tabular}
\end{sidewaystable*}

\section{Radial velocities}

Table~\ref{tab:radial_velocities} contains the radial velocities obtained at Apache Point and the Haute-Provence Observatories.


\section{Isochrone fitting of the systems}

\begin{figure*}
\centering
\includegraphics[width=0.31\textwidth,angle=0]{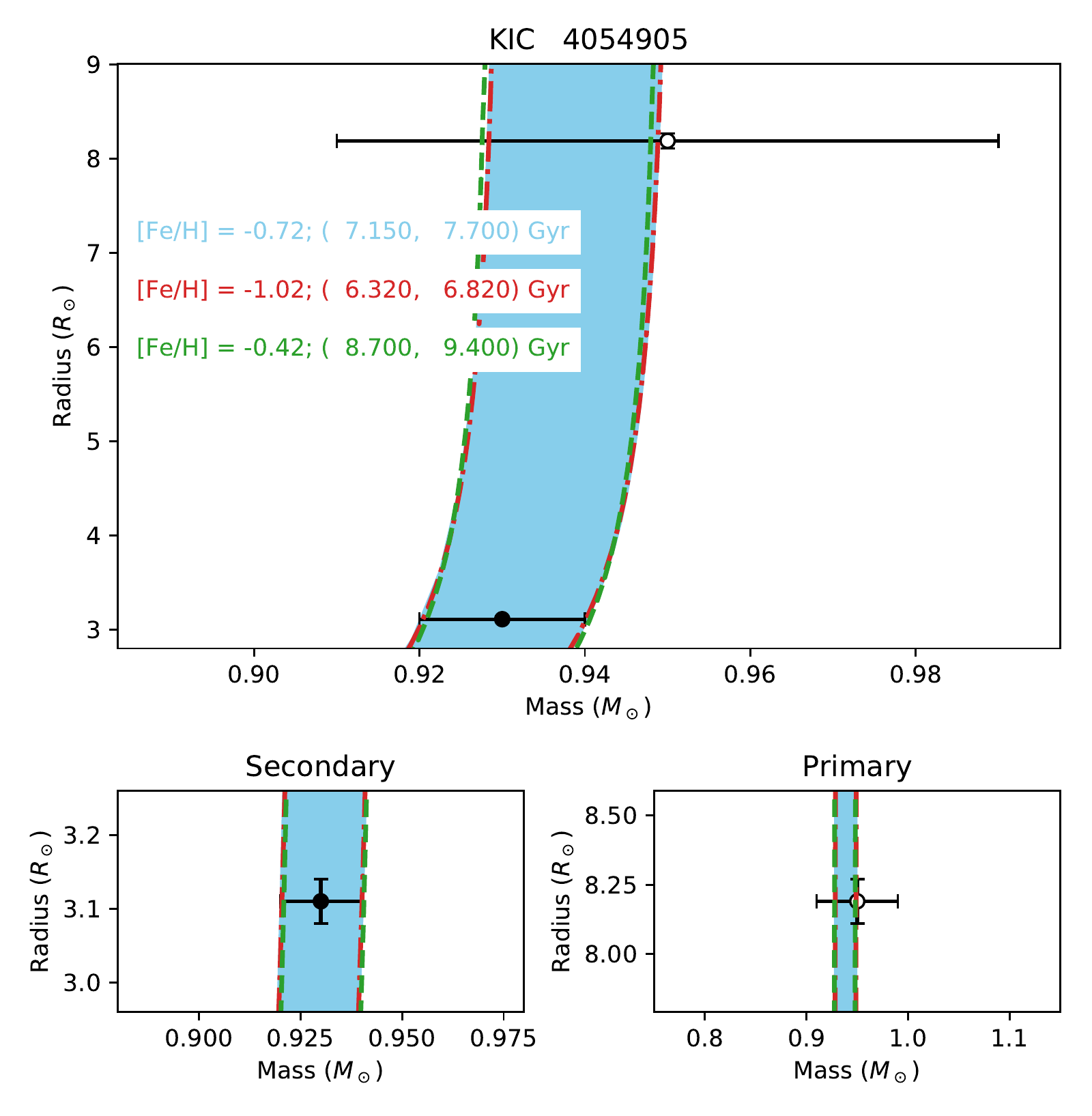}
\includegraphics[width=0.31\textwidth,angle=0]{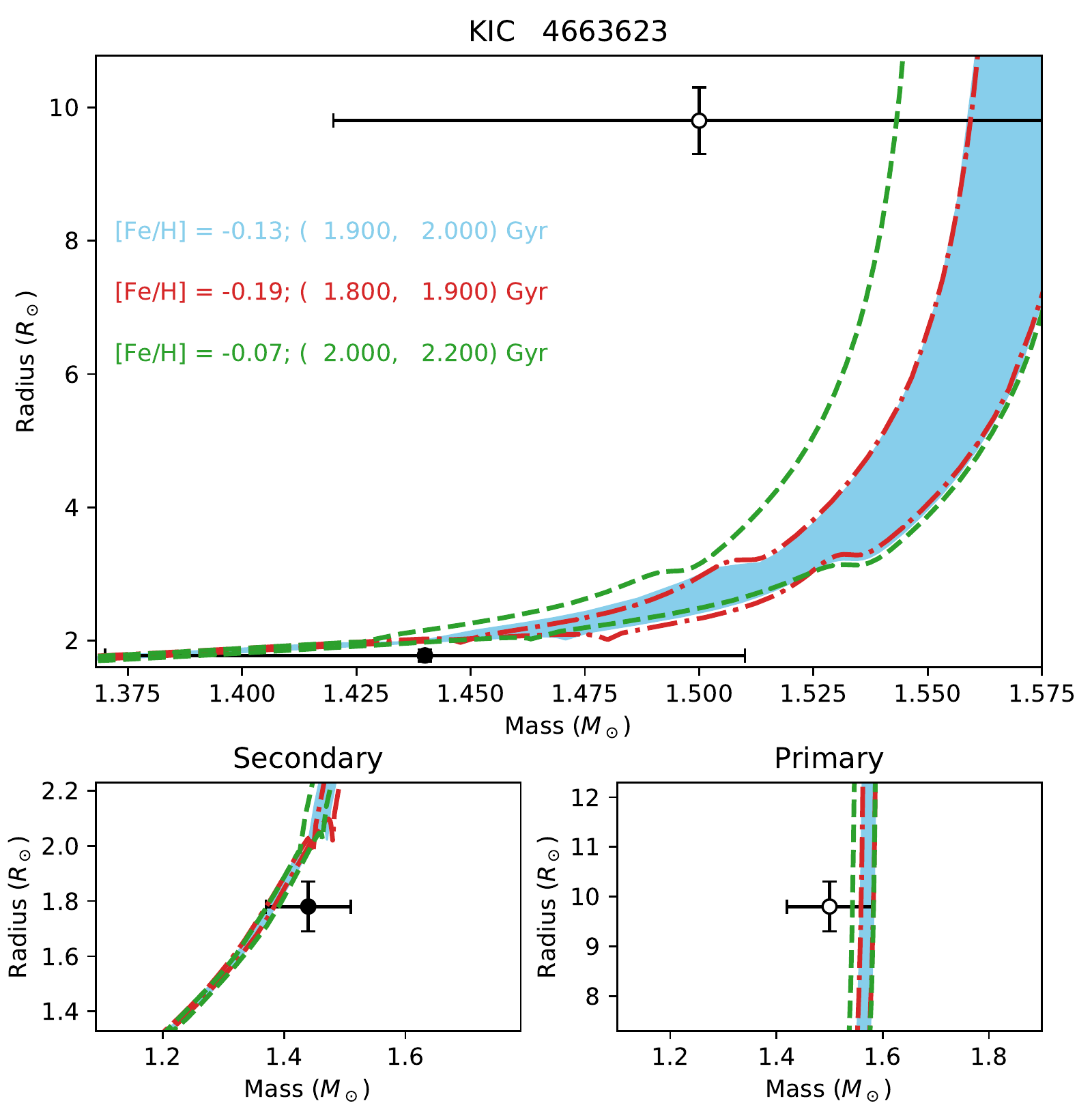}
\includegraphics[width=0.31\textwidth,angle=0]{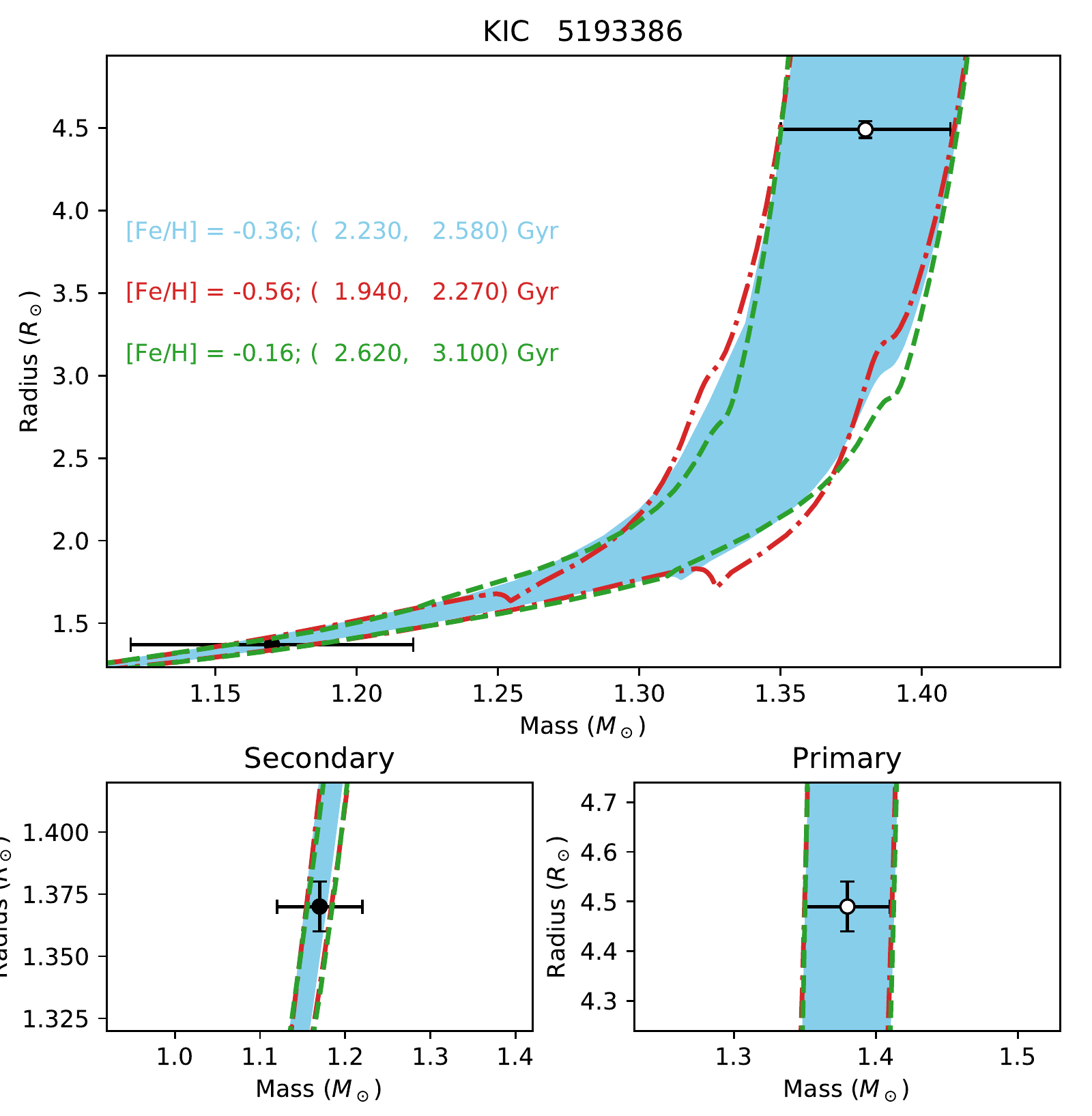}
\includegraphics[width=0.31\textwidth,angle=0]{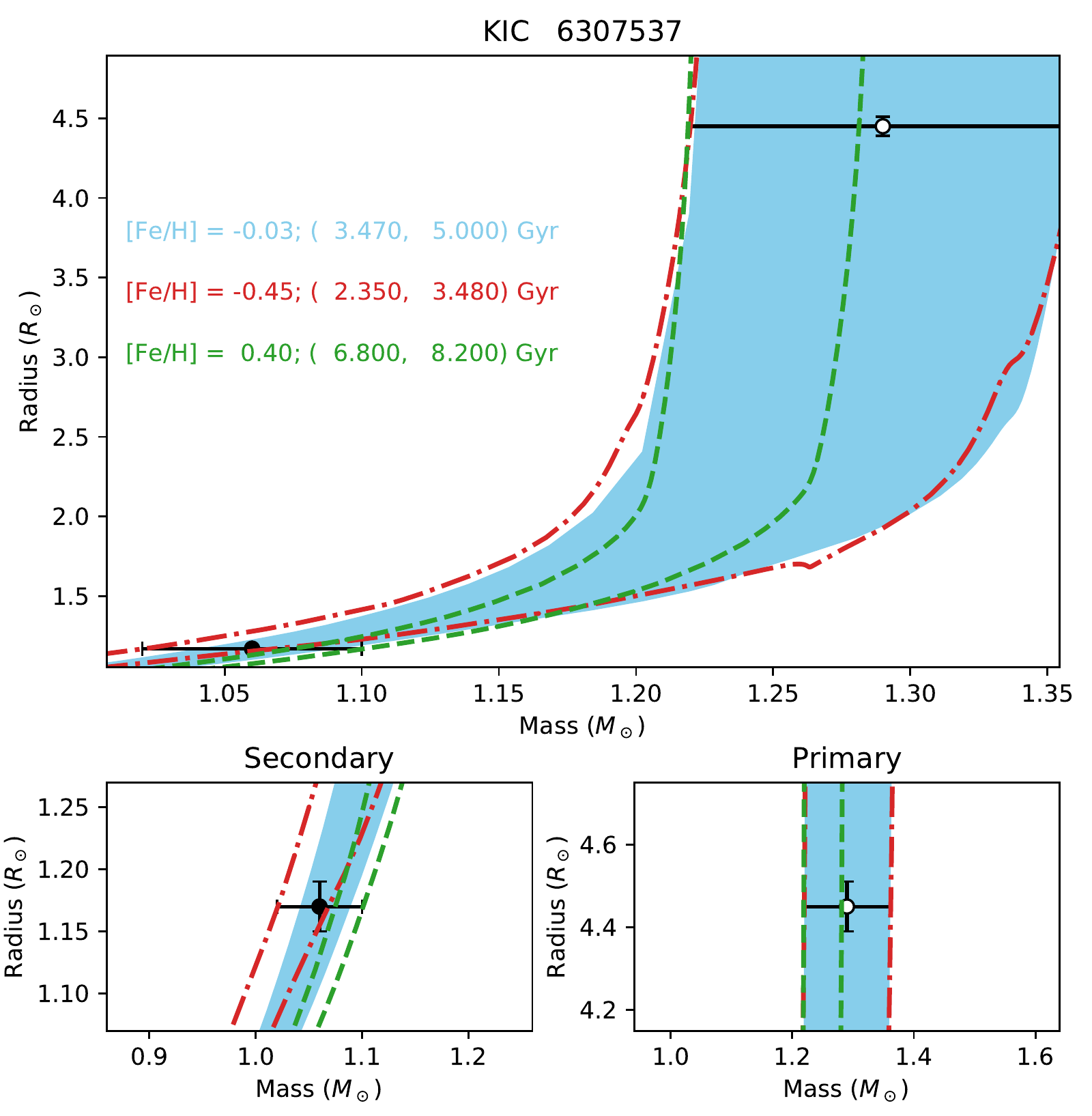}
\includegraphics[width=0.31\textwidth,angle=0]{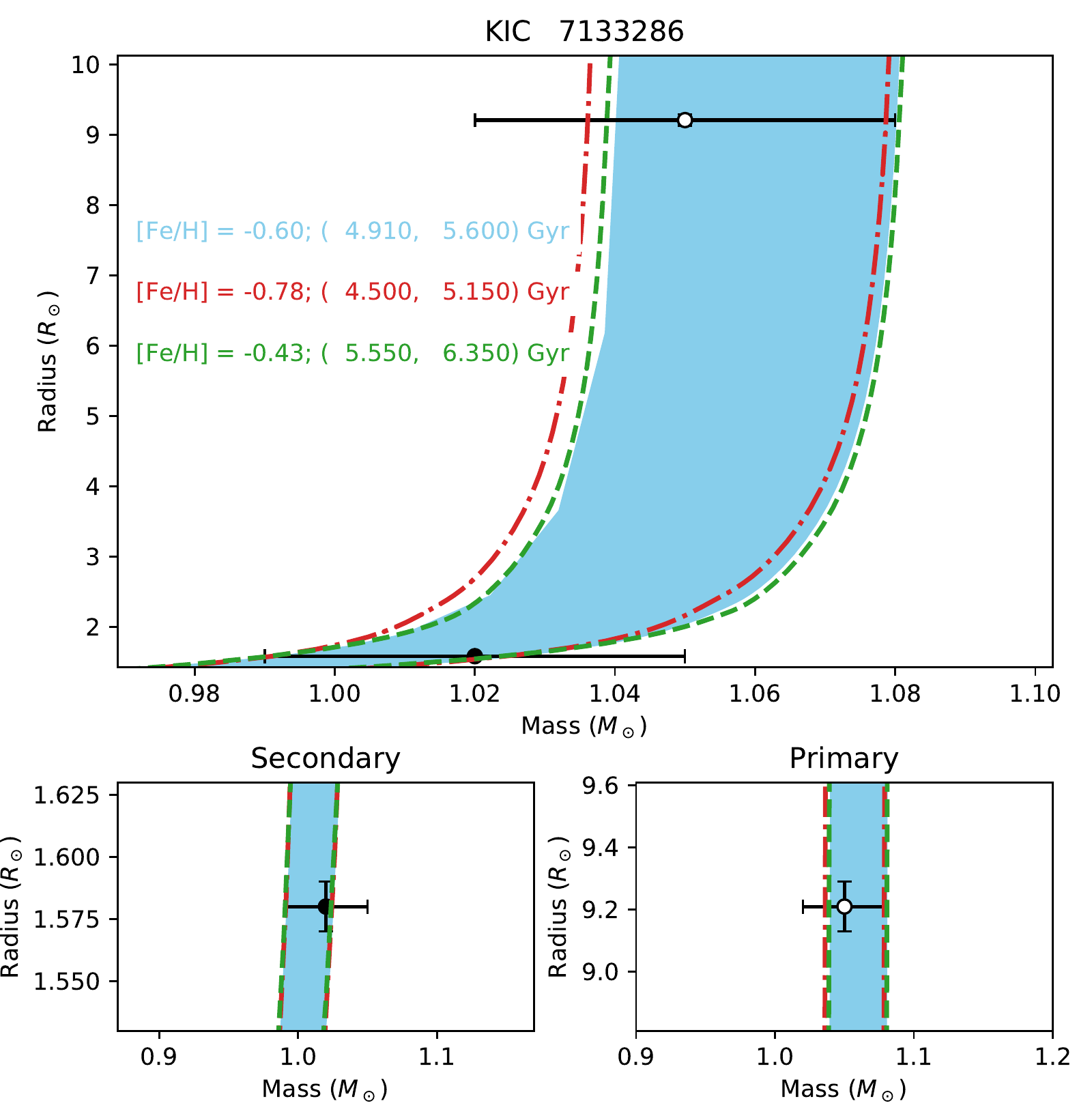}
\includegraphics[width=0.31\textwidth,angle=0]{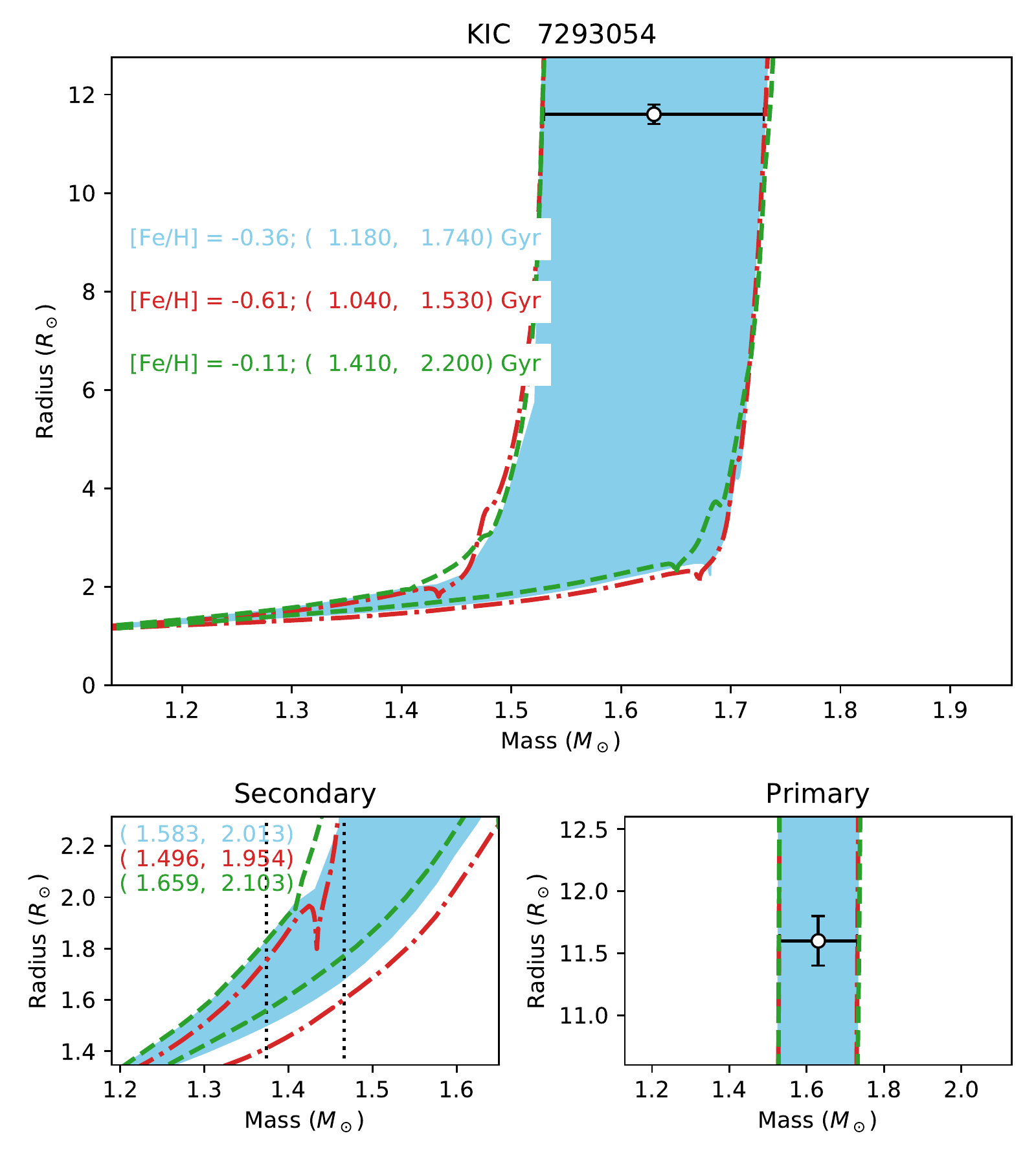}
\includegraphics[width=0.31\textwidth,angle=0]{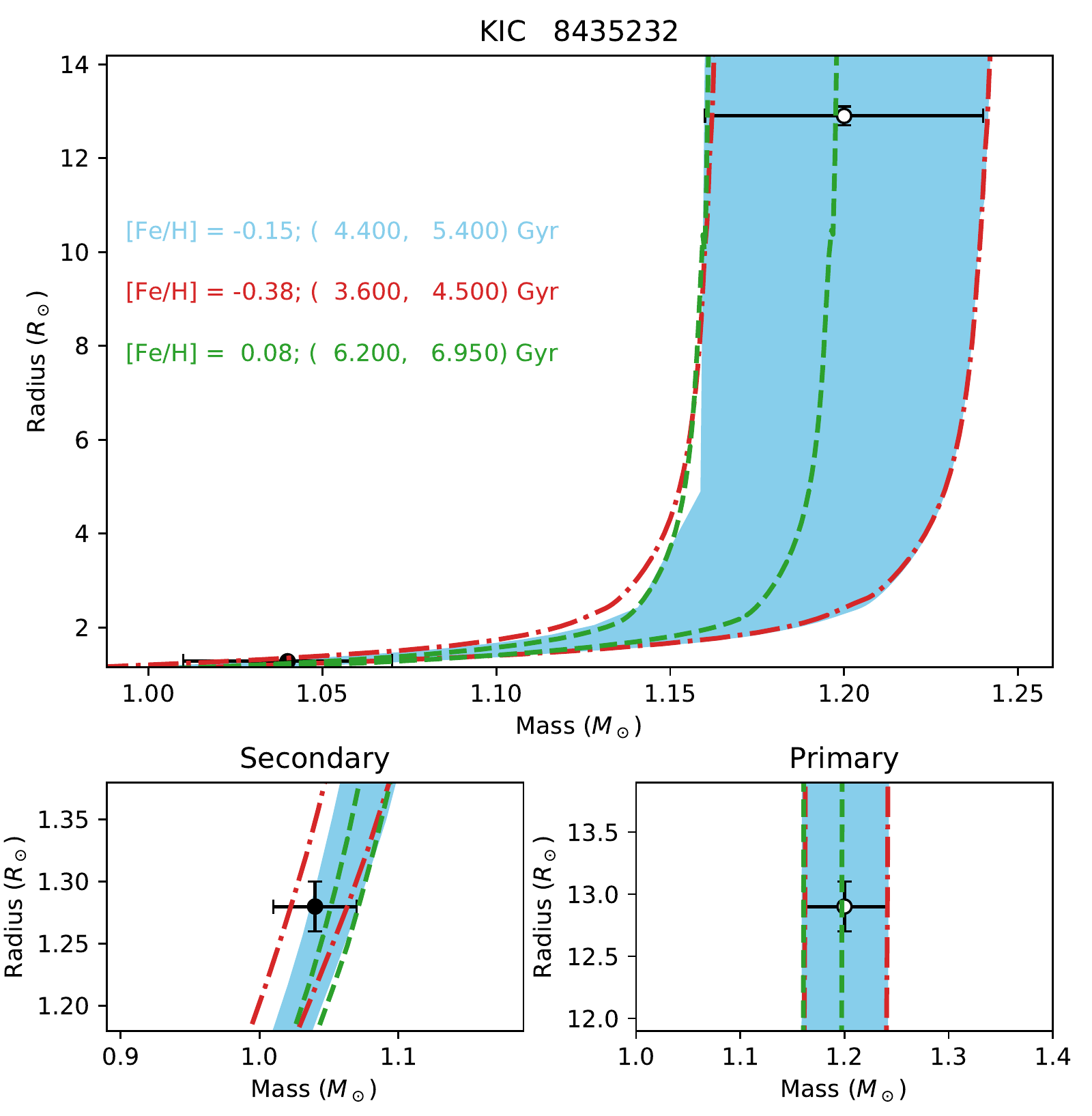}
\includegraphics[width=0.31\textwidth,angle=0]{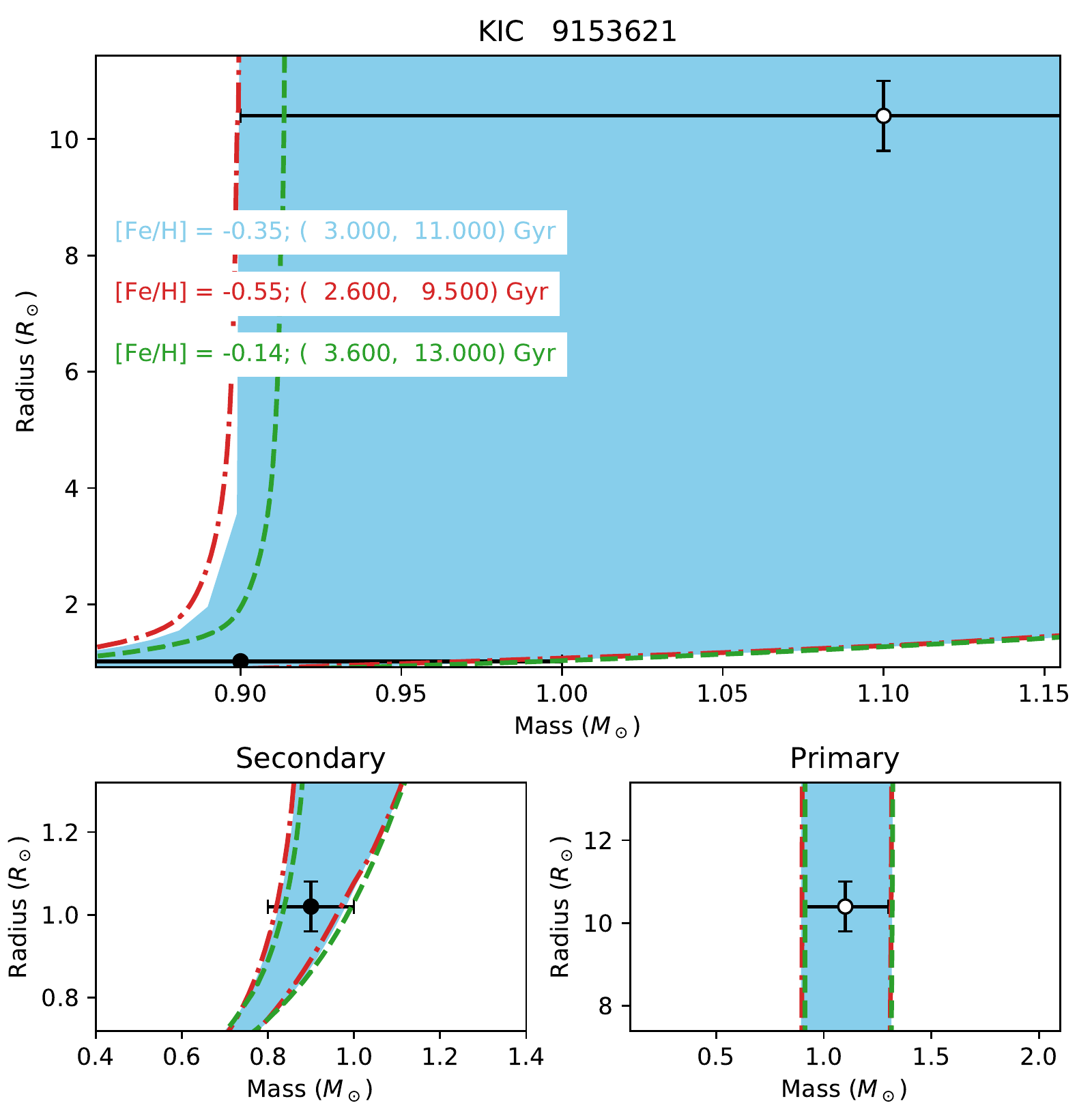}
\includegraphics[width=0.31\textwidth,angle=0]{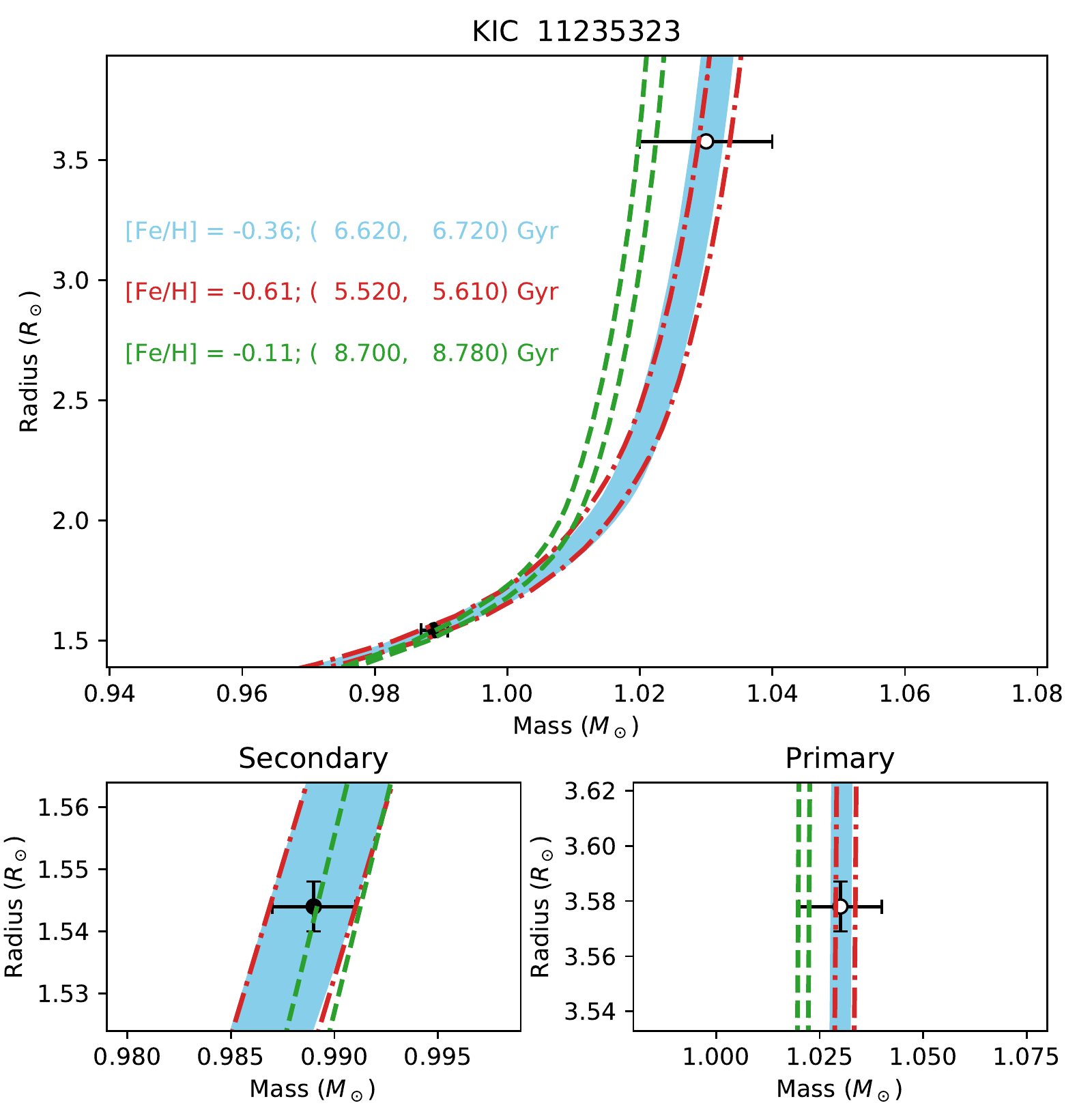}
\caption{Isochrone fitting of our SB2 systems (cf. Figure \ref{fig:RV_eclipses_SB2}) with YaPSI isochrones. Each panel corresponds to one system. For each panel, the top subpanel show the giant component (white dot) and the secondary (black dot), the bottom-right and bottom-left subpanels are zoom-ins of the area around the primary and secondary components, respectively. The isochrones are computed for three values of metallicity (see Sect.~\ref{subsec:age_stellar_model}).}
\label{fig:iso_fit1}
\end{figure*}

\begin{figure*}
\centering
\includegraphics[width=0.31\textwidth,angle=0]{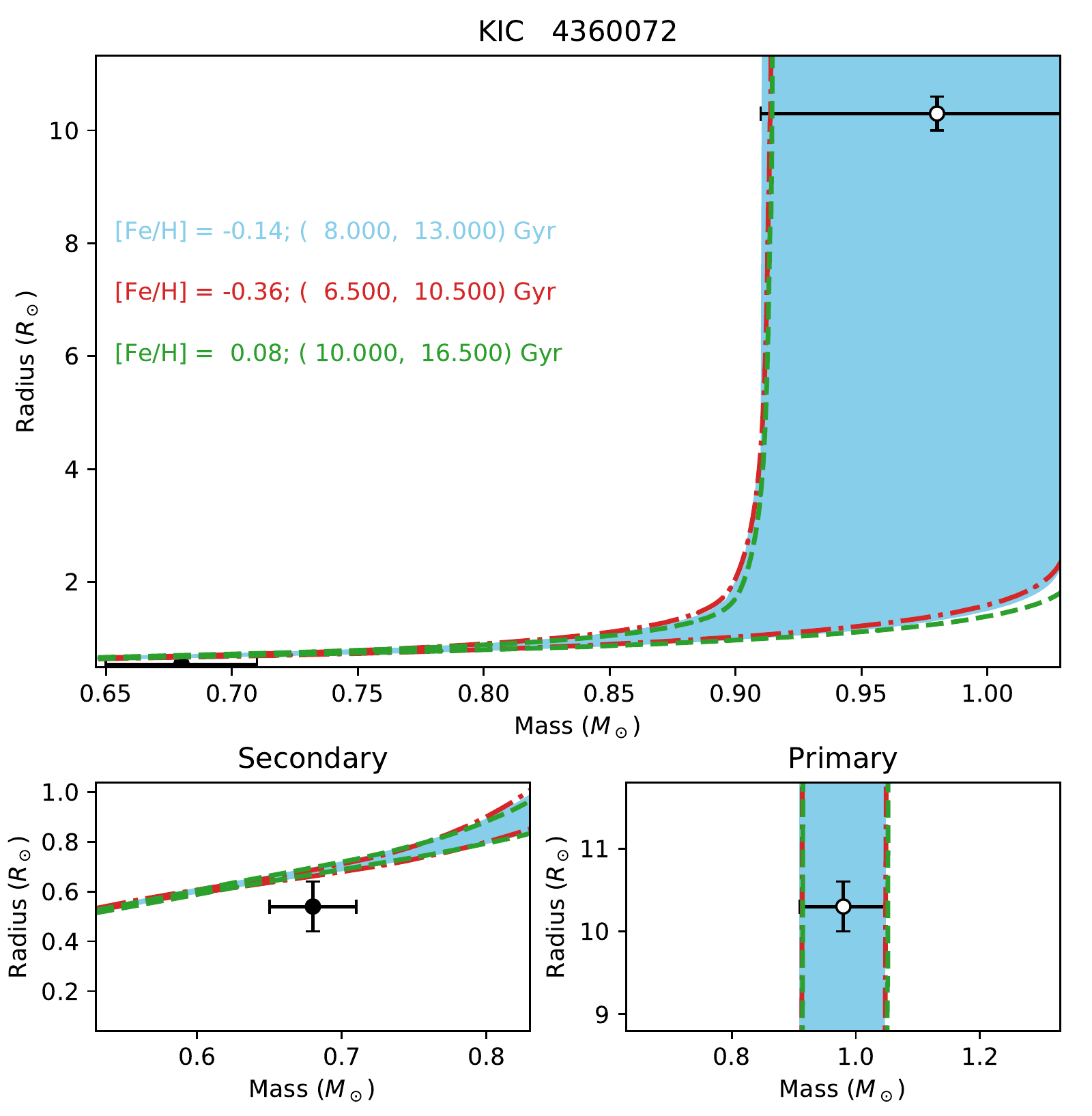}
\includegraphics[width=0.31\textwidth,angle=0]{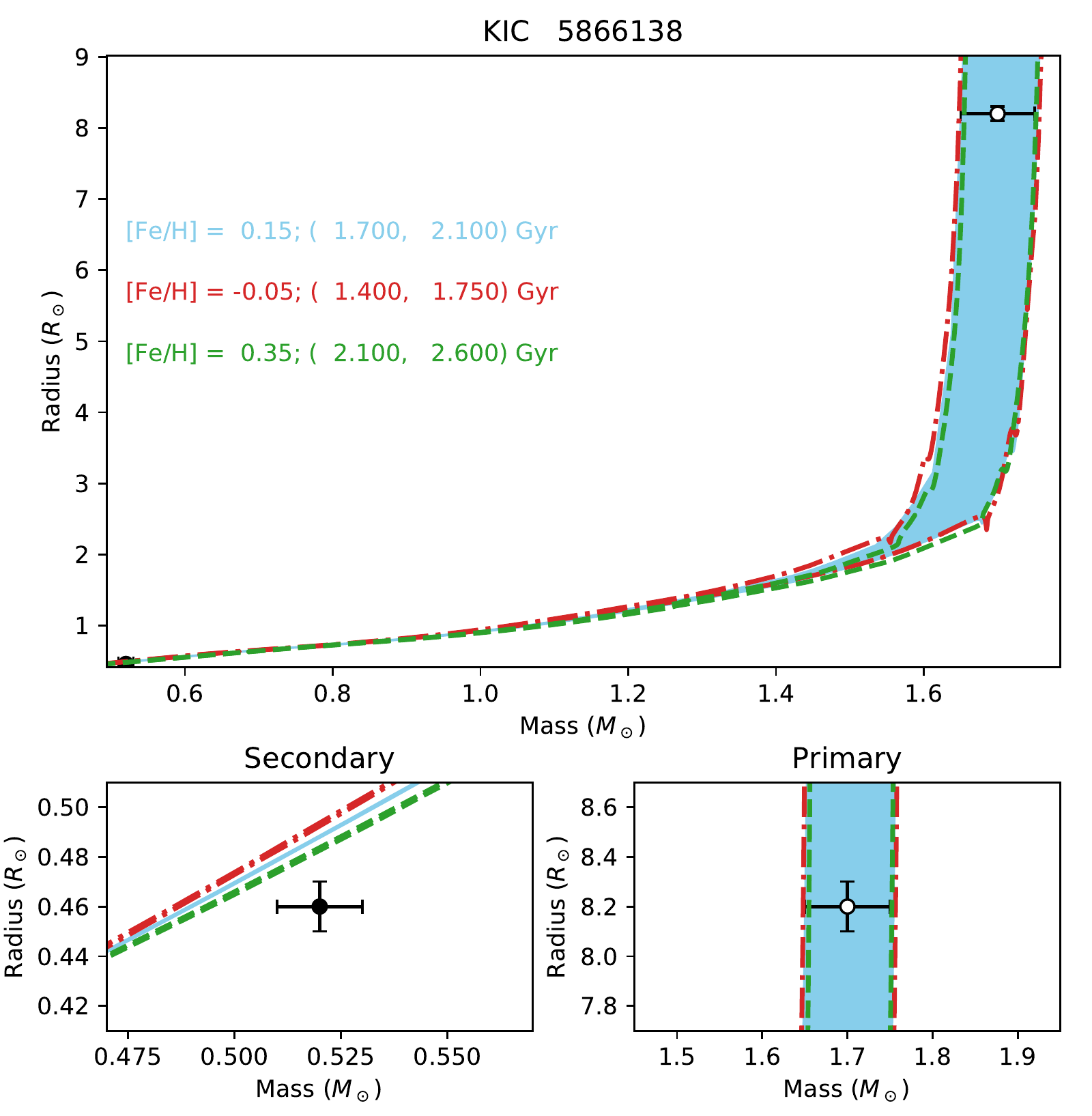}
\includegraphics[width=0.31\textwidth,angle=0]{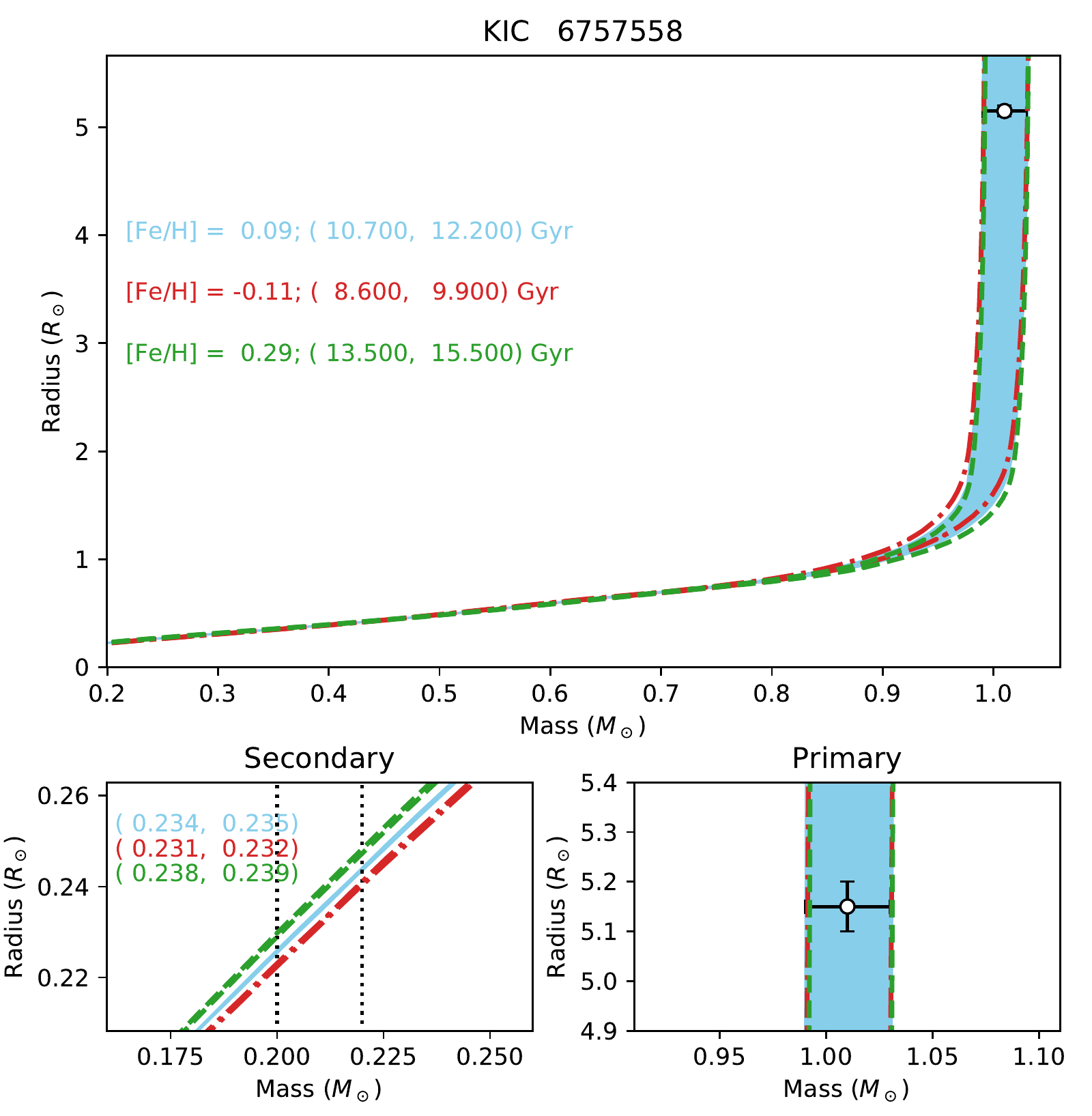}
\includegraphics[width=0.31\textwidth,angle=0]{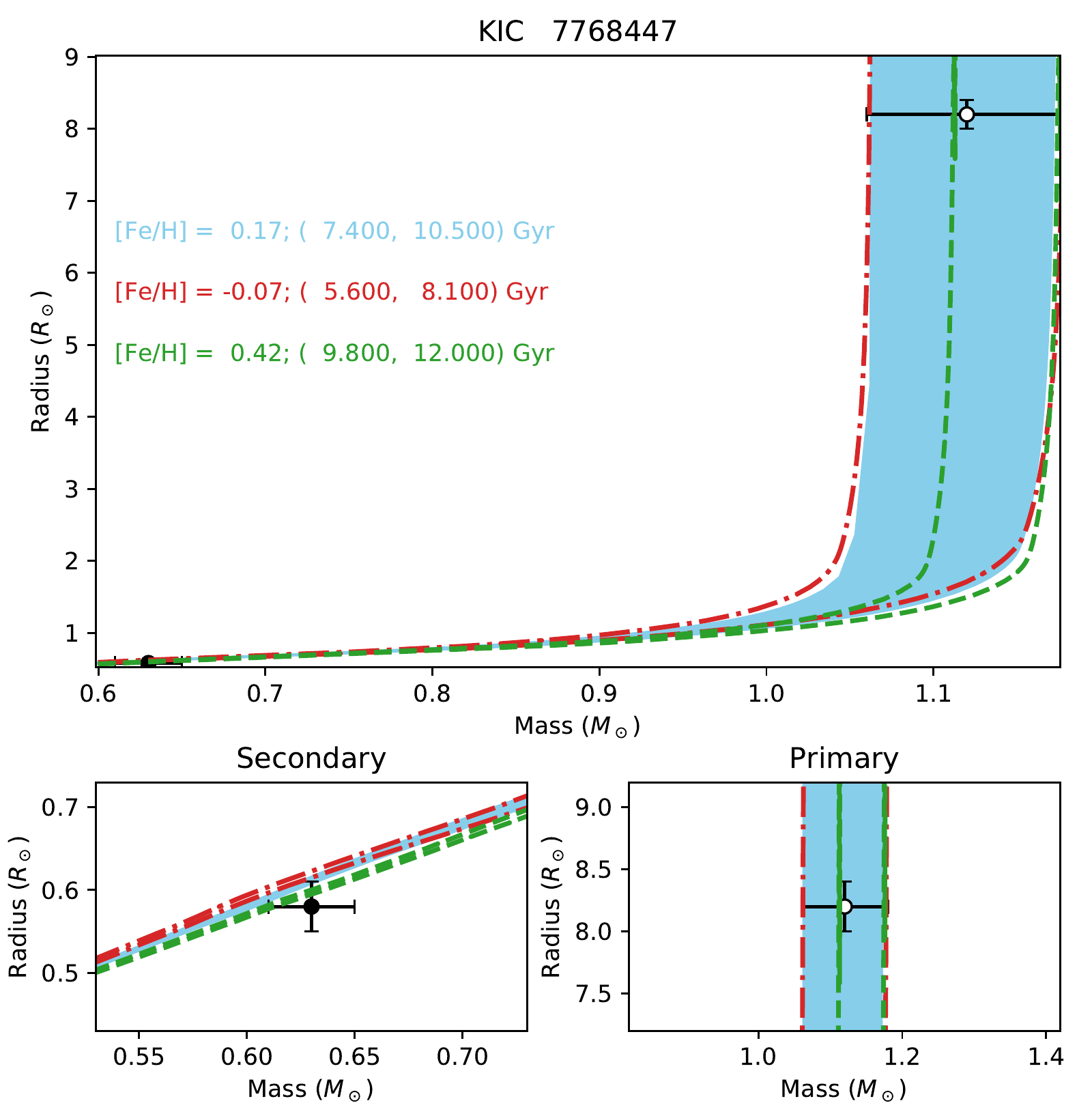}
\includegraphics[width=0.31\textwidth,angle=0]{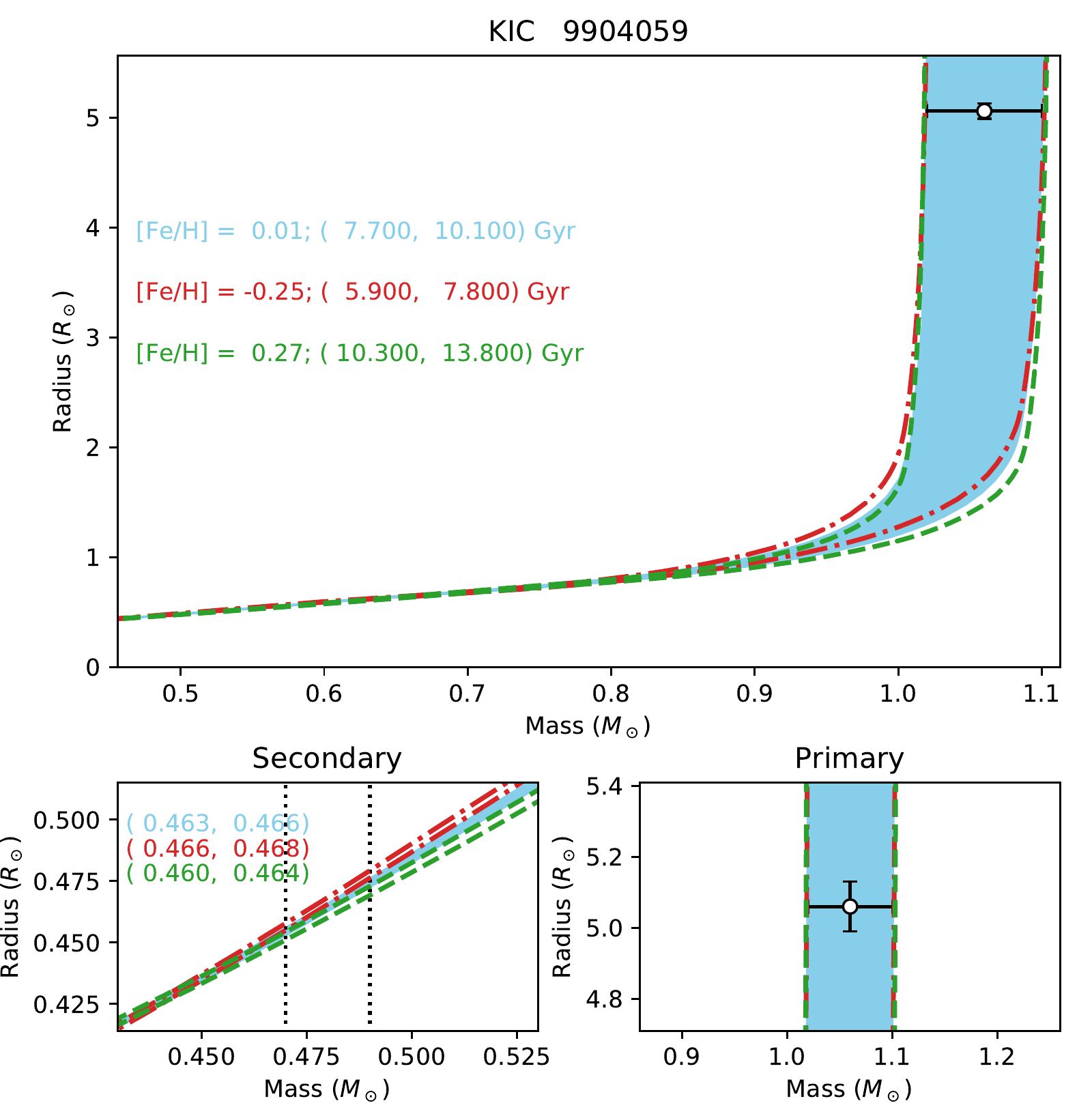}
\includegraphics[width=0.31\textwidth,angle=0]{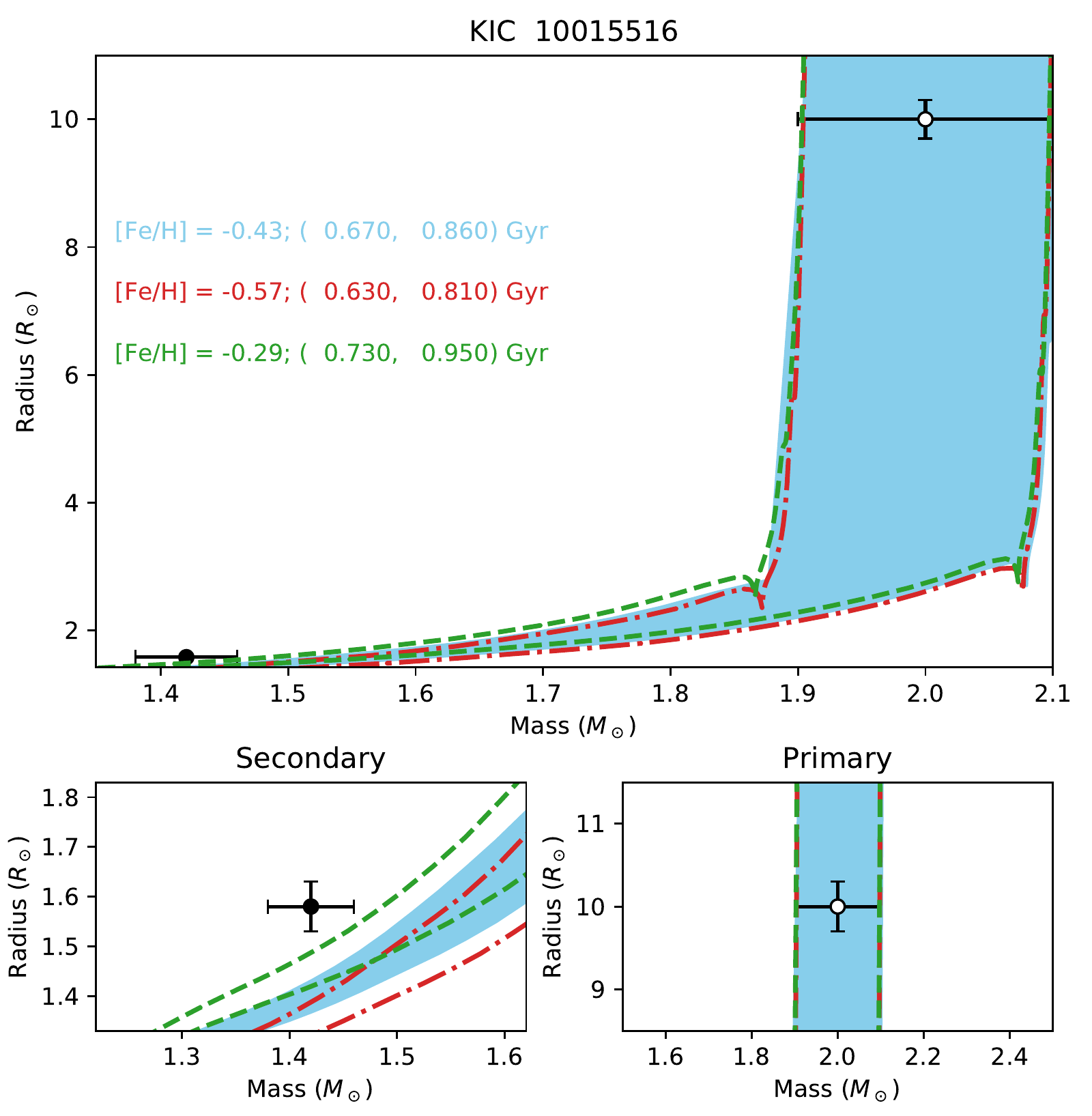}
\includegraphics[width=0.31\textwidth,angle=0]{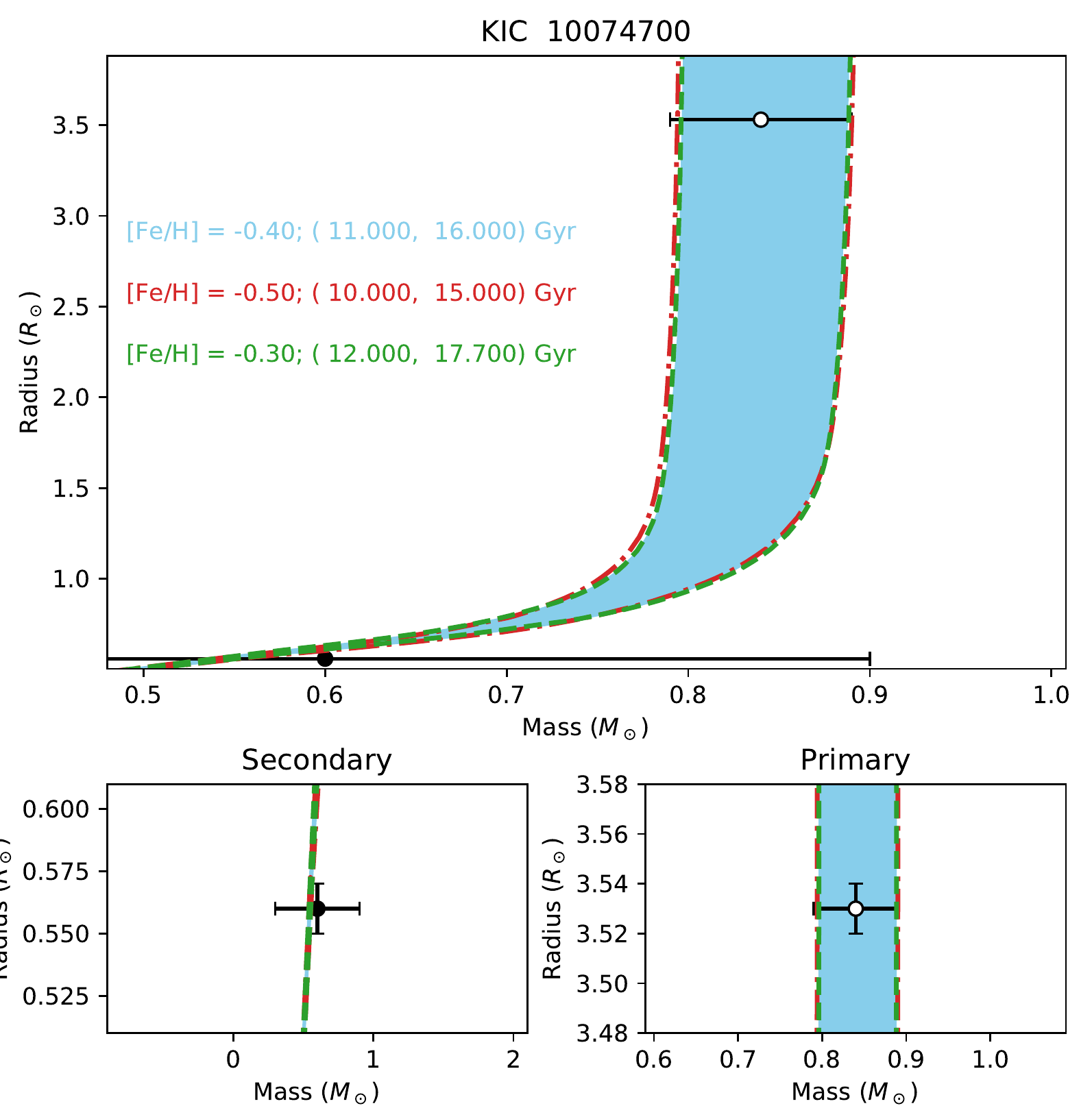}
\caption{Same as Fig.~\ref{fig:iso_fit1} for our SB1 systems (cf. Figure \ref{fig:RV_eclipses_SB1}).}
\label{fig:iso_fit2}
\end{figure*}

\clearpage
\onecolumn

\begin{longtable}{lll}
\caption{
Radial velocity data corresponding to the spectra obtained for the present paper. Dates are mid-exposure times expressed in \textit{Kepler} Julian dates (KJD), which are barycentric Julian dates BJD with an offset: KJD = BJD $-$ 2,454,833 days. The first spectrum was taken on March 25, 2016 (KJD $=2639.9$), and the last on November 15, 2020 (KJD$=4336.5$).  The least significant digit in brackets after the value indicates the statistical uncertainty arising from the position of the peaks in the BF. For the APO data, the dispersion of the data point with respect to the best-fit models indicate that the actual RV uncertainties are actually about 0.5 km s$^{-1}$ for the giant component and 1 km s$^{-1}$ for the companion. The statistical error on the BF does not include instrumental effect and atmospheric seeing. The empirical error bars were used to model our data.
Dates with a $\star$ symbol correspond with data taken at OHP.
The exact timestamps are provided in the electronic table associated to the paper. 
}
\label{tab:radial_velocities}\\
\hline\hline
Date (KJD) & RV1 ($\mathrm{km.s}^{-1}$) & RV2 ($\mathrm{km.s}^{-1}$) \\
\hline
\endfirsthead
\caption{Continued.} \\
\hline\hline
Date (KJD) & RV1 ($\mathrm{km.s}^{-1}$) & RV2 ($\mathrm{km.s}^{-1}$) \\
\hline
\endhead
\hline
\endfoot
\hline
\endlastfoot
\hline
\multicolumn{3}{c}{4054905}            \\
\hline
2639.9411 &   -2.474(28) &    26.00(18) \\
2670.9044 &   -4.058(26) &    26.47(15) \\
2736.8762 &    4.562(28) &    18.46(14) \\
2787.6329 &   18.855(29) &     2.62(19) \\
2817.8751 &   32.108(30) &   -12.81(11) \\
2841.5760 &   38.295(34) &   -17.03(12) \\
2848.6794 &   36.071(27) &   -13.26(16) \\
2850.5840 &   33.567(26) &   -13.05(16) \\
2850.6727 &   35.244(26) &   -11.20(17) \\
3027.9970 &     8.04(52) &    14.68(84) \\
3178.8424 &   -1.292(28) &    23.71(13) \\
3219.7737 &   -4.936(28) &    25.81(11) \\
3444.9214 &    1.591(27) &    21.73(16) \\
3446.4615$^\star$ &   0.8774(34) &  21.2368(34) \\
3450.4808$^\star$ &   -0.2895(34) &  21.2368(34) \\
3454.9107 &   -0.991(26) &    24.79(16) \\
3551.6723 &    3.716(27) &    21.50(12) \\
\hline
\multicolumn{3}{c}{4360072}            \\
\hline
2639.9215 &   67.012(23) &          ... \\
2670.9419 &   68.593(22) &          ... \\
2736.9122 &   68.603(22) &          ... \\
2787.6678 &   67.392(25) &          ... \\
2841.6178 &   65.674(31) &          ... \\
2848.5954 &   62.316(22) &          ... \\
2850.6165 &   64.849(23) &          ... \\
3028.8859 &   55.935(21) &          ... \\
3090.9026 &   53.876(21) &          ... \\
3182.6077 &   50.740(22) &          ... \\
3203.5560 &   49.593(26) &          ... \\
3216.6201 &   49.716(21) &          ... \\
3219.7552 &   48.225(22) &          ... \\
3398.8290 &   48.800(22) &          ... \\
3416.9320 &   50.094(22) &          ... \\
3444.8169 &   49.769(21) &          ... \\
3568.3692$^\star$ &  56.7605(34) &          ... \\
\hline
\multicolumn{3}{c}{4473933}             \\
\hline
2640.8954 &   29.534(56) &          ... \\
2670.8507 &   16.835(56) &          ... \\
2841.6618 &    25.92(16) &          ... \\
2848.7129 &   29.225(57) &          ... \\
2868.5606 &   24.067(56) &          ... \\
2872.6013 &   21.525(57) &          ... \\
3027.8276 &  -16.210(54) &          ... \\
3060.8789 &   29.585(59) &          ... \\
3195.6454 &    8.771(63) &          ... \\
3216.6350 &  -27.886(55) &          ... \\
3217.7197 &  -29.736(57) &          ... \\
3219.7902 &  -34.563(90) &          ... \\
3448.6011$^\star$ &   3.1167(42) &          ... \\
3449.4791$^\star$ &   5.3440(44) &          ... \\
3450.3877$^\star$ &  7.6257(41) & ... \\
3568.4190$^\star$ & 26.6542(41) & ... \\
\hline
\multicolumn{3}{c}{4663623}              \\
\hline
3398.8718 &    14.19(12) &   -30.71(47) \\
3416.9001 &   12.745(58) &   -30.86(39) \\
3444.9399 &    6.997(63) &   -26.68(47) \\
3446.5064$^\star$ &   7.7468(32) &   1.0176(32) \\
3454.9435 &    5.804(64) &   -24.89(43) \\
3549.7307 &  -12.912(74) &    -1.4(2.5) \\
3557.7885 &  -14.781(77) &          ... \\
3559.7878 &  -14.947(67) &     0.78(46) \\
3617.5587 &  -24.004(56) &     7.21(46) \\
\hline
\multicolumn{3}{c}{5193386}              \\
\hline
2639.9662 &  -67.682(70) &    -3.67(24) \\
2850.7793 &  -27.684(75) &   -50.05(17) \\
3028.9068 &  -85.374(66) &    14.14(17) \\
3189.7110 &    3.533(66) &   -91.43(21) \\
3203.7668 &   -43.01(22) &          ... \\
3216.7735 &  -65.991(72) &   -13.41(19) \\
3217.5705 &  -74.129(71) &    -0.52(18) \\
3219.5748 &  -86.034(62) &    16.03(18) \\
3220.5822 &  -86.341(70) &    18.21(19) \\
3398.9234 &   -9.994(63) &   -74.73(21) \\
3416.8781 &  -54.906(71) &   -19.71(16) \\
\hline
\multicolumn{3}{c}{5866138}            \\
\hline
2639.8914 &  -18.388(28) &          ... \\
2670.9688 &  -17.399(26) &          ... \\
2841.7104 &   -42.02(12) &          ... \\
2848.6626 &  -44.002(27) &          ... \\
2850.7166 &  -42.176(27) &          ... \\
3028.9251 &  -18.048(26) &          ... \\
3189.6639 &  -43.789(28) &          ... \\
3203.6149 &  -30.679(30) &          ... \\
3216.5454 &   -25.91(24) &          ... \\
3217.7478 &  -25.621(30) &          ... \\
3398.9026 &  -18.010(28) &          ... \\
3416.9484 &  -17.933(27) &          ... \\
3447.3921$^\star$ & -19.0039(43) &          ... \\
3448.5436$^\star$ & -19.0010(36) &          ... \\
3454.9293 &  -18.582(26) &          ... \\
3568.3620$^\star$ & -23.2137(34) &          ... \\
\hline
\multicolumn{3}{c}{6307537}              \\
\hline
2639.9993 &   -6.294(26) &   -27.75(13) \\
2841.6912 &   25.673(44) &   -63.39(24) \\
2848.7657 &  -12.458(47) &   -24.09(22) \\
3028.9415 &  -26.591(26) &    -5.91(14) \\
3060.9340 &  -43.524(24) &    14.76(13) \\
3182.6513 &  -56.267(26) &    30.67(13) \\
3189.7300 &  -32.230(26) &    -0.68(12) \\
3203.6636 &    4.597(26) &   -41.56(11) \\
3216.6066 &  -51.340(25) &    23.69(13) \\
3217.6647 &  -44.899(28) &    16.53(13) \\
3219.6895 &  -29.593(24) &    -1.83(12) \\
3445.4894$^\star$ & -27.6056(38) &  -5.1037(35) \\
3449.4222$^\star$ & -53.6294(37) &  26.7354(37) \\
3568.3835$^\star$ &  -53.5431(34) & 26.5214(34) \\
\hline
\multicolumn{3}{c}{6757558}              \\
\hline
2639.8539 &  -16.570(11) &          ... \\
2848.6461 &  -6.0567(96) &          ... \\
2850.6995 &  -5.2152(99) &          ... \\
3028.9576 & -17.1852(92) &          ... \\
3182.6856 & -10.1128(90) &          ... \\
3195.6606 &  -9.3797(92) &          ... \\
3216.6995 &  -8.3930(84) &          ... \\
3219.7256 &  -6.9203(95) &          ... \\
3398.8133 & -16.4188(91) &          ... \\
3416.8138 & -16.8464(97) &          ... \\
3446.5357$^\star$  & -16.7122(20) &          ... \\
\hline
\multicolumn{3}{c}{7133286}             \\
\hline
2640.9115 &  -62.651(84) &   -45.36(49) \\
2848.7485 &  -70.447(72) &   -38.01(48) \\
3027.8437 &  -74.521(72) &   -35.45(32) \\
3060.9173 &  -40.101(69) &   -68.84(33) \\
3182.7575 &  -78.431(69) &   -28.90(42) \\
3189.7462 &  -91.840(68) &   -15.26(33) \\
3203.6311 &  -22.981(79) &   -86.75(37) \\
3219.5968 &  -69.679(69) &   -38.18(31) \\
3220.5595 &  -74.694(71) &   -32.76(30) \\
3398.9435 &  -16.480(73) &   -94.01(34) \\
3416.8574 &  -89.530(72) &   -16.39(44) \\
3445.5692$^\star$ & -37.7055(36) & -71.1040(36) \\
\hline
\multicolumn{3}{c}{7293054}              \\
\hline
2640.9293 &  -21.018(11) &          ... \\
2670.9540 &  -23.405(11) &          ... \\
2736.9583 &  -25.230(11) &          ... \\
2787.6844 &  -27.162(18) &          ... \\
2841.6369 &  -31.594(15) &          ... \\
2848.6102 &  -34.627(12) &   -13.66(13) \\
2850.6338 &  -32.986(11) &   -11.02(14) \\
3027.8610 &   -3.172(11) &  -46.650(77) \\
3090.9186 &   -9.315(11) &   -39.55(15) \\
3182.5937 &  -14.884(11) &   -33.69(12) \\
3189.6495 &  -15.744(11) &  -34.209(97) \\
3195.6016 &  -16.260(11) &   -33.91(10) \\
3216.5686 &  -17.170(14) &   -33.41(17) \\
3219.7415 &  -17.758(12) &   -35.07(12) \\
3444.8321 &  -27.831(11) &          ... \\
3445.4274$^\star$ & -27.4945(35) &          ... \\
3446.3874$^\star$ & -27.5233(34) &          ... \\
3446.5996$^\star$ & -27.5093(34) &          ... \\
3542.6564 &  -34.451(11) &    -9.11(12) \\
3549.7466 &  -34.422(12) &    -8.49(11) \\
3557.7730 &  -35.512(11) &    -7.85(11) \\
3559.7715 &  -35.874(12) &    -8.17(11) \\
3568.3228$^\star$ & -37.0890(24) & -10.6922(24) \\
3616.5340 &  -44.920(11) &    -1.02(14) \\
3617.5391 &  -45.190(11) &     0.79(14) \\
3843.8042 &  -14.308(12) &   -34.08(11) \\
3856.9345 &  -15.541(12) &  -34.724(98) \\
4255.6295 &  -39.746(12) & -5.69(15) \\
4264.5895 &  -40.650(12) & -6.04(15) \\
4285.5934 &  -43.428(12) &  0.27(15) \\
4290.5714 &  -45.448(12) & -0.04(15) \\
4291.5834 &  -45.622(12) & 1.15(15) \\
4309.5624 &  -50.544(12) & 6.50(15) \\
4312.5662 &  -51.787(12) & 7.54(15) \\
4319.5615 &  -56.589(12) & 11.44(15) \\
4322.5473 &  -56.708(12) & 15.05(15) \\
4336.5349 &  -50.169(12) & 5.53(15) \\
\hline
\multicolumn{3}{c}{7768447}            \\
\hline
2640.9441 &  -49.256(17) &          ... \\
2817.9294 &  -61.837(20) &          ... \\
2848.6962 &  -76.860(18) &          ... \\
2850.7329 &  -74.748(17) &          ... \\
3027.8761 &  -46.108(17) &          ... \\
3060.8636 &  -58.282(17) &          ... \\
3182.7741 &  -57.015(16) &          ... \\
3203.7440 &  -85.316(21) &          ... \\
3216.7326 &  -76.033(17) &          ... \\
3217.7345 &  -74.636(19) &          ... \\
3220.6397 &  -71.364(18) &          ... \\
3398.9629 &  -45.035(17) &          ... \\
3416.9619 &  -50.547(17) &          ... \\
3448.5590$^\star$ & -85.0634(30) &          ... \\
3450.4275$^\star$ & -85.0187(34) &          ... \\
\hline
\multicolumn{3}{c}{8435232}           \\
\hline
2640.9638 &  -58.817(50) &    13.53(37) \\
2848.7298 &  -47.126(55) &     2.58(23) \\
2868.5793 &    9.214(52) &   -63.34(34) \\
2872.6238 &   -0.841(56) &   -53.87(55) \\
3027.8946 &  -27.263(49) &          ... \\
3060.8960 &    6.159(49) &   -62.64(65) \\
3182.6257 &  -50.342(49) &     2.79(49) \\
3189.7674 &  -60.643(55) &    15.86(74) \\
3216.7517 &    5.905(51) &   -61.04(25) \\
3217.6813 &    5.393(51) &   -58.07(16) \\
3219.6181 &   -0.033(50) &   -53.66(69) \\
3220.6057 &   -3.399(50) &   -50.60(33) \\
3446.5692$^\star$ & -35.3111(34) &          ... \\ 
\hline
\multicolumn{3}{c}{9153621}            \\
\hline
2670.9263 &   47.087(20) &          ... \\
2736.8934 &   36.482(21) &          ... \\
2787.6519 &   30.816(23) &          ... \\
2817.8933 &   26.609(28) &          ... \\
2841.5956 &   26.993(31) &          ... \\
2848.5805 &   23.190(21) &          ... \\
2850.6016 &   25.887(21) &          ... \\
3027.9172 &   37.495(20) &          ... \\
3060.8244 &   32.605(20) &          ... \\
3090.8860 &   30.616(19) &          ... \\
3182.7894 &   24.581(20) &          ... \\
3189.7854 &   24.062(20) &          ... \\
3203.7080 &   26.298(24) &          ... \\
3217.5926 &   31.742(23) &          ... \\
3219.6417 &   34.107(20) &          ... \\
3220.6237 &   34.665(21) &          ... \\
3454.8280 &   27.171(20) &    52.1(1.6) \\
3542.6362 &   74.288(20) &    -3.49(91) \\
3549.7009 &   72.042(21) &    -4.92(54) \\
3450.5511$^\star$ &  26.9340(34) & 47.5210(33) \\
3551.6484 &   70.522(20) &    -3.52(66) \\
3557.7332 &   63.727(21) &     4.36(60) \\
3559.7313 &   61.839(22) &     5.3(1.1) \\
\hline
\multicolumn{3}{c}{9904059}              \\
\hline
2641.0155 &   -4.351(89) &          ... \\
2817.9114 &  -40.027(19) &          ... \\
2848.6284 &  -6.2598(97) &          ... \\
2850.6525 &  -3.0982(97) &          ... \\
3027.9350 & -38.0051(98) &          ... \\
3060.8412 &  -5.3891(99) &          ... \\
3090.9355 & -20.4216(94) &          ... \\
3182.7077 & -14.0052(99) &          ... \\
3195.6860 & -21.7066(99) &          ... \\
3216.6563 & -33.8285(97) &          ... \\
3217.6999 &  -34.064(12) &          ... \\
3398.9792 &  -19.478(11) &          ... \\
\hline
\multicolumn{3}{c}{10015516}             \\
\hline
2670.8859 &  -33.780(28) &          ... \\
2736.8589 &  -37.545(28) &          ... \\
2780.6425 &  -24.117(28) &          ... \\
2787.6152 &  -40.485(28) &          ... \\
2817.8599 &   -1.359(29) &          ... \\
2841.5575 &   -3.135(33) &          ... \\
2848.5457 &  -26.015(29) &          ... \\
2850.5685 &  -29.245(28) &          ... \\
2850.7987 &  -29.321(28) &          ... \\
2872.7479 &  -37.060(30) &          ... \\
3027.9546 &   12.687(28) &          ... \\
3060.8097 &  -45.670(27) &          ... \\
3090.8642 &    4.124(28) &          ... \\
3445.3837$^\star$ &   9.4094(41) &          ... \\
3445.4556$^\star$ &   9.3139(37) &          ... \\
3446.4008$^\star$ &   7.5174(32) &          ... \\
3449.3888$^\star$ &   0.4134(32) &          ... \\
3568.4058$^\star$ &  11.6788(31) &          ... \\
\hline
\multicolumn{3}{c}{11235323}              \\
\hline
2639.8726 &    -6.44(12) &   -19.22(16) \\
2872.6573 &   32.503(68) &   -59.55(11) \\
3028.9765 &   37.677(60) &  -62.053(90) \\
3182.6681 &   11.894(58) &  -33.987(92) \\
3189.6800 &    8.907(58) &  -32.181(97) \\
3195.6233 &  -60.756(59) &   38.890(84) \\
3203.6797 &   26.840(58) &  -51.318(98) \\
3216.5852 &  -56.590(59) &   36.538(90) \\
3217.6130 &  -49.089(64) &    27.99(11) \\
3219.7063 &  -23.940(62) &   -0.573(85) \\
\end{longtable}

\end{document}